\documentclass[12pt]{article}
\pdfoutput=1
\usepackage{graphicx,amssymb,amsmath,empheq,accents}
\usepackage{authblk}
\usepackage{amsthm}
\usepackage{units}
\usepackage{empheq}
\usepackage{booktabs}
\usepackage{subfig} 
\usepackage{bigints}
\usepackage[mathscr]{euscript}
\numberwithin{equation}{section}

\usepackage{hyperref}
\hypersetup{colorlinks = true, linkcolor=black, citecolor=black, urlcolor=blue}
\usepackage{url}
\usepackage{bbm}
\theoremstyle{plain}

\usepackage{tikz}
\usepackage{tikz-cd}
\usetikzlibrary{arrows}
\usetikzlibrary{intersections}
\usetikzlibrary{shapes.geometric}
\usetikzlibrary{decorations.pathmorphing, patterns,shapes}
\usetikzlibrary{decorations.markings}

\pgfdeclarepatternformonly{south west lines}{\pgfqpoint{-0pt}{-0pt}}{\pgfqpoint{3pt}{3pt}}{\pgfqpoint{3pt}{3pt}}{
        \pgfsetlinewidth{0.4pt}
        \pgfpathmoveto{\pgfqpoint{0pt}{0pt}}
        \pgfpathlineto{\pgfqpoint{3pt}{3pt}}
        \pgfpathmoveto{\pgfqpoint{2.8pt}{-.2pt}}
        \pgfpathlineto{\pgfqpoint{3.2pt}{.2pt}}
        \pgfpathmoveto{\pgfqpoint{-.2pt}{2.8pt}}
        \pgfpathlineto{\pgfqpoint{.2pt}{3.2pt}}
        \pgfusepath{stroke}}

\tikzset{
  mid arrow/.style={postaction={decorate,decoration={
        markings,
        mark=at position .575 with {\arrow[#1]{stealth}}
      }}},
  near arrow/.style={postaction={decorate,decoration={
        markings,
        mark=at position .275 with {\arrow[#1]{stealth}}
      }}},
   far arrow/.style={postaction={decorate,decoration={
        markings,
        mark=at position .800 with {\arrow[#1]{stealth}}
      }}},
       midd arrow/.style={postaction={decorate,decoration={
        markings,
        mark=at position .50 with {\arrow[#1]{stealth}}
      }}},
}

\usepackage[nosort]{cite}

\renewcommand{\bar}{\overline}
\renewcommand{\tilde}{\widetilde}
\renewcommand{\hat}{\widehat}
\renewcommand{\leq}{\leqslant}
\renewcommand{\geq}{\geqslant}
\renewcommand{\Re}{\operatorname{Re}}
\renewcommand{\Im}{\operatorname{Im}}

\newcommand{\Tr}{\operatorname{Tr}}

\newcommand{\const}{\operatorname{const}}
\newcommand{\Diff}{\operatorname{Diff}}

\newcommand{\UU}{\operatorname{U}}

\newcommand{\SO}{\operatorname{SO}}
\newcommand{\SL}{\operatorname{SL}}
\newcommand{\tSL}{\widetilde{\SL}}

\newcommand{\PSL}{\operatorname{PSL}}

\newcommand{\RR}{\mathbb{R}}

\newcommand{\ZZ}{\mathbb{Z}}

\newcommand{\calA}{\mathcal{A}}

\newcommand{\calE}{\mathscr{E}}
\newcommand{\calF}{\mathcal{F}}
\newcommand{\calG}{\mathcal{G}}

\newcommand{\calJ}{\mathcal{J}}

\newcommand{\calL}{\mathcal{L}}

\newcommand{\calQ}{\mathcal{Q}}

\newcommand{\calS}{\mathcal{S}}

\newcommand{\rA}{\textrm{A}}
\newcommand{\rS}{\textrm{S}}

\newcommand{\beq}{\begin{equation}}
\newcommand{\eeq}{\end{equation}}
\def\bea{\begin{eqnarray}}
\def\eea{\end{eqnarray}}

\newcommand{\Sch}{\operatorname{Sch}}
\newcommand{\sgn}{\operatorname{sgn}}
\newcommand{\eff}{\text{eff}}

\newcommand{\Res}{\operatorname{Res}}

\newcommand{\sz}{\calS}
\newcommand{\fz}{f}
\newcommand{\gz}{\calG}
\newcommand*{\antisym}[1]{\,\calA\!\left\{#1\right\}}
\newcommand{\one}{\mathbf{1}}
\newcommand{\spooky}{\text{sp}}

\newlength{\fighskip} \fighskip=2pt
\newlength{\figvskip} \figvskip=3pt

\newcommand*{\figbox}[2]{{
  \def\figscale{#1}
  \def\arraystretch{0.8}
  \arraycolsep=0pt
  \begin{array}{c}
    \vbox{\vskip\figscale\figvskip
      \hbox{\hskip\figscale\fighskip
        \includegraphics[scale=\figscale]{#2}}}
  \end{array}}}

\title{
Notes on the complex Sachdev-Ye-Kitaev model
}

\author[1]{Yingfei Gu}
\author[2,3]{Alexei Kitaev}
\author[1]{Subir Sachdev}
\author[1]{Grigory Tarnopolsky}

\affil[1]{\normalsize\it Harvard University, Cambridge, MA 02138, USA}
\affil[2]{\normalsize\it California Institute of Technology, Pasadena, CA 91125, USA}
\affil[3]{\normalsize\it Institute for Advanced Study, Princeton, NJ 08540, USA}

\date{March 22, 2022}

\begin{document}

\maketitle

\begin{abstract}
We describe numerous properties of the Sachdev-Ye-Kitaev model for complex fermions with $N\gg 1$ flavors and a global $\UU(1)$ charge. We provide a general definition of the charge in the $(G,\Sigma)$ formalism, and compute its universal relation to the infrared asymmetry of the Green function. The same relation is obtained by a renormalization theory. 
The conserved charge contributes a compact scalar field to the effective action, from which we derive the many-body density of states and extract the charge compressibility. We compute the latter via three distinct numerical methods and obtain consistent results.  
Finally, we present a two dimensional bulk picture with free Dirac fermions for the zero temperature entropy.


\end{abstract}
\setcounter{tocdepth}{2}

\vspace{30pt}
\tableofcontents

\newpage
\section{Introduction}

The Sachdev-Ye-Kitaev (SYK) models \cite{SaYe93,Kit.KITP.2} of fermions with random interactions have been the focus of much recent attention in both the quantum gravity and the condensed matter literature. The majority of this work has focused on the model with Majorana fermions, which has no globally conserved charge, other than the Hamiltonian itself. In this paper, we direct our attention to the model with $N\gg 1$ complex fermions \cite{SS15}, a.k.a. the complex SYK model: 
\begin{equation}
\hat{H} = \sum_{\substack{j_1<\ldots<j_{q/2},\\ k_1<\ldots<k_{q/2}}} J_{j_1\ldots j_{q/2}\,, k_1 \ldots k_{q/2}} \antisym{\hat{\psi}^\dagger_{j_1} \ldots \hat{\psi}^\dagger_{j_{q/2}}  \hat{\psi}_{k_1} \ldots \hat{\psi}_{k_{q/2}}}
\label{eq: Hamiltonian}
\end{equation}
where $\antisym{\cdots}$ denotes the antisymmetrized product of operators. 
The couplings $J_{j_1\ldots j_{q/2}\,, k_1 \ldots k_{q/2}}$ are independent random complex variables with zero mean and the following variance: 
\begin{equation}
\bar{\left| J_{j_1\ldots j_{q/2}\,, k_1 \ldots k_{q/2}} \right|^2}
= J^2\, \frac{(q/2)!\,(q/2-1)!}{N^{q-1}} \,.
\end{equation} 
One advantage of the antisymmetrized Hamiltonian is that it makes the particle-hole symmetry manifest. 
For example at $q=4$, the Hamiltonian has the following form
\begin{equation}
\qquad \hat{H}= \sum_{j_1<j_{2}\,, k_1<k_{2}} J_{j_1j_{2}, k_1 k_{2} }\Big(\hat{\psi}^\dagger_{j_1}  \hat{\psi}^\dagger_{j_{2}}  \hat{\psi}_{k_1}  \hat{\psi}_{k_{2}}+\hat{C}_{j_1j_2,k_1k_2}
\Big)\,, 
\label{phsymham}
\end{equation}
where $\hat{C}_{j_1j_2,k_1k_2}$ collects various terms arising from anti-commuting fermion operators; more explicitly,
\begin{equation}
\hat{C}_{j_1j_2,k_1k_2} = 
\frac{1}{2}\Bigl(\delta_{j_{1}k_{1}}\hat{\psi}^\dagger_{j_{2}} \hat{\psi}_{k_2}-\delta_{j_{1}k_{2}}\hat{\psi}^\dagger_{j_{2}} \hat{\psi}_{k_1}-\delta_{j_{2}k_{1}}\hat{\psi}^\dagger_{j_{1}} \hat{\psi}_{k_2}+
\delta_{j_{2}k_{2}}\hat{\psi}^\dagger_{j_{1}} \hat{\psi}_{k_1}
+ \frac{1}{2} \delta_{j_1k_2} \delta_{j_2k_1} -\frac{1}{2} \delta_{j_1k_1} \delta_{j_2k_2}\Bigr)\,.
\end{equation}
Without $\hat{C}_{j_1j_2,k_1k_2}$ term, the Hamiltonian is not invariant under the particle-hole symmetry $\hat{\psi}^\dagger_j \leftrightarrow \hat{\psi}_j$. Using the same notation, we define the globally conserved $\UU(1)$ charge $\hat{Q}$  by
\begin{equation}
\hat{Q} =  \sum_{j} \antisym{\hat{\psi}_j^\dagger \hat\psi_j}
=\sum_{j} \hat{\psi}_j^\dagger \hat\psi_j - \frac{N}{2} \,.
\label{chargeoper}
\end{equation}
It is related to the ultraviolet (UV) asymmetry of the Green function
\begin{equation}
G(\tau_1,\tau_2) =  - \langle {\rm T} \hat{\psi}(\tau_1)\hat\psi^\dagger (\tau_2) \rangle\,, \quad G(0^+) = -\frac{1}{2} + \calQ \,, \quad G(0^-) = \frac{1}{2} + \calQ\,,\quad \calQ=\frac{\langle \hat{Q}\rangle }{N} \,.
\label{charge Green function}
\end{equation}
In the infrared (IR), the spectral asymmetry is characterized by the long-time behavior of the Green function 
\begin{equation}
G_{\beta=\infty}(\pm \tau)
\sim \mp e^{\pm \pi \calE}  \tau^{-2\Delta} \qquad \text{for} \quad \text{and} \quad \tau \gg J^{-1}\,,
 \label{GE}
\end{equation}
or equivalently the small frequency behavior
\begin{equation}
G_{\beta=\infty}(\pm i \omega) \sim \mp i e^{\mp i \theta} \omega^{2\Delta -1} \qquad \text{for} \quad \text{and} \quad 0<  \omega \ll J\,,
\end{equation}
where $\beta= T^{-1}$ is the inverse temperature, $\Delta = 1/q$ is the scaling dimension of the fermion operator, $\calE \in (-\infty,+\infty)$ and $\theta\in (-\Delta \pi, \Delta \pi)$ are the spectral asymmetry parameters 
related by the following formula 
\begin{equation}
    e^{2 \pi \calE} = \frac{\sin(\pi \Delta + \theta)}{\sin(\pi \Delta - \theta)}\,.
\label{calEtheta}
\end{equation}
Note the value of $(\calE, \theta)$ can not be fixed by the IR equations; so there is a one-parameter family of solutions in the scaling limit \cite{SaYe93}. 
Ultimately, the actual value of $(\calE,\theta)$ is set by the value of specific charge $\calQ$. Although charge is a UV property of the system, the relationship between $(\calE,\theta)$ and $\calQ$ is universal and independent of UV details \cite{GPS01,SS15}:
\begin{equation}
\calQ = - \frac{\theta}{\pi} - \left( \frac{1}{2} -\Delta \right) \frac{\sin (2 \theta)}{\sin (2\pi \Delta)}\,.
\label{charge theta intro}
\end{equation}
We will provide new derivations of Eq.~(\ref{charge theta intro}) here (see Eqs.~(\ref{charge theta}), (\ref{dQdtheta}) and Section~\ref{sec:bulk}). This universal relation is analogous to the Luttinger relation of Fermi liquid theory, which relates the size of the Fermi surface (an IR quantity) to the total charge (a UV quantity).

The form of Eq.~(\ref{GE}) also applies to fermionic fields with unit $\UU(1)$ charge in asymptotically AdS$_2$ black holes, as was computed by Faulkner et al. \cite{Faulkner09}; the parameter $\calE$ is then a dimensionless measure of the electric field near the AdS$_2$ horizon \cite{SS15} (see Appendix~\ref{app:em} and Eq.~(\ref{defE})). For both SYK models and black holes, fields with $\UU(1)$ charge $p$ have the asymmetry factor $e^{\pm \pi p \calE}$.

Another key feature of the SYK models is the presence of a non-zero entropy in the zero temperature limit \cite{GPS01}:
\begin{equation}
\lim_{\beta \rightarrow \infty}\lim_{N\to\infty} \frac{S(N, N\calQ, \beta^{-1})}{N} = \sz(\calQ) > 0\,,
\label{defS0}
\end{equation}
The function $\sz(\calQ)$ is {\it universal}, in that it is determined only by the structure of the low energy conformal theory, and is independent of the UV perturbations to the Hamiltonian which are irrelevant to the low energy. 
Such a zero temperature entropy is {\it not\/} associated with an exponentially large ground state degeneracy. Instead, it signals an exponentially small many-body energy level spacing  down to the ground state; see Section~\ref{sec:dos}. For each given $N$, the entropy does go to zero at exponentially low temperatures.
We will present a new derivation of $\sz(\calQ)$ in Section~\ref{sec:bulk} using a two dimensional bulk picture involving massive Dirac fermions on the hyperbolic plane.

At finite but sufficiently low temperatures, the dynamics of the Majorana SYK model is governed by a collective mode with the Schwarzian action \cite{Kit.KITP.2,MS16-remarks,KS17-soft}. An analogous effective theory of the complex SYK model also includes a $\UU(1)$ mode \cite{Davison17}
\begin{equation}
I_{\eff} [\varphi, \lambda]
= \frac{NK}{2} \int_0^{\beta} d \tau
\bigl(\lambda'(\tau) + i\calE\varphi'(\tau)\bigr)^2
-  \frac{N\gamma}{4 \pi^2} \int_0^{\beta} d \tau \,
\Sch\left( \tan\frac{\varphi(\tau)}{2},\, \tau\right),
\label{Seff}
\end{equation}
where $\varphi(\tau)$ is a
monotonic time reparameterization obeying $\varphi(\tau + \beta) = \varphi(\tau) + 2\pi$, and $\lambda (\tau)$ is a phase field  obeying $\lambda (\tau + \beta) = \lambda (\tau) + 2 \pi n$ with integer winding number $n$ conjugate to the total charge $Q$.  Notation 
$\Sch ( f(x), x )$ stands for the Schwarzian derivative
\begin{equation}
\Sch ( f(x), x ) := \frac{f'''}{f'} - \frac{3}{2} \left( \frac{f''}{f'} \right)^2 \,.
\end{equation}
In this effective theory, the $\UU(1)$ and $\SL(2,\RR)$ freedom are actually decoupled, which can be demonstrated by the variable change $\lambda(\tau) =\tilde{\lambda}(\tau) +i\calE\bigl(\frac{2\pi}{\beta}\tau-\varphi(\tau)\bigr)$.

The action~\eqref{Seff} is characterized by two parameters, $K$ and $\gamma$, and these can be specified by their connection to thermodynamics. They depend upon the specific charge $\calQ$ (or the chemical potential $\mu$), but this dependence has been left implicit. 
The leading low temperature correction to the entropy in Eq.~(\ref{defS0}) at fixed $N$ and $Q$ is
\begin{equation}
\frac{S}{N} = \sz  + \gamma\, \beta^{-1} + O(\beta^{-2})\,, \label{defgamma}
\end{equation}
and so $\gamma$ is the $T$-linear coefficient of the specific heat at fixed charge, as in Fermi liquid theory. The parameter $K$
is the zero temperature compressibility
\begin{equation}
K = \frac{d\calQ}{d\mu} \quad, \quad \text{at} \quad \beta=\infty\,. \label{defK}
\end{equation}
Unlike $\sz$ and $\calE$, the parameters $K$ and $\gamma$ are not universal, and depend upon details of the microscopic Hamiltonian and not just the low energy conformal field theory. 

The zero temperature entropy in Eq.~(\ref{defS0}) and the pair of soft modes as in Eq.~(\ref{Seff}) are also pertinent to higher dimensional charged black holes with AdS$_2$ horizons, and this is discussed elsewhere \cite{SS10,SS15,JMDS16b,KJ16,HV16,Davison17,Gaikwad:2018dfc,Nayak:2018qej,Moitra:2018jqs,Chaturvedi:2018uov,Sachdev19,Moitra:2019bub}. Key aspects of such black holes are summarized in Appendix~\ref{app:em}. We also note that supersymmetric higher dimensional black holes with AdS$_2$ horizons obtained from string theory have integer values for $e^{N\sz}$ \cite{Dabholkar:2004yr,Dabholkar:2014ema}, as does the SYK model with $\mathcal{N}=2$ supersymmetry \cite{Fu:2016vas} (which we do not consider here).

An important property of both complex SYK and charged black holes with AdS$_2$ horizons is the following relationship between the entropy $\sz(\calQ)$ in Eq.~(\ref{defS0}) and the parameter $\calE$:
\beq
\frac{d \sz(\calQ)}{d\calQ} = 2 \pi \calE\,. \label{dSdQ}
\eeq
This relationship first appeared in the study of SYK-like models by Georges et al. \cite{GPS01}, building upon large $N$ studies of the multichannel Kondo problem \cite{PGKS97}.
Independently, this relationship appeared as a general property of black holes with AdS$_2$ horizons in the work of Sen \cite{Sen05,Sen08}, where $\calE$ is identified with the electric field on the horizon \cite{Faulkner09}, as noted above. It was only later that the identity of this relationship between the SYK and black hole models was recognized \cite{SS15}. We will obtain a deeper understanding of Eq.~(\ref{dSdQ}) in the present paper, based on the global $\UU(1)$ symmetry associated with the conserved charge and the locality of effective action.

Let us summarize our notation for thermodynamic quantities. These quantities are of the order of $N$: the total charge $Q$ (which is integer for $N$ even, and half-integer for $N$ odd), action $I$, entropy $S$, and the associated free energy and grand potential. $N$-independent quantities include: the temperature $T=\beta^{-1}$, chemical potential $\mu$, spectral asymmetry parameter $\calE$, specific charge $\calQ$, zero-temperature entropy $\sz$, charge compressibility $K$, and the $T$-linear coefficient in the specific heat $\gamma$. Except the first two, they are defined in the large $N$ limit.

\subsection{Outline of the paper} 
We begin Section~\ref{sec:charge} by setting up the formalism for the complex SYK model as a path integral over the two-time Green function and self energy. We introduce a definition of the conserved charge $\calQ$ suitable for this formalism and then derive the known universal relation between $\calQ$ and $\calE$ (Eq.~(\ref{charge theta})). 
In Section~\ref{sec:generalform}, we find a general form of a local effective action $I_{\eff}$ and derive Eq.~(\ref{Seff}). 
Section~\ref{sec:thermo} is concerned with thermodynamic quantities and a discussion of what parameters come from the UV. In Section~\ref{sec:dos}, we evaluate the path integral over $\lambda$ and $\varphi$ with action $I_{\eff}$ exactly, which yields new results for the many-body density of states. 

Section~\ref{secRG} sets up a renormalization theory of the complex SYK model. This will enable us to obtain another derivation of the relationship between the specific charge $\calQ$ and the spectral asymmetry $\calE$.

In Section~\ref{sec:compress}, we turn to the calculation of the parameters of the effective action, in particular charge compressibility $K$. We present three numerical computations that yield values of $K$ in good agreement with each other. These computations and our analysis show that all energy scales contribute to the charge compressibility. A low energy analysis based on linear coupling, mentioned in Section~\ref{sec:thermo}, or conformal perturbation theory (see Appendix~\ref{app:GrishaK}) does not yield the correct value of $K$, even though such methods work \cite{MS16-remarks,KS17-soft} for the Schwarzian mode.

Section~\ref{sec:bulk} presents a two dimensional bulk derivation of the zero temperature entropy of the complex SYK model. We show that the $\calE$-dependent value of the zero temperature entropy per fermion $\sz$ can be obtained from a Euclidean path integral over massive Dirac fermions on hyperbolic plane ${\rm H^2}$. We show that the appropriate quantity is the ratio of fermionic determinants with different boundary conditions on the boundary of ${\rm H^2}$. Another bulk interpretation of the entropy appears in Appendix~\ref{app:em}, where we recall the connection to higher-dimensional black holes. In $d+2$ dimensions ($d \geq 2$), the AdS$_2$ arises as a factor in the near-horizon geometry of a near-extremal charged black hole. In this picture, $\sz$ is related to the horizon area in the $d$ extra dimensions, and, as we noted above, this $\sz$ also obeys the differential relation~(\ref{dSdQ}).

\section{Low temperature properties} 
\label{sec:charge}

In this section, we analyze the complex SYK model based on the $(G,\Sigma)$ action. We provide a general definition of charge in this framework and prove its universal relation to the IR asymmetry of the Green function. Furthermore, we find the general form of  effective action and evaluate the path integral over the low energy fluctuations, which yield new results for the many-body density of states.

\subsection{Preliminaries}
We start with a review of the basics. 
For convenience, we measure time in units of $J^{-1}$, which is equivalent to setting $J$ to 1. For the Hamiltonian \eqref{eq: Hamiltonian}, we may consider either the partition function for a fixed charge or the grand partition function. The latter can be obtained from the $(G,\Sigma)$ action:
\begin{equation}
\begin{aligned}
\frac{I}{N} = - \ln \det \left( - \sigma - \Sigma \right) &- \int d\tau_1 d\tau_2 \left[ 
\Sigma(\tau_1,\tau_2) G(\tau_2,\tau_1) + \frac{1}{q} \left(- G(\tau_1,\tau_2) G(\tau_2,\tau_1) \right)^{\frac{q}{2}}
\right] \,, \\
&\text{where} \quad \sigma(\tau_1,\tau_2) = \delta' (\tau_1-\tau_2) - \mu \delta (\tau_1-\tau_2) \,.
\end{aligned}
\label{Gsigma action}
\end{equation}
The Schwinger-Dyson equations are as follows: 
\begin{equation}
-\underbrace{(\Sigma+\sigma)}_{\tilde{\Sigma}} G = 1, \quad \Sigma(\tau_1,\tau_2) = G(\tau_1,\tau_2)^{\frac{q}{2}} (-G(\tau_2,\tau_1))^{\frac{q}{2}-1} \,.  \label{SchwDyseq}
\end{equation}
The general idea of solving these equations in the IR limit is to ignore $\sigma$, which is localized at short times. However, care should be taken because the Fourier transform of $\sigma$ contains the non-negligible, $\omega$-independent term $-\mu$. Fortunately, this term is absent from $\tilde{\Sigma}$, so we will use $G$ and $\tilde{\Sigma}$ as independent variables. Thus, $\sigma$ moves to the second equation in~\eqref{SchwDyseq}, where it can be safely ignored as the equation is solved in the time representation.

Since the IR equations do not depend on $\mu$, we get a family of solutions parametrized by a formally independent variable $\calE$. At zero temperature,  
\begin{equation}
\begin{gathered}
G_{\beta=\infty}(\pm \tau)
= \mp e^{\pm \pi \calE} b^{\Delta} \tau^{-2\Delta}\,, \quad
\tilde{\Sigma}_{\beta=\infty}(\pm \tau) 
= \mp e^{\pm \pi \calE} b^{1-\Delta} \tau^{-2 (1-\Delta)}
\qquad \text{for} \quad \tau \gg 1 \\
\text{where} \qquad b = \frac{1-2\Delta}{2\pi} \cdot \frac{\sin (2\pi \Delta)}{2 \cos \pi (\Delta+i\calE ) \cos \pi (\Delta-i \calE) } \,.
\end{gathered}
\label{zeroT Green}
\end{equation}
We can also introduce a parameter $\theta$ to characterize the spectral asymmetry in the frequency domain:
\begin{equation}
\begin{aligned}
G_{\beta=\infty}(\pm i \omega) &= \mp i e^{\mp i \theta} \sqrt{ 
\frac{\Gamma(2-2\Delta)}{\Gamma(2\Delta)}} b^{\Delta - \frac{1}{2}} \omega^{2\Delta -1} \\
\tilde{\Sigma}_{\beta=\infty}(\pm i \omega) &= \mp i e^{\pm i \theta} \sqrt{ 
\frac{\Gamma(2\Delta)}{\Gamma(2-2\Delta)}} b^{ \frac{1}{2}- \Delta } \omega^{1-2\Delta }
\end{aligned} \qquad  \text{for} \quad 0 < \omega \ll 1 \,.
\label{zeroT Green frequency}
\end{equation}
The spectral asymmetry parameters $\calE$ and $\theta$ are related by the equations
\begin{equation}
e^{-2i\theta} =
\frac{\cos(\pi(\Delta+i\calE))}{\cos(\pi(\Delta-i\calE))}\,,\qquad
  e^{2 \pi \calE} = \frac{\sin(\pi \Delta + \theta)}{\sin(\pi \Delta - \theta)}\,.
\label{calEtheta2}
\end{equation}
Using these relations, we can also express the prefactor $b$ as a function of $\theta$:
\begin{equation}
    b= \frac{1-2\Delta}{2\pi } \cdot \frac{2 \sin (\pi \Delta + \theta) \sin (\pi \Delta - \theta)}{\sin (2\pi \Delta)} \,.
\end{equation}
The zero-temperature solutions can be extended to finite temperature:
\begin{equation}
\begin{aligned}
G(\tau) &\approx G_c(\tau) = - b^{\Delta } \left( \frac{\beta}{\pi} \sin \frac{\pi \tau}{\beta} \right)^{-2\Delta} e^{2\pi \calE \left( \frac{1}{2}-\frac{\tau}{\beta}  \right)}  \\ 
\tilde{\Sigma}(\tau) &\approx \tilde{\Sigma}_c(\tau) = - b^{1-\Delta } \left( \frac{\beta}{\pi} \sin \frac{\pi \tau}{\beta} \right)^{-2(1-\Delta)} e^{2\pi \calE \left( \frac{1}{2}-\frac{\tau}{\beta}  \right)} 
\end{aligned}
\qquad \begin{array}{c}\text{for} \quad 0<\tau<\beta,\\[5pt]
\tau \gg 1\,\, \text{and} \,\, \beta -\tau \gg 1 \,.
\end{array}
\label{Gconf}
\end{equation} 
The subscript $c$ here means ``conformal''. 
In the frequency domain with Matsubara frequencies $ \omega_n= \frac{2\pi}{\beta} \left( n + \frac{1}{2} \right)  \ll 1$, the Green function and self energy have the following form,
\begin{equation}
\begin{aligned}
G(\pm i \omega_n) &\approx G_c(\pm i \omega_n) = \mp i e^{\mp i \theta} \sqrt{ 
\frac{\Gamma(2-2\Delta)}{\Gamma(2\Delta)}}  \left( \frac{2\pi}{\beta} \sqrt{b} \right)^{2\Delta - 1} \frac{\Gamma\left( n + \frac{1}{2} + \Delta \pm i \calE  \right)}{\Gamma
\left( 
n + \frac{3}{2} - \Delta \pm i \calE
\right)
 }\,,  \\
\tilde{\Sigma}(\pm i \omega_n) & \approx\tilde{\Sigma}_c(\pm i \omega_n) = \mp i e^{\pm i \theta} \sqrt{ 
\frac{\Gamma(2\Delta)}{\Gamma(2-2\Delta)}} \left( \frac{2\pi}{\beta} \sqrt{b}\right)^{1-2\Delta } 
\frac{\Gamma\left( n + \frac{3}{2} - \Delta \pm i \calE  \right)}{\Gamma
\left( 
n + \frac{1}{2} + \Delta \pm i \calE
\right)
 }\,.
\end{aligned} 
\end{equation}

Given these exact solutions to the IR equations, it remains to be checked whether they extrapolate at higher energies to a solution to the full UV equations which depend upon $\mu$. This has been examined numerically \cite{SaYe93,GPS99,GPS01,Fu:2016yrv,Azeyanagi2018}, and a consistent extrapolation exists for $|\mu| \lesssim 0.24$ \cite{Azeyanagi2018,planckian19}. The IR parameter $\calE$ can be determined as a smooth, odd function of the UV parameter $\mu$ over this regime.

In addition to the emergent reparameterization symmetry that is present in the low energy limit of the Majorana SYK model, the complex SYK model has an extra emergent symmetry related to phase fluctuation:
\begin{equation}
\begin{aligned}
    G(\tau_1,\tau_2) &\rightarrow \varphi' (\tau_1)^{\Delta} \varphi' (\tau_2)^{\Delta} e^{i (\lambda(\tau_1)-\lambda(\tau_2))} G (\varphi(\tau_1),\varphi(\tau_2)) \\
    \tilde{\Sigma}(\tau_1,\tau_2) &\rightarrow \varphi' (\tau_1)^{1-\Delta} \varphi' (\tau_2)^{1-\Delta} e^{i (\lambda(\tau_1)-\lambda(\tau_2))} \tilde{\Sigma} (\varphi(\tau_1),\varphi(\tau_2)) \\
\end{aligned}\,,
\label{symmetry}
\end{equation}
where $\varphi(\tau)$ is a monotonic time reparameterization with winding number $1$ and $\lambda(\tau)$ is a phase fluctuation with possibly arbitrary integer winding number. The symmetries are not exact in the presence of $\sigma$ term in the $(G,\Sigma)$ action \eqref{Gsigma action}. To make this point more transparent, it is useful to rewrite the action in terms of $(G,\tilde{\Sigma})$: 
\begin{equation}
\begin{aligned}
\frac{I}{N} = -\ln \det \bigl( - \tilde{ \Sigma} \bigr) - &\int d\tau_1 d\tau_2 \left[ \tilde{\Sigma}(\tau_1,\tau_2)  G(\tau_2,\tau_1) + \frac{1}{q} \left(  - G(\tau_1,\tau_2) G(\tau_2,\tau_1) \right)^{\frac{q}{2}} \right]  \\
+& \int d\tau_1 d\tau_2~ \sigma(\tau_1,\tau_2) G(\tau_2,\tau_1)
\,.
\end{aligned}
\label{GSigma tilde action}
\end{equation}
Now the first line of the r.h.s. of Eq.~\eqref{GSigma tilde action} is invariant under the symmetry transformation \eqref{symmetry}, while the second line changes. This point will be further discussed in Section~\ref{sec:generalform}.

\subsection{Charge}
\label{section: charge}

For an explicit UV source field $\sigma$ (cf.\ Eq.~\eqref{Gsigma action}) that arises from a microscopic Hamiltonian, the charge is conventionally defined by the UV asymmetry of the Green function as Eq.~\eqref{charge Green function}. In this section we will derive a formula for charge in $(G,\tilde{\Sigma})$ action framework for general source field $\sigma$ (without assuming time translation symmetry) using ideas similar to those in Appendix C of Ref.~\cite{kitaev2006anyons}. 

\subsubsection{``Flow'' of Green function $G$}
 Let us consider the action \eqref{GSigma tilde action} 
 with 
 $\sigma(\tau_1,\tau_2)$ depending on both times, not just on 
 $\tau=\tau_1-\tau_2$. If $(G,\tilde{\Sigma})$ is stationary (i.e. satisfies the Schwinger-Dyson equations) and $\beta=\infty$, then
\begin{equation}
\int_{-\infty}^{+\infty} \left( \sigma (\tau_1,\tau_0) G(\tau_0,\tau_1) - \sigma(\tau_0,\tau_1) G(\tau_1,\tau_0)   \right) d\tau_1 =0 \,.
\label{conserved current}
\end{equation} 
This can be established by considering an infinitesimal variation $\delta\lambda(\tau)$ and the corresponding variations
\begin{equation}
\begin{aligned}
\delta G(\tau_1,\tau_2) &= i \left(\delta\lambda (\tau_1)-\delta\lambda (\tau_2) \right) G(\tau_1,\tau_2) \,, \\ 
\delta \tilde\Sigma(\tau_1,\tau_2) &= i \left( \delta\lambda (\tau_1)-\delta\lambda (\tau_2) \right) \tilde\Sigma(\tau_1,\tau_2) \,.
\end{aligned}
\end{equation}
Only the $\sigma G$ term in \eqref{GSigma tilde action} has a non-trivial variation, which is proportional to the l.h.s.\ of \eqref{conserved current} if $\delta\lambda(\tau)\propto\delta(\tau-\tau_0)$. On the other hand, the variation of the action must be zero since $(G,\tilde{\Sigma})$ is stationary.

Following the ideas in Appendix C of Ref.~\cite{kitaev2006anyons}, we may call
\begin{equation}
  j(\tau_1,\tau_2)=\sigma(\tau_1,\tau_2)G(\tau_2,\tau_1) - \sigma(\tau_2,\tau_1)G(\tau_1,\tau_2)  : \quad
  \begin{tikzpicture}[scale=1, baseline={([yshift=-4pt]current bounding box.center)}]
\draw [thick, ->,>=stealth] (30pt,0pt) -- (120pt,0pt) ;
\filldraw  (50pt,0pt)  circle (1pt) node[below] {$\tau_1$} ;
\filldraw  (100pt,0pt)  circle (1pt) node[below] {$\tau_2$} ;
\draw[mid arrow] (50pt,0pt) .. controls (60pt,20pt) and (90pt,20pt).. (100pt,0pt);
\end{tikzpicture}
\label{eq: current0}
\end{equation}
the ``current'' flowing from $\tau_1$ to $\tau_2$. 
To make a closer analogy to the aforementioned reference, let us substitute the expression $\sigma(\tau_1,\tau_2) = \tilde{\Sigma}(\tau_1,\tau_2) - G(\tau_1,\tau_2)^{\frac{q}{2}}(- G(\tau_2,\tau_1))^{\frac{q}{2}-1}$ (obtained from the Schwinger-Dyson equations) into the current formula:
\begin{equation}
     j(\tau_1,\tau_2)=\tilde{\Sigma}(\tau_1,\tau_2)G(\tau_2,\tau_1) - \tilde{\Sigma}(\tau_2,\tau_1)G(\tau_1,\tau_2)\,.
     \label{eq: current of G}
\end{equation}
Treating $G$ and $\tilde{\Sigma}$ as matrices indexed by $(\tau_1,\tau_2)$, we have $\tilde{\Sigma} = -G^{-1}$. If $G$ were a unitary quasidiagonal matrix, the results in Appendix C of Ref.~\cite{kitaev2006anyons} would apply, and certain quantities would be quantized. However, here Green function $G$ has non-trivial IR asymptotics violating the conditions of being quasidiagonal. Nevertheless, we will use similar ideas and definitions as the aforementioned reference.

\begin{figure}[t]
\center
\subfloat[Conservation law]{
\begin{tikzpicture}[scale=1.2, baseline={(current bounding box.center)}]
\draw [->,>=stealth,thick] (-20pt,0pt) -- (120pt,0pt) ;
\draw [white, dashed] (80pt,-3pt) -- (80pt,30pt); 
\filldraw  (50pt,0pt)  circle (1pt);
\draw[near arrow] (50pt,0pt) .. controls (60pt,20pt) and (90pt,20pt).. (100pt,0pt);
\draw[near arrow] (50pt,0pt) .. controls (60pt,10pt) and (70pt,10pt).. (80pt,0pt);
\draw[far arrow] (20pt,0pt) .. controls (30pt,10pt) and (40pt,10pt).. (50pt,0pt);
\draw[far arrow] (0pt,0pt) .. controls (10pt,20pt) and (40pt,20pt).. (50pt,0pt);
\draw [thick, blue, densely dashed]  (50pt,0pt)  circle (6pt);
\node[below] at (50pt,-5pt) {$\tau_0$};
\end{tikzpicture}}
\hspace{20pt}
\subfloat[Total current through a cross section]{
\begin{tikzpicture}[scale=1.2, baseline={(current bounding box.center)}]
\draw [->,>=stealth, thick] (-10pt,0pt) -- (170pt,0pt) ;
\draw[thick, mid arrow] (110pt,0pt) .. controls (120pt,10pt) and (130pt,10pt).. (140pt,0pt);
\draw[thick, mid arrow] (90pt,0pt) .. controls (100pt,10pt) and (110pt,10pt).. (120pt,0pt);
\draw[thick, mid arrow] (70pt,0pt) .. controls (80pt,10pt) and (90pt,10pt).. (100pt,0pt);
\draw[thick, mid arrow] (50pt,0pt) .. controls (60pt,10pt) and (70pt,10pt).. (80pt,0pt);
\draw[thick, mid arrow] (30pt,0pt) .. controls (40pt,10pt) and (50pt,10pt).. (60pt,0pt);
\draw [thick, blue, dashed] (80pt,-5pt) -- (80pt,30pt); 
\draw[red, mid arrow] (95pt,0pt) .. controls (115pt,20pt) and (135pt,20pt).. (155pt,0pt);
\draw[red, mid arrow] (65pt,0pt) .. controls (85pt,20pt) and (105pt,20pt).. (125pt,0pt);
\draw[red, mid arrow] (35pt,0pt) .. controls (55pt,20pt) and (75pt,20pt).. (95pt,0pt);
\draw[red, mid arrow] (5pt,0pt) .. controls (25pt,20pt) and (45pt,20pt).. (65pt,0pt);
\node[below] at (80pt,-5pt) {$\tau_0$};
\end{tikzpicture}}
\caption{(a) Conservation law: the total current through a closed (dashed) circle is zero; (b) Flow $\calQ$, as the total current through a cross section $\tau_0$ (blue dashed line), is independent of the position $\tau_0$. In general, there are contributions from all time scales; longer scale currents are shown in red.}
\label{fig: definition of Q}
\end{figure}
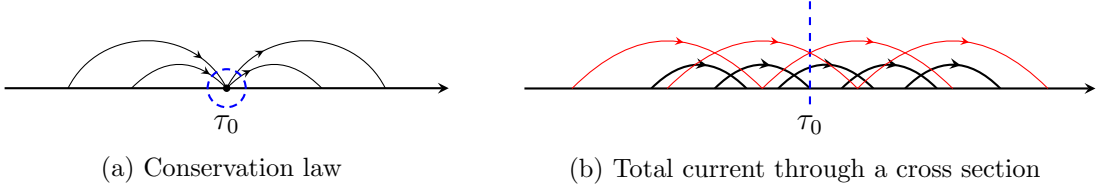

Note that Eq.~\eqref{conserved current} can be interpreted as the conservation of the current at each point $\tau_0$:
\begin{equation}
    \int_{-\infty}^{+\infty}  j(\tau_1,\tau_0)d\tau_1 = 0,
\end{equation}
as illustrated in Fig.~\ref{fig: definition of Q}\,(a). It follows that the total current through a cross section $\tau_0$,
\begin{equation}
    \calQ = \int_{-\infty}^{\tau_0} d\tau_1 \int_{\tau_0}^{+\infty} d\tau_2 ~j(\tau_1,\tau_2) \,,
\label{eq: charge0}
\end{equation}
(see Fig.~\ref{fig: definition of Q}\,(b)) is independent of $\tau_0$. As explained below, this quantity is a natural generalization of the specific charge $Q/N$ to general sources. We may call $\calQ$ the ``flow'' of the matrix $G$ as it depends solely on $G$ through Eq.~\eqref{eq: current of G} with $\tilde{\Sigma}=-G^{-1}$. We also remark that the definition of the flow does not rely on the time translation symmetry. That is, the source $\sigma(\tau_1,\tau_2)$, and the Green function $G(\tau_1,\tau_2)$, may depend on both $\tau_1$ and $\tau_2$ rather than just $\tau=\tau_1-\tau_2$.

We now explain the interpretation of the flow $\calQ$ as charge. Plugging the definition \eqref{eq: current0} of the current $j$ into Eq.~\eqref{eq: charge0}, we get
\begin{equation}
\calQ = \int_{-\infty}^{\tau_0} d\tau_1 \int_{\tau_0}^{+\infty} d\tau_2 \bigl( 
\sigma (\tau_1,\tau_2) G(\tau_2,\tau_1) - \sigma(\tau_2,\tau_1) G(\tau_1,\tau_2)
\bigr)\,.
\label{eq: defQ 2 var}
\end{equation}
This formula reduces to a simpler form when the source has the time translation symmetry, i.e.\ for $\sigma(\tau_1,\tau_2)=\sigma(\tau)$, where $\tau=\tau_1-\tau_2$:
\begin{equation}
\calQ = \int_{0}^{+\infty} d\tau \int_{0}^{\tau} d\tau_2 \bigl(
\sigma (-\tau) G(\tau) - \sigma(\tau) G(-\tau)
\bigr)=-\int_{-\infty}^{+\infty} d\tau~ \tau \sigma(\tau) G(-\tau) \,.
\label{defQ}
\end{equation}
The last expression in turn reduces to the conventional definition of the charge when $\sigma(\tau) = \delta'(\tau) - \mu \delta (\tau)$. In this case, for the Green function $G(\tau)$ that is discontinuous at $\tau=0$, we use the average $\frac{1}{2}(G(0^+)+G(0^-))$ to define its value at $\tau=0$. Thus,
\begin{equation}
\calQ = - \int_{-\infty}^{+\infty} d\tau~\tau \delta'(\tau) G(-\tau)  = \int_{-\infty}^{+\infty} d\tau \delta(\tau) G(-\tau) = \frac{ G(0^+)+G(0^-)}{2} \,,
\end{equation}
in agreement with Eq.~\eqref{charge Green function}.
For extremely local UV sources such as $\delta'(\tau)$ and $\delta(\tau)$, the charge is a local quantity. However, if we consider a general source $\sigma(\tau)$, the r.h.s. of Eq.~\eqref{defQ} includes contributions from all scales; see Fig.~\ref{fig: definition of Q}\,(b) for a cartoon.  

\subsubsection{Invariance of the charge}
We will show that the charge $\calQ$ depends only on the UV and 
IR asymptotics of $G(\tau_1,\tau_2)$ and $\tilde{\Sigma}(\tau_1,\tau_2)$ (where $\tilde{\Sigma} = - G^{-1}$) as well as some topological data. The UV asymptotics is determined by the $\delta'(\tau_1-\tau_2)$ term in $\tilde{\Sigma}$. To formulate the invariance, let $(G_1,\tilde{\Sigma}_1)$ and $(G_2, \tilde{\Sigma}_2)$ have the same asymptotics and in addition, let the following ``relative winding number'' be zero:
\begin{equation}
\nu (G_1,G_2) = \frac{1}{2\pi i} \int_0^{+\infty} \frac{d}{d\omega} \left( 
\ln \frac{G_1(i\omega)}{G_2(i\omega)}
\right) d\omega \,.
\end{equation}
If $\nu(G_1,G_2)=0$, then $(G_1,\tilde{\Sigma}_1)$ can be continuously deformed into $(G_2,\tilde{\Sigma}_2)$. Here it is important to consider the winding number in frequency domain rather than time domain, because the Schwinger-Dyson equation $ \tilde{\Sigma}(\omega) = - G(\omega)^{-1}$ guarantees that a smooth path in $(G,\tilde{\Sigma})$ space will disallow both singularities and zeros of $G(\omega)$. This will not work for $G(\tau)$, since the other equation ${\Sigma}(\tau)= G(\tau)^{\frac{q}{2}} (-G(-\tau))^{\frac{q}{2}-1}$ does not constrain  zeros of $G(\tau)$.

To prove that the charge is invariant under such deformation, it is sufficient to consider infinitesimal, asymptotically trivial deformations. Let us use the formula
\begin{equation}
\calQ = \int_{-\infty}^{+\infty} d\tau_1 \int_{-\infty}^{+\infty} d\tau_2\, (f(\tau_2)-f(\tau_1)) \sigma(\tau_1,\tau_2) G(\tau_2,\tau_1)\,,
\end{equation}
where $f$ is an arbitrary function such that 
\begin{equation}
    \lim_{\tau\rightarrow +\infty} f(\tau) =1 \,, \quad \text{and} \quad
    \lim_{\tau \rightarrow -\infty} f(\tau)=0 \, :\qquad 
    \begin{tikzpicture}[scale=0.9,baseline={([yshift=-4pt]current bounding box.center)}]
    \draw [->,>=stealth] (50pt,0pt) -- (50pt,30pt) node[right]{$f$};
    \draw[->,>=stealth] (0pt,-1pt) -- (110pt,-1pt) node[right]{$\tau$};
    \draw[thick, blue] (0pt,0pt)--(40pt,0pt) .. controls (50pt,0pt) and (50pt,10pt) .. (60pt,10pt) -- (90pt,10pt) ;
    \draw [dashed] (30pt,10pt) -- (100pt,10pt) node[right]{$1$};
    \end{tikzpicture}
\end{equation}
This formula coincides with Eq.~\eqref{eq: defQ 2 var} for the step function $f(\tau) = \theta(\tau-\tau_0)$. 
The integral does not depend on the details of $f$ because of the conservation law, namely Eq.~\eqref{conserved current}.\footnote{More explicitly, if we vary $f$ without changing its asymptotics, the corresponding variation of the charge is proportional to the l.h.s. of the Eq.~\eqref{conserved current} and therefore vanishes.} More intuitively,  $f$ may be interpreted as a linear combination of step functions $\theta(\tau-\tau_0)$ with some weights for each cross section position $\tau_0$. In other words, we can blur the cross section, and this will not affect the flow. 

We proceed by anti-symmetrizing the integrand,
\begin{equation}
\calQ= \frac{1}{2} \int d\tau_1 d\tau_2\, (f(\tau_2)-f(\tau_1)) \left( 
\sigma(\tau_1,\tau_2) G(\tau_2,\tau_1) - \sigma (\tau_2,\tau_1) G(\tau_1,\tau_2)
\right)\,.
\end{equation} 
Since 
$\sigma(\tau_1,\tau_2)= \tilde\Sigma(\tau_1,\tau_2)- G(\tau_1,\tau_2)^{\frac{q}{2}} (- G(\tau_2,\tau_1))^{\frac{q}{2}-1}$, we also get this equation:
\begin{equation}
\calQ = \frac{1}{2} \int d\tau_1 d\tau_2\, (f(\tau_2)-f(\tau_1)) \left(
\tilde{ \Sigma}(\tau_1,\tau_2) G(\tau_2,\tau_1) - \tilde{\Sigma}(\tau_2,\tau_1) G(\tau_1,\tau_2)
\right)\,.
\end{equation}
Note that the two terms cannot be integrated separately because the corresponding integrals are not absolutely convergent (since $G(\tau_1,\tau_2) \sim |\tau_1-\tau_2|^{-2\Delta}$, $\Sigma\sim |\tau_1-\tau_2|^{-2+2\Delta}$, there is a logarithmic divergence in IR). 
Now, consider an infinitesimal variation $\delta G$ and
$\delta \tilde{\Sigma} = \tilde{\Sigma} \left( \delta G \right) \tilde{\Sigma}$ such that 
$\delta G(\tau_1,\tau_2)$ and $\delta \tilde{\Sigma}(\tau_1,\tau_2)$ decay sufficiently fast as $\tau_1-\tau_2 \rightarrow \infty$:
\begin{equation}
\begin{aligned}
\delta\calQ & = \int d\tau_1 d\tau_2 \left( f(\tau_2)-f(\tau_1)\right) 
\delta \left(
\tilde{\Sigma}(\tau_1,\tau_2) G(\tau_2,\tau_1) 
\right)  \\
&= \int d\tau_1 d\tau_2 \left( f(\tau_2)-f(\tau_1)\right) G(\tau_2,\tau_1) 
\delta 
\tilde{\Sigma}(\tau_1,\tau_2)   \\&+ \int d\tau_3 d\tau_4 \left( f(\tau_3)-f(\tau_4)\right) 
\tilde{\Sigma}(\tau_4,\tau_3) \delta G(\tau_3,\tau_4)  \,.
    \end{aligned}
\end{equation}
Substituting $\delta \tilde{\Sigma} = \tilde{\Sigma} \left( \delta G \right) \tilde{\Sigma}$ into the first line of the last expression and using 
$\tilde{\Sigma}(\tau_4,\tau_3)
= - \int d\tau_2 d\tau_1 \tilde{\Sigma}(\tau_4,\tau_2) G(\tau_2,\tau_1) \tilde{\Sigma}(\tau_1,\tau_3) $ in the second line, we get
\begin{equation}
    \delta\calQ = \int d^4\tau \left(
f(\tau_2) - f(\tau_3)+ f(\tau_4)-f(\tau_1)
 \right)
 \tilde{\Sigma} (\tau_4,\tau_2) G(\tau_2,\tau_1) \tilde{\Sigma} (\tau_1,\tau_3) \delta G(\tau_3,\tau_4)\,.
\end{equation}
Now, we can regroup and integrate the terms containing $f(\tau_2)-f(\tau_3)$ and $f(\tau_4)-f(\tau_1)$ separately:
\begin{equation}
\begin{aligned}
    &\int d^3\tau \left(
f(\tau_2) - f(\tau_3)
 \right) \int d\tau_1~
 \tilde{\Sigma} (\tau_4,\tau_2) G(\tau_2,\tau_1) \tilde{\Sigma} (\tau_1,\tau_3) \delta G(\tau_3,\tau_4) = 0 \,, \\
 & \int d^3\tau \left(
f(\tau_4) - f(\tau_1)
 \right) \int d\tau_2~
 \tilde{\Sigma} (\tau_4,\tau_2) G(\tau_2,\tau_1) \tilde{\Sigma} (\tau_1,\tau_3) \delta G(\tau_3,\tau_4) = 0 \,.
\end{aligned}
\end{equation}
Each integral contains a delta function that annihilates $f(\tau_2)-f(\tau_3)$ in the first case and $f(\tau_4)-f(\tau_1)$ in the second case. Therefore, we conclude that $\delta \calQ=0$ for asymptotically and topologically trivial deformations.

\subsubsection{Calculation of the charge}
In fact, we can calculate the charge in the complex SYK model using the IR asymptotics. (The result for $q=4$ was originally derived in Ref.~\cite{GPS01} using a different method, see Appendix~\ref{app:GPS} for a detailed discussion.) We will use the antisymmetrized version of Eq.~\eqref{defQ} to express $\sigma$ in terms of $\tilde{\Sigma}$ as we have done before:
\begin{equation}
\begin{aligned}
\calQ& = - \int_{-\infty}^{+\infty} \tau \sigma(\tau) G(-\tau) d\tau = - \frac{1}{2} \int_{-\infty}^{+\infty} \tau \bigl( \sigma(\tau) G(-\tau) - \sigma(-\tau) G(\tau) \bigr) d\tau \\
&= - \frac{1}{2} \int_{-\infty}^{+\infty} \tau \left( \tilde{\Sigma} (\tau) G(-\tau) - \tilde{\Sigma} (-\tau) G(\tau )\right) d\tau \,.
\label{charge formula}
\end{aligned}
\end{equation}
The two terms in the last expression almost cancel each other at $\tau \gg 1$, but individually, the corresponding integrals are logarithmically divergent. To proceed, let us replace $G(\tau)$ with 
\begin{equation}
G_{\eta}(\tau) = \begin{cases}
G(\tau) & \text{for} \quad |\tau| \lesssim 1 \\
G(\tau) |\tau|^{-2\eta} & \text{for} \quad |\tau| \gg 1
\end{cases}\,,
\label{G_eta}
\end{equation}
where $\eta$ is a small positive number. This change has little effect on the integrand in \eqref{charge formula}, but the two terms can now be separated. The corresponding integrals are equal to each other due to the symmetry $\tau\to -\tau$. Thus,
\begin{equation}
\begin{aligned}
\calQ  = \lim_{\eta \rightarrow 0} \left[ - \int_{-\infty}^{+\infty} \tau \tilde{\Sigma}(\tau) G_{\eta}(-\tau) d\tau \right]= \lim_{\eta \rightarrow 0} \left[ \frac{1}{2\pi i} \int_{-\infty}^{+\infty} \left( \partial_\omega G(i\omega)^{-1}\right) G_{\eta}(i\omega) d\omega
\right] \,.
\end{aligned}
\label{charge integral}
\end{equation}
It is important that the symmetric-in-time regularization \eqref{G_eta} is not symmetric in frequency, which has nontrivial consequences. The Fourier transform of $G_{\eta}$ is 
\begin{equation}
\begin{pmatrix}
G_{\eta}(i \omega) \\
G_{\eta} (-i\omega)
\end{pmatrix} 
= \Gamma(1-2\Delta') 
\begin{pmatrix}
i^{1-2\Delta'} & i^{-1+2\Delta'} \\ 
i^{-1+2\Delta'} & i^{1-2\Delta'}
\end{pmatrix}
\begin{pmatrix}
- b^\Delta e^{\pi \calE} \\
b^{\Delta} e^{-\pi \calE}
\end{pmatrix} \omega^{-1+2\Delta'}\,, \quad 0 < \omega \ll 1 \,,
\end{equation}
where $\Delta' = \Delta+\eta$.
 Expanding the $\omega$-independent coefficients in this expression to the first order in $\eta$, we explicitly see the asymmetry:
 \begin{equation}
 G_{\eta} (\pm i\omega) \approx \omega^{2\eta} G(\pm i \omega) \left[
 1+ \eta \bigl(
 -2\psi(1-2\Delta) - \pi \tan \pi (\Delta \pm i\calE)
 \bigr)
 \right]\,,
 \end{equation}
 where $\psi(x)=\Gamma'(x)/\Gamma(x)$ is the digamma function. 
Now, we are in a position to evaluate the integral over $d\omega$ in \eqref{charge integral}, which should be understood as a principal value integral. For frequencies above some threshold $\omega_0$ (such that $\omega_0 \ll 1$ but $\eta \ln \frac{1}{\omega_0} \ll 1$), the difference between $G$ and $G_{\eta}$ may be neglected. On the other hand, if $|\omega|<\omega_0$, then $G(i\omega)\sim |\omega|^{-1+2\Delta}$, and hence, $\partial_\omega G(i\omega)^{-1} \approx (1-2\Delta) \omega^{-1} G(i\omega)^{-1}$. Using these approximations, we get
\begin{equation}
\begin{aligned}
\int_{-\infty}^{+\infty} &\left( \partial_\omega G(i\omega)^{-1}\right) G_\eta(i\omega) d\omega \\
& \approx \int_{\omega_0}^{+\infty} \partial_\omega \ln \frac{G(-i\omega)}{G(i\omega)} ~d\omega    + (1-2\Delta)\int_0^{\omega_0} \omega^{-1} \left(
\frac{G_{\eta}(i\omega)}{G(i\omega)} - \frac{G_{\eta}(-i\omega)}{G(-i\omega)}
 \right) d\omega \\
 & \approx -2i \theta + (1-2\Delta) \eta \bigl(
-\pi \tan \pi (\Delta+i\calE) + \pi \tan \pi (\Delta-i\calE) 
  \bigr) \int_0^{\omega_0} \omega^{-1+2\eta} d\omega \,.
\end{aligned}
\label{anomalous phase shift}
\end{equation} 
The last integral is simply ${1}/({2\eta})$, so the result is independent of $\eta$. Including the factor ${1}/({2\pi i})$ from \eqref{charge integral} we obtain:
\begin{equation}
\begin{aligned}
\calQ &= - \frac{\theta}{\pi} - \frac{1-2\Delta}{4i }\bigl(
\tan \pi (\Delta+i\calE) - \tan \pi (\Delta -i \calE)
\bigr) \\
& = - \frac{\theta}{\pi} - \left( \frac{1}{2} -\Delta \right) \frac{\sin (2 \theta)}{\sin (2\pi \Delta)}\,,
\end{aligned}
\label{charge theta}
\end{equation}
which reproduces the result in Ref.~\cite{Davison17}. 

\subsubsection{Analogy with field-theoretic anomalies} 
In some sense, the calculation of the charge performed here (also see Appendix~\ref{app:GPS} for parallel discussions based on symmetric-in-frequency regularizations) has a flavor of perturbative anomalies in quantum field theory. Both describe a mismatch between the IR and UV. In the case of anomaly, there is no consistent UV cutoff respecting the symmetry; in our case, the UV behavior is well-defined but quantifiably different from the IR. By analogy with the Fermi liquid theory, one might expect the charge to be given by the first term in Eq.~\eqref{charge theta}. However, that is not correct due to the non-trivial effect of regularization, which produces the second term, 
\begin{equation}
    -\left( \frac{1}{2}-\Delta \right)
    \frac{\sin (2\theta)}{\sin (2\pi \Delta)} \,.
\end{equation}
In Appendix~\ref{app:GPS}, we will further comment on this term and relate it to the Luttinger-Ward term in the standard analysis\cite{luttinger1960ground}.

\subsection{Covariant formalism and the effective action}
\label{sec:generalform}

At the low temperatures, $\beta \gg 1$, the action~\eqref{GSigma tilde action} is almost invariant under the emergent symmetry \eqref{symmetry}. In other words, we can generate ``quasi-solutions'' of the Schwinger-Dyson equations by applying $(\varphi,\lambda)$ transformations to the actual solution $(G_*,\tilde{\Sigma}_*)$:
\begin{equation}
\begin{aligned}
    G(\tau_1,\tau_2)& = \varphi' (\tau_1)^{\Delta} \varphi' (\tau_2)^{\Delta} e^{i (\lambda(\tau_1)-\lambda(\tau_2))} G_*(\varphi(\tau_1),\varphi(\tau_2))\,, \\
    \tilde{\Sigma}(\tau_1,\tau_2) &= \varphi' (\tau_1)^{1-\Delta} \varphi' (\tau_2)^{1-\Delta} e^{i (\lambda(\tau_1)-\lambda(\tau_2))} \tilde{\Sigma}_* (\varphi(\tau_1),\varphi(\tau_2)) \,.
\end{aligned}
\label{approximate solutions}
\end{equation}
Such quasi-solutions cost a small increase in action,\footnote{Eq.~\eqref{approximate solutions} defines the IR part of a quasi-solution, while the UV part should be tuned to minimize the cost.} which we call the ``effective action''. More exactly, $I_{\eff}[\varphi,\lambda]= I[G,\tilde{\Sigma}]-\const$, where the constant depends on convention: we may set it to $I[G_*,\tilde{\Sigma}_*]$ or use a different base value. The goal of this section is to determine the general form of $I_{\eff}[\varphi,\lambda]$ to leading orders.

\subsubsection{Covariant formalism} 
The form of the approximate solutions coincides with the transformation laws for functions changing from one frame to another. The latter is described by a diffeomorphism of the time circle together with a gauge transformation. For instance, a field $\psi(x)$ with scaling dimension $\Delta$ and charge $p$ defined in frame ``$x$'' can be transformed to frame $(y,\phi)$ by the following formula:
\begin{equation}
   \psi_{y,\phi}(y)= \left(\frac{dy}{dx}\right)^{-\Delta}  e^{-ip \phi(x)} \psi(x)\,. \label{covtrans}
\end{equation}
It is also straightforward to define the transformation laws for $G$ and $\tilde{\Sigma}$, i.e. functions of two variables. 
Taking this viewpoint, the ``quasi-solution'' \eqref{approximate solutions} may be interpreted as follows. In order to generate a quasi-solution $(G,\tilde{\Sigma})$ in the physical frame $x=\tau$, we start with the frame $(y,\phi)=(\varphi,\lambda)$, where the Green function is given by $G_*$. Then we pull back to the physical frame using the inverse transformation of \eqref{covtrans}, namely
\begin{equation}
    \psi(x) =  \left(\frac{dy}{dx}\right)^{\Delta}  e^{ip \phi(x)} \psi_{y,\phi}(y(x))\,.
\end{equation}
From this perspective, the effective action $I_{\eff}[\varphi,\lambda]$ in fact measures the failure of $(\varphi,\lambda)$ to be the physical frame. 

At first glance, introducing the notion of ``frame'' in this problem seems redundant because in the end we should write all results in the physical frame. However, the condition that the action is invariant under frame transformations (if we also transform $\sigma$) is helpful in determining the general form of the effective action $I_{\eff}[\varphi,\lambda]$.

Now, let us consider expressing the $(G,\tilde{\Sigma})$ action in a general frame $(\varphi,\lambda)$. (In this setting, $G$ and $\tilde{\Sigma}$ are arbitrary and not related to $\varphi$ and $\lambda$ in any particular way.) In the new frame, the fields are defined as follows:
\begin{equation}
    \begin{aligned}
    G_{\varphi,\lambda}(\varphi_1,\varphi_2)& := \varphi' (\tau_1)^{-\Delta} \varphi' (\tau_2)^{-\Delta} e^{-i (\lambda(\tau_1)-\lambda(\tau_2))} G (\tau_1, \tau_2)\,, \\
    \tilde{\Sigma}_{\varphi,\lambda}(\varphi_1,\varphi_2) &:= \varphi' (\tau_1)^{-1+\Delta} \varphi' (\tau_2)^{-1+\Delta} e^{-i (\lambda(\tau_1)-\lambda(\tau_2))} \tilde{\Sigma} (\tau_1,\tau_2) \,,\\
    \sigma_{\varphi,\lambda}(\varphi_1,\varphi_2) &:= \varphi' (\tau_1)^{-1+\Delta} \varphi' (\tau_2)^{-1+\Delta} e^{-i (\lambda(\tau_1)-\lambda(\tau_2))} \sigma (\tau_1,\tau_2) \,.
\end{aligned}
\end{equation}
Representing $G$, $\tilde{\Sigma}$, $\sigma$ in terms of $G_{\varphi,\lambda}$, $\tilde{\Sigma}_{\varphi,\lambda}$, $\sigma_{\varphi,\lambda}$ and plugging into Eq.~\eqref{GSigma tilde action}, we get
\begin{equation}
    \begin{aligned}
\frac{I}{N} = -\ln \det \bigl( - \tilde{ \Sigma}_{\varphi,\lambda} \bigr) - &\int d\varphi_1 d\varphi_2 \left[ \tilde{\Sigma}_{\varphi,\lambda}(\varphi_1,\varphi_2)  G_{\varphi,\lambda}(\varphi_2,\varphi_1) + \frac{1}{q} \left(  - G_{\varphi,\lambda}(\varphi_1,\varphi_2) G_{\varphi,\lambda}(\varphi_2,\varphi_1) \right)^{\frac{q}{2}} \right]  \\
+& \int d\varphi_1 d\varphi_2~ \sigma_{\varphi,\lambda}(\varphi_1,\varphi_2) G_{\varphi,\lambda}(\varphi_2,\varphi_1)
\,. 
\end{aligned}
\label{action new frame}
\end{equation}
Naively, the $\ln\det$ term transforms nontrivially under $\varphi$. However, the determinant needs some UV regularization anyway, and we will use a particular regularization to make it frame-independent:\footnote{The regularized determinant depends on UV details of $\tilde{\Sigma}$. This issue is not important for the present discussion, but it can be mitigated by the use of a different regularization \cite{KS17-soft}: $\det (-\tilde{\Sigma}) \rightarrow \det(-\tilde{\Sigma})/\det(-\tilde{\Sigma}_*)$.}
\begin{equation}
\det (-\tilde{\Sigma}) \rightarrow \frac{\det (-\tilde{\Sigma})}{\det (-\sigma)} \cdot 2 \cosh \frac{\beta \mu}{2} \,.
\label{eq: regdet}
\end{equation}
The second factor is the free fermion partition function (formally equal to $\det(-\sigma)$).

The expression \eqref{action new frame} for the $(G,\tilde{\Sigma})$ action in frame $(\varphi,\lambda)$ has the same general form as in the physical frame, but the UV source in different:
\begin{equation}
\begin{aligned}
\sigma_{\varphi,\lambda}(\varphi_1,\varphi_2)&:=  \varphi'(\tau_1)^{\Delta-1} \varphi'(\tau_2)^{\Delta-1} e^{-i (\lambda(\tau_1)-\lambda(\tau_2))}(\delta'(\tau_1-\tau_2)-\mu(\tau_1) \delta(\tau_1-\tau_2))
\\ &=\varphi'(\tau_1)^{\Delta}
\varphi'(\tau_2)^{\Delta} \left( \delta' (\varphi_1-\varphi_2) - \varphi'(\tau_1)^{-1} \mu_{\lambda} (\varphi_1) \delta (\varphi_1-\varphi_2)
\right)\,,
\end{aligned}
\label{source}
\end{equation}
where 
\begin{equation}
\mu_{\lambda}(\varphi) = \mu(\tau) - i \frac{d\lambda(\tau)}{d\tau}\,.
\label{chemical potential}
\end{equation}
We will make a few comments on this transformation.

\begin{itemize}
\item First of all, the non-trivial change of the source lifts the degeneracy of quasi-solutions and induces the effective action $I_{\eff}[\varphi,\lambda]$, which can be explained as follows. If   $\sigma_{\varphi,\lambda}$ were given by the same expression as the standard source, $\sigma_{\text{std}}(\varphi_1,\varphi_2) =\delta'(\varphi_1-\varphi_2)-\mu \delta(\varphi_1-\varphi_2)$, then $(G_{\varphi,\lambda},\tilde{\Sigma}_{\varphi,\lambda}) =(G_*,\tilde{\Sigma}_*)$ would be a stationary point of the action \eqref{action new frame} in frame $(\varphi,\lambda)$, and therefore, its pullback $(G,\tilde{\Sigma})$ would satisfy the Schwinger-Dyson equations in the physical frame. Thus, the actual distinction between solutions and quasi-solutions can be attributed to the difference between $\sigma_{\varphi,\lambda}$ and $\sigma_{\text{std}}$. In the first approximation, the effective action is obtained by integrating $\sigma_{\varphi,\lambda}-\sigma_{\text{std}}$ against $G_*$.

\item Following Ref.~\cite{KS17-soft}, let us define a field
$\varepsilon_{\varphi}(\varphi) = \varphi'(\tau)$ which sets the length scale for the frame $\varphi$. Using this notation, we have
\begin{equation}
    \sigma_{\varphi, \lambda} (\varphi_1,\varphi_2) = \varepsilon_\varphi(\varphi_1)^{\Delta} \varepsilon_\varphi(\varphi_2)^{\Delta} \left( \delta' (\varphi_1-\varphi_2) - \varepsilon_\varphi(\varphi_1)^{-1} \mu_{\lambda} (\varphi_1) \delta (\varphi_1-\varphi_2)
\right)\,. 
\label{length scale}
\end{equation}
Let us try to understand the meaning of the powers of $\varepsilon$. We will use this terminology: a field that scales as $[\text{length}]^{-\alpha}$ and transforms in the corresponding way under diffeomorphisms is said to have dimension $\alpha$ and called an ``$\alpha$-form''. Thus, $\varepsilon$ has dimension $-1$ and $\mu$ has dimension $0$ (because its transformation law \eqref{chemical potential} does not contain $d\varphi/d\tau$). The function $\sigma(\varphi_1,\varphi_2)$ transforms as a $(1-\Delta,1-\Delta)$ form; as such, it is comparable with $\varepsilon_\varphi(\varphi_1)^{\Delta-1} \varepsilon_\varphi(\varphi_2)^{\Delta-1}\sim \varepsilon^{-2(1-\Delta)}$. The actual powers of $\varepsilon$ in Eq.~\eqref{length scale} may be written as $\varepsilon^{h-2(1-\Delta)}$, where $h$ is associated with the remaining factor, that is, $\delta'(\varphi_1-\varphi_2)$ or $\mu_{\lambda} (\varphi_1) \delta(\varphi_1-\varphi_2)$. In fact, $\delta'(\varphi_1-\varphi_2)$ is diffeomorphism-invariant if treated as a $(1,1)$ form, i.e.\ its total dimension is $h=2$, whereas $\delta(\varphi_1-\varphi_2)$ corresponds to $h=1$. So everything is consistent. In a conventional field theory, the source \eqref{length scale} would be represented by the term
\begin{equation}
\int(\varepsilon\Phi+\mu\Psi)\,d\varphi
\end{equation}
in the action, where the fields $\Phi=\Phi(\varphi)$ and $\Psi=\Psi(\varphi)$ have dimensions $2$ and $1$, respectively. (The exponent of $\varepsilon$ in \eqref{length scale} differs by $2\Delta-1$ because in the $(G,\tilde{\Sigma})$ action, $\sigma$ is multiplied by $G$ and integrated over two variables rather than one.)

\item The expression \eqref{chemical potential} for the chemical potential in the $(\varphi,\lambda)$ frame may be interpreted as {\bf gauge invariance}. It sets a non-trivial constraint on the effective action; namely, the dependence on the soft mode $\lambda$ is tied to its dependence on $\mu$:
\begin{equation}
    i \frac{\delta I_{\eff}[\varphi,\lambda]}{\delta \lambda'} = \frac{\delta I_{\eff}[\varphi,\lambda]}{\delta \mu}\,.
\end{equation}
\end{itemize}

\subsubsection{Diffeomorphism-invariant effective action}
Now we are ready to determine the general form of the effective action. 
Let us consider the following quasi-solution:
\begin{equation}
    G(\tau_1,\tau_2) = \varphi' (\tau_1)^{\Delta} \varphi' (\tau_2)^{\Delta} e^{i (\lambda(\tau_1)-\lambda(\tau_2))} G_*(\varphi(\tau_1),\varphi(\tau_2))\,,
    \label{fLambda quasisolution}
\end{equation}
where $\varphi$ maps the time circle to the standard circle of length $2\pi$ (i.e.\ $\varphi(\tau+\beta)=\varphi(\tau)+2\pi$) and $G_*$ is the IR solution of the Schwinger-Dyson equations with $\beta$ formally set to $2\pi$:
\begin{equation}
\qquad G_*(\varphi_1,\varphi_2)
= - b^{\Delta} \left(2 \sin \frac{\varphi_1-\varphi_2}{2} \right)^{-2\Delta} e^{\calE (\pi-\varphi_1+\varphi_2)},\qquad
0<\varphi_1-\varphi_2<2\pi.
\end{equation}
To separate the $\UU(1)$ and $\SL(2,\RR)$ degrees of freedom, let $\lambda(\tau) =\tilde{\lambda}(\tau) +i\calE\bigl(\frac{2\pi}{\beta}\tau-\varphi(\tau)\bigr)$ so that
\begin{equation}
G(\tau_1,\tau_2)=-b^{\Delta}
e^{i(\tilde{\lambda}(\tau_1)-\tilde{\lambda}(\tau_2))}
\Biggl(\frac{\sqrt{\varphi'(\tau_1)\varphi'(\tau_2)}}
{2\sin\frac{\varphi(\tau_1)-\varphi(\tau_2)}{2}}\Biggr)^{2\Delta}
e^{2\pi\calE\left(\frac{1}{2}-\frac{\tau_1-\tau_2}{\beta}\right)}\,.
\label{qsol1}
\end{equation}

In general, the effective action contains local and non-local terms.\footnote{For a non-local correction to the Schwarzian effective action in the Majorana SYK model, see Ref.~\cite{KS17-soft}. The non-local correction is subleading in the $\beta^{-1}$ expansion and will not be studied in this paper.} The local part describes interactions between the UV sources $\varepsilon$, $\mu$ and some IR data. We have discussed the origin of the fields $\varepsilon$, $\mu$ in last section; now let us search for the IR fields by the intermediate asymptotic expression of $G$ for $1\ll |\tau_1-\tau_2|\ll \beta$:
\begin{equation}
    G(\tau_1,\tau_2) \approx G_{\beta=\infty} (\tau_1,\tau_2) \bigl(1+ A(\tau_+)(\tau_1-\tau_2) + B(\tau_+) (\tau_1-\tau_2)^2+\ldots \bigr)\,, 
    \label{eq: IR expansion}
\end{equation}
where $\tau_+ = \frac{\tau_1+\tau_2}{2}$, and the coefficients $A$, $B$ are obtained by Taylor expanding the quasi-solution \eqref{qsol1} w.r.t. small time separation $\tau_1-\tau_2$. These coefficient have the following explicit form:
\begin{equation}
\label{ABtau}
A(\tau) =  i\tilde{\lambda}'(\tau)-\frac{2\pi}{\beta}\calE\,, \qquad
B(\tau) = \frac{\Delta}{6} \Sch\bigl(e^{i \varphi(\tau)},\tau \bigr) + \frac{1}{2} A(\tau)^2 \,.
\end{equation}
Thus, all relevant IR information is contained in the fields
\begin{equation}
A(\tau) = i\tilde{\lambda}'(\tau)-\frac{2\pi}{\beta}\calE \,,\qquad
O(\tau) = \Sch \bigl(e^{i  \varphi(\tau)},\tau\bigr) \,.
\end{equation}

The local part of the action should have a covariant expression in an arbitrary frame $x$. We aim to find the effective action to $\beta^{-1}$ order. In other words, the action should have the accuracy to recover the free energy or grand potential to $T^2$ order and capture, for example, the linear dependence of the specific heat. With the UV source $(\varepsilon,\mu)$ and IR data $(A,O)$, the most general diffeomorphism- and gauge-invariant action is
\begin{equation}
    \frac{I_{\eff}[\calE, \varphi,\tilde{\lambda}]}{N}
= \int \varepsilon_x^{-1} \fz(\mu-A)\, dx - \gz(\calE)
- \alpha_S \int \left( \varepsilon_x O_x -\rho_x \right) dx \,.
    \label{diffeoinv action}
\end{equation}
Let us discuss some of its features as well as defining the field $\rho$.

\begin{itemize}
    \item In addition to the fluctuating fields $\varphi$ and $\lambda$, the action depends on the global parameter $\calE$. Its equilibrium value will be determined by finding an extremum (actually, a maximum) of $I_{\eff}[\calE, \varphi,\tilde{\lambda}]$ in $\calE$ with fixed external parameter $\beta$, $\mu$.
    
\item The function $\fz$ is a general even function characterizing the charge response to the chemical potential. The gauge invariance \eqref{chemical potential} requires that any dependence on $\mu$ be through the combination $\mu-A$. The $\gz(\calE)$ term is of order $1$ and related to the zero temperature entropy. 

    \item We have expressed the action in a general frame $x$ to emphasize its covariant properties. In particular, we have used the notation $O_x$ and introduced a field $\rho_x$ (see \cite{KS17-soft}, Eq.~(167)):
    \begin{equation}
    O_x(x) = \Sch(e^{i\varphi(\tau(x))},x) \,, \quad
        \rho_x = \frac{(\partial_x \varepsilon_x)^2}{2\varepsilon_x} - \partial_x^2 \varepsilon_x\,  
    \end{equation}
    to form an invariant expression $\int \left( \varepsilon_x O_x -\rho_x \right) dx$. (Recall that $\varepsilon_x = x'(\tau)$ is the field setting the local length scale, where $\tau$ is the physical frame.)
    To show the diffeomorphism invariance, it is essential to check the transformation laws of the local operators $O,\rho,\varepsilon$. The transformation law of $O$ is given by the chain rule of the Schwarzian derivative:
    \begin{equation}
    O_y(y) = \left(y'(x) \right)^{-2} \left( O_x(x)- \Sch(y,x) \right)\,.
    \end{equation}
    This can be further summarized in a matrix form. In fact, the pair $(1, O)$ forms a representation of $\Diff(S^1)$, 
    \begin{equation}
        \begin{pmatrix}
         1 \\
         O_y
        \end{pmatrix} = \begin{pmatrix}
        1 & 0 \\
        -y'(x)^{-2} \Sch(y,x) & y'(x)^{-2}
        \end{pmatrix}
        \begin{pmatrix}
        1 \\
        O_x
        \end{pmatrix}\,.
    \end{equation}
    Similarly, the pair $(\varepsilon,\rho)$ also forms a representation,
    \begin{equation}
        \begin{pmatrix}
        \varepsilon_y \\
        \rho_y
        \end{pmatrix} = 
        y'(x)
        \begin{pmatrix}
        1 &  0 \\
        -y'(x)^{-2} \Sch(y,x) &  y'(x)^{-2}
        \end{pmatrix}
        \begin{pmatrix}
        \varepsilon_x \\
        \rho_x 
        \end{pmatrix} \,.
    \end{equation}
Thus, the following combination transform as a $1$-form
\begin{equation}
    \varepsilon_y O_y - \rho_y = y'(x)^{-1} (\varepsilon_x O_x - \rho_x)\,,
\end{equation}
which further implies the diffeomorphism invariance of the action~\eqref{diffeoinv action}.

The form of the $\rho$ field may look obscure; however, it will naturally arise 
when we try to transform the Schwarzian action from the physical frame $\tau$ to a general frame $x(\tau)$ using the chain rule and express everything in the new frame
\begin{equation}
    O_\tau(\tau) = \varepsilon_x^2 O_x (x) + \Sch(x,\tau) = \varepsilon_x^2 O_x(x) - \varepsilon_x \rho_x(x)\,.
\end{equation}
In other words, in the physical frame $\varepsilon=1$,\, $\rho=0$,  and the last term in the effective action \eqref{diffeoinv action} is just the familiar Schwarzian action
$
    -\alpha_S \int \Sch(x,\tau) d\tau 
$.
\end{itemize}

Now, let us restrict to the physical frame $x=\tau$. Expanding $\fz(\mu-A)$ to the second order in $A=i\tilde{\lambda}'-\frac{2\pi}{\beta}\calE$, we get
\begin{equation}
\begin{aligned}
\frac{I_{\eff} [\calE,\varphi,\tilde{\lambda}]}{N} &= \beta \fz(\mu) + 2\pi(\calE-in) \fz'(\mu) - \gz(\calE)\\
&+  \int \left[
-\frac{\fz''(\mu)}{2} \left( \tilde{\lambda}'(\tau) + i\frac{2\pi}{\beta}\calE\right)^2 - \alpha_S \Sch(e^{i\varphi(\tau)},\tau ) 
 \right] d\tau \,,
\end{aligned}
\label{full effective action}
\end{equation}
where $n$ is the winding number of the function $\tilde{\lambda}$, defined by the generalized periodicity condition $\tilde{\lambda}(\tau+\beta) =\tilde{\lambda}(\tau)+2\pi n$. The second line in the above equation is equivalent to the action \eqref{Seff}, originally derived in Ref.~\cite{Davison17}, with
\begin{equation}
K = - \fz'' (\mu)  \,, \qquad \gamma = 4 \pi^2 \alpha_S \,,
\end{equation}
To further simplify the effective action, let
\begin{equation}
\tilde{\lambda}(\tau)=\bar{\lambda}(\tau)+\frac{2\pi n}{\beta}\tau \,.
\end{equation}
where $\bar{\lambda}$ has zero winding number. Then
\begin{equation}
\frac{I_{\eff} [\calE,n,\varphi,\bar{\lambda}]}{N}
= \beta \fz\Bigl(\mu+\frac{2\pi}{\beta}(\calE-in)\Bigr) - \gz(\calE)
+  \int \left[
-\frac{\fz''(\mu)}{2} \,\bar{\lambda\kern1pt}'^2
- \alpha_S \Sch(e^{i\varphi},\tau ) \right] d\tau \,.
\label{effact1}
\end{equation}

\subsection{Thermodynamics}
\label{sec:thermo}

We now use the effective action $I_{\eff}[\calE,n,\varphi,\bar{\lambda}]$ to compute the low temperature expansion of the grand potential $\Omega(\beta^{-1},\mu)$. If $N$ is large, we may use the saddle point field configurations, $\varphi(\tau)=\frac{2\pi}{\beta}\tau$,\, $\bar{\lambda}(\tau)=0$,\, $n=0$, and find the extremum $I_*$ of $I_{\eff}[\calE, \varphi,\lambda]$ with respect to $\calE$:
\begin{equation}
\frac{\Omega}{N} = \frac{I_*}{\beta N}  = \fz(\mu_0) - \gz(\calE)\beta^{-1} - 2\pi^2 \alpha_S \beta^{-2}
\label{grandpotential}
\end{equation}
where
\begin{equation}
\mu_0=\mu+ \frac{2\pi}{\beta} \calE \,.
\end{equation}
The saddle point condition for $\calE$ requires that
\begin{equation}
\gz'(\calE) = 2\pi \fz'(\mu_0) = -2\pi\calQ.
\label{g and Q}
\end{equation}
where $\calQ= -N^{-1}{\partial \Omega(\beta^{-1},\mu)}/{\partial \mu} = - \fz'(\mu_0) $. Eq.~\eqref{g and Q} implies that $\calQ$ is a function of $\calE$. This function has been calculated in Eq.~\eqref{charge theta}, so one can compute $\gz(\calE)$ as well.

\subsubsection{Free energy and entropy}
 We can also find the free energy $F(\beta^{-1},Q)= \Omega + \mu Q$:
\begin{equation}
\frac{F}{N} = \calF_0(\calQ) - \sz(\calQ)\beta^{-1} - 2\pi^2 \alpha_S \beta^{-2} \,,
\end{equation}
where $\sz= \sz(\calQ)$ is the ``zero temperature entropy'' per site and
\begin{alignat}{3}
\calF_0(\calQ) &= \fz(\mu_0) + \mu_0\calQ \qquad
&&\text{for} \quad
&\fz'(\mu_0)&= -\calQ \,,
\displaybreak[0]\label{Legendre0}\\[3pt]
\sz(\calQ) &= \gz(\calE)+ 2\pi \calE \calQ \qquad
&&\text{for} \quad
&\gz'(\calE)&= - 2\pi\calQ \,.
\label{Legendre}
\end{alignat}
The formula \eqref{Legendre} says that $\sz(\calQ)$ and $\gz(\calE)$ are related by the Legendre transformation, where $\calQ$ and $2\pi\calE$ are the conjugate variables. It leads to the fundamental relation (\ref{dSdQ}) between the entropy and the particle-hole asymmetry. Equation \eqref{Legendre0} is the usual Legendre duality between the free energy and the grand potential at zero temperature; it implies that $\mu_0=d\calF_0(\calQ)/d\calQ$. At finite temperature, the chemical potential receives a definite correction:
\begin{equation}
    \mu(\beta^{-1},\calQ) = \mu_0(\calQ) - \frac{2\pi}{\beta} \calE(\calQ).
    \label{eq: rule}
\end{equation}

Similar relations hold for the low energy limit of charged black holes \cite{Faulkner09,Sachdev19}, as reviewed in Appendix~\ref{app:em}. The low $T$ limit must be taken at fixed $Q$ with $\mu$ obeying (\ref{dmudt}), to obtain a near-horizon metric that is conformally equivalent to AdS$_2$.

\subsubsection{Charge compressibility}
\label{sec: RG charge compressibility}
As discussed in Refs.~\cite{MS16-remarks, KS17-soft}, the specific heat is determined by the prefactor $\alpha_S$ of the Schwarzian action, which is related to the magnitude $\alpha_G$ of the leading UV-sourced correction to the IR Green function. Specifically,
\begin{equation}
    \alpha_S = \frac{-k_c'(2)(1-\Delta)b}{6}\alpha_G \,,
\end{equation}
where $k_c'(2)$, $\Delta$ and $b$ are all IR data that can be obtained in the conformal limit. 

Now, for the complex SYK model we have one more thermodynamic coefficient to determine, namely the charge compressibility $K$. A natural question is whether the charge compressibility can be determined in a similar way by the same UV parameter $\alpha_G$. This possibility is based on the observation that the IR degrees of freedom $A(\tau)$, $B(\tau)$ in Eqs.~\eqref{eq: IR expansion}, \eqref{ABtau} satisfy the relation
\begin{equation}
B(\tau) = \frac{\Delta}{6} \Sch\bigl(e^{i \varphi(\tau)},\tau \bigr) + \frac{1}{2}A(\tau)^2,
\label{BA relation}
\end{equation}
which might constrain the form of the effective action. Or is the charge compressibility independent of $\alpha_S$ and requires a separate numerical study?

To answer this question, let us think about possible couplings between the IR degrees of freedom and some UV data. The idea of renormalization theory, as used in Ref.~\cite{KS17-soft}, is not to solve the actual problem in the UV (which is hard) but to replace it with a more tractable model with sufficiently many parameters that would reproduce the leading IR behavior and any possible corrections to it. The simplest term to include in the $(G,\tilde{\Sigma})$ action is the linear coupling $\int d\tau_1 d\tau_2\, \sigma(\tau_1,\tau_2) G(\tau_2,\tau_1)$ of the UV source to the Green function, where the latter is represented by the asymptotic expression \eqref{eq: IR expansion} at intermediate time intervals $\tau_1-\tau_2$ with coefficients $A\bigl(\frac{\tau_1+\tau_2}{2}\bigr)$ and $B\bigl(\frac{\tau_1+\tau_2}{2}\bigr)$. By smearing the actual, very singular source, nonlinear effects can be reduced. In this approximation, the effective action is a sum of terms proportional to $\int A(\tau)\, d\tau$ and $\int B(\tau)\, d\tau$. Any contribution of the form $\int A(\tau)^2\, d\tau$ is due to the second term via Eq.~\eqref{BA relation}. But on the other hand,
\begin{equation}
\frac{I_{\eff}[\calE, \varphi,\lambda]}{N} = \int \fz(\mu-A)\, d\tau - \gz(\calE)- \alpha_S \int\Sch\bigl(e^{i \varphi(\tau)},\tau \bigr)\,d\tau \,,
\end{equation}
see Eq.~\eqref{diffeoinv action}. Therefore, the linear model predicts the following value of the charge compressibility $K=-\fz''(\mu)$:
\begin{equation}
    K_{\rm linear}= \frac{6 \alpha_S}{\Delta} = \frac{3}{2\pi^2 \Delta} \gamma \,. \label{Kgamma relation}
\end{equation}

However, a nonlinear coupling of the form\footnote{In contrast, we won't worry about non-linear contributions to the specific heat because they are subleading in temperature.} $\int s(\tau_1,\tau_2,\tau_3,\tau_4) G(\tau_1,\tau_2)  G(\tau_3,\tau_4)\, d^4\tau$ can also generate a term proportional to $\int A(\tau)^2\, d\tau$. Let us denote this additional contribution by $K_{\rm non-linear}$ so that
\begin{equation}
    K = K_{\rm linear}+ K_{\rm non-linear}. 
\end{equation}
Its actual value is not accessible without numerics. In Section~\ref{sec:compress} we will present numerical computations for the total $K$ at half filling, namely $\mu=0$ and $\calQ=0$. 

We would like to make a final remark on the ratio $K_{\rm linear}/\gamma$ in (\ref{Kgamma relation}). It agrees with Eq.~(\ref{Kovergamma}) obtained from a different analysis following the perturbation theory done in Ref.~\cite{MS16-remarks}. This analysis relies on the UV parameter $\alpha_G$, see Appendix~\ref{app:GrishaK} for details. Similarly to $K_{\rm linear}$, it does not include the additional non-universal UV contributions to the compressibility.

\subsection{Partition function at low temperatures and the density of states}
\label{sec:dos}

We first overview some relevant results for the Majorana SYK model. An interesting time scale in the problem is given by the coefficient of the Schwarzian action, $\alpha_SN$. If the inverse temperature $\beta$ is of this order of magnitude or greater, quantum fluctuations are strong. This regime was originally studied in Ref.~\cite{BAK16} (see also \cite{Altland:2019czw}). The density of states (DOS) and the partition function for the pure Schwarzian model are as follows \cite{Cotler:2016fpe,Garcia-Garcia:2017pzl,BaAlKa17,Stanford:2017thb}, where the energy $E$ and the temperature $\beta^{-1}$ are measured in units of $(\alpha_SN)^{-1}$:
\begin{equation}
\rho_{\Sch}(E)=\sinh\biggl(2\pi\sqrt{2E}\biggr),\qquad
Z_{\Sch}(\beta^{-1})= \int e^{-\beta E}\rho_{\Sch}(E)\,dE
=\frac{1}{2}\biggl(\frac{2\pi}{\beta}\biggr)^{3/2} e^{2\pi^2/\beta}\,.
\end{equation}
These functions are defined up to an overall factor that depends on the normalization of the integration measure.

The DOS and the partition function for the Majorana SYK model contain some additional factors. Up to a common overall constant,
\begin{equation}
\rho(E)\sim \alpha_S N^{-1/2}e^{N\sz}
\rho_{\Sch}\bigl(\alpha_SN(E-E_0)\bigr)\,,\quad\:
Z(\beta^{-1})\sim 
N^{-3/2}e^{-E_0\beta+N\sz}Z_{\Sch}\bigl(\alpha_SN\beta^{-1}\bigr) \,,
\label{rho_Maj}
\end{equation}
where, $E_0$ is the ground state energy. The factor $N^{-3/2}$ in the partition function has been introduced to obtain the correct asymptotic behavior for $\beta\gg 1$ fixed and $N$ going to infinity:
\begin{equation}
\ln Z = N\bigl(-\calF_0\beta+\sz+2\pi^2\alpha_S\beta^{-1}
+\cdots\bigr) +N^0\biggl(-\const\cdot\beta-\frac{3}{2}\ln\beta+\cdots\biggr)
+O(N^{-1}).
\end{equation}
Note the absence of a $\ln N$ term. Indeed, the logarithm of the partition function admits a $1/N$ expansion, where different terms correspond to different classes of Feynman diagrams. In particular, the $N^0$ term is given by the sum of ladders closed into a loop, yielding the expression $-\frac{1}{2}\Tr\ln(1-K_{G})$. Here $K_G$ is the exact ladder kernel; it has ${}\sim\beta$ eigenvalues that are not too small, whereas the $-\frac{3}{2}\ln\beta$ contribution is due to the eigenvalues close to $1$.

There is one more thing to take into account---variations between different samples:
\begin{equation}
\ln\overline{Z}-\overline{\ln Z}
\approx \frac{1}{2}\overline{(\delta\ln Z)^2} \,,\quad\,
\text{where }\, \delta\ln Z=\ln Z-\overline{\ln Z} \,.
\end{equation}
In Eq.~\eqref{rho_Maj}, $E_0$ should be understood as the ground state energy for a particular realization of disorder. This may or may not be important, depending on the parameter $q$. Indeed, the sample-to-sample fluctuations are dominated by a particular Feynman diagram that contributes to $\ln\overline{Z}$ but not to $\overline{\ln Z}$ \cite{KS17-soft}. Assuming that $N\gg\beta\gg1$, its value can be estimated as follows:
\begin{equation}
\ln\overline{Z}-\overline{\ln Z} \:\approx\:
\figbox{0.8}{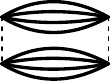} \:\sim\: N^{2-q}\beta^2 \,.
\end{equation}
Therefore, the fluctuations of the free energy are of the order of $\delta F\sim N^{1-q/2}$ with no singular temperature dependence. We conclude that for $\beta\sim N$, the sample-to-sample fluctuations are significant if $q=4$ but not at larger values of $q$.\medskip

For the complex SYK model, the density of states is a function of two conserved quantities, charge and energy:
\begin{equation}
\rho(E,Q)=\Tr\Bigl(\Pi_{Q}\,\delta\bigl(\hat{H}-E\kern1pt\hat{1}\bigl)\Bigr),
\end{equation}
where $\hat{H}$ and $\hat{Q}$ are defined in Eqs.~(\ref{eq: Hamiltonian}) and~(\ref{chargeoper}), respectively, and $\Pi_Q$ is the projector onto the subspace with a given value of $Q$. For simplicity, we assume that $N$ is even so that $Q$ takes on integer values. The partition function for a fixed $Q$ and the grand partition function are as follows:
\begin{equation}
Z_Q(\beta^{-1})=\int e^{-\beta E} \rho(E,Q)\, dE \,,\qquad
Z(\beta^{-1},\mu)=\sum_{Q=-\infty}^{\infty}e^{\beta\mu Q}Z_Q(\beta^{-1})\,.
\end{equation}
In analytical calculations, we will be interested in the case where $E$ is close to $E_0(Q)$, the lowest eigenvalue of $\hat{H}$ with charge $Q$. We will show that
\begin{equation}
\rho(E,Q)\sim \alpha_SN^{-1}e^{N\sz(Q/N)}
\rho_{\Sch}\bigl(\alpha_SN(E-E_0(Q))\bigr)
\label{DQE}
\end{equation}
up to a constant factor, or equivalently,
\begin{equation}
\ln Z_{Q}(\beta^{-1}) \approx -\beta E_0(Q) + N\sz(Q/N)
+\ln\bigl(N^{-2}Z_{\Sch}(\alpha_SN\beta^{-1})\bigr)
\label{lnZQ}
\end{equation}
up to a constant term. However, $E_0(Q)$ is difficult to compute with sufficient (say, $1/N$) precision; for $q=4$, it depends on the realization of disorder. A simple though not very accurate approximation is as follows:
\begin{equation}
E_0(Q)=N\calF_0(Q/N)+\const+O(N^{-1}) \,.
\end{equation}

We now derive Eq.~\eqref{lnZQ} from the effective action. Note that the integration measure is defined up to some $N$-dependent factor.\footnote{The $(G,\tilde{\Sigma})$ action is free from such ambiguity. However, we have lost track of normalization when eliminating $\tilde{\Sigma}$ and ``hard'' degrees of freedom in $G$.} We will use the factor $N^{-3/2}$ that comes with $Z_{\Sch}$ as previously explained. The additional normalization factors in our calculation are reasonably well motivated but not trustworthy, so the overall power of $N$ in front of $Z_{\Sch}$ will be checked independently.

Using the effective action \eqref{effact1}, the grand partition function is expressed as follows:
\begin{equation}
\begin{aligned}
Z(\beta^{-1},\mu)&=\sum_{n=-\infty}^{\infty}
\int D\calE\,\frac{D\bar{\lambda}}{\UU(1)}\,\frac{D\varphi}{\PSL(2,\RR)}\,
\exp\bigl(-I_\eff[\calE,n,\varphi,\bar{\lambda}]\bigr)\\[2pt]
&=Z_{\UU(1)}(\beta^{-1},\mu)\cdot N^{-3/2}Z_{\Sch}(\alpha_SN\beta^{-1}) \,.
\end{aligned}
\end{equation}
Let us focus on $Z_{\UU(1)}$, which involves the variables $n$, $\calE$, and $\tilde{\lambda}$. The notation ${D\bar{\lambda}}/{\UU(1)}$ indicates that we consider $\bar{\lambda}(\tau)$ up to an additive constant. The corresponding integral,
\begin{equation}
\int\frac{D\bar{\lambda}}{\UU(1)}\,
\exp\biggl(-\frac{NK}{2}\int\bar{\lambda}'^2\,d\tau\biggr)
=\sqrt{\frac{NK}{2\pi\beta}} \,,
\end{equation}
may be interpreted as the partition function (per unit length) of a free particle with mass $NK$. The integral over $\calE$ is evaluated using the method of steepest descent. Since
\begin{equation}
\frac{\partial^2 I_\eff}{\partial\calE^2} \approx -\gz''(\calE)N
=2\pi\frac{\partial\calQ}{\partial\calE}N <0,
\end{equation}
the integration path is parallel to the imaginary axis, and the symbol $D\calE$ is understood as ${d\calE}/{i}$ up to some real factor of the order of $1$. Thus,
\begin{equation}
Z_{\UU(1)}(\beta^{-1},\mu) \sim\sqrt{\frac{NK}{2\pi\beta}}
\sum_{n=-\infty}^{\infty}\int_{-i\infty}^{i\infty} \frac{d\calE}{i}\,
e^{-\tilde{I}_\eff} \,,\quad \text{where }\,
\frac{\tilde{I}_\eff}{N} =\beta\fz\Bigl(\mu+\frac{2\pi}{\beta}(\calE-in)\Bigr)
- \gz(\calE)\,.
\end{equation}

For each value of the winding number $n$, the effective action attains its extremum at the value of $\calE$ determined by the equation $\gz'(\calE) =2\pi \fz'\bigl(\mu+\frac{2\pi}{\beta}(\calE-in)\bigr)$. Replacing the right-hand side with $2\pi\fz'(\mu)$ introduces an $O(\beta^{-1})$ error in $\calE$ and an $O(\beta^{-2})$ error in $\tilde{I}_\eff$; the latter is within the precision of the effective action model.\footnote{Here we have assumed that $n\lesssim 1$, which is true if $\beta\lesssim N$. But even in the opposite limit, the error in $\tilde{I}_\eff$ is relatively small.} The value of ${\partial^2 \tilde{I}_\eff}/{\partial\calE^2}$ at the extremum is also almost independent of $n$. Applying the method of steepest descent and choosing the order $1$ factor for future convenience, we get
\begin{equation}
\begin{aligned}
Z_{\UU(1)}(\beta^{-1},\mu)
&\sim \sqrt{\frac{2\pi K}{\beta}}
\sum_{n=-\infty}^{\infty} \exp\biggl(-\beta\tilde{\Omega}\Bigl(\beta^{-1},\,
\mu-i\frac{2\pi}{\beta}n\Bigr)\biggr)\\
&\approx\sqrt{\frac{2\pi K}{\beta}}\,e^{-\beta\tilde{\Omega}(\beta^{-1},\mu)}
\sum_{n=-\infty}^{\infty}
\exp\biggl(-2\pi i\calQ Nn-\frac{2\pi^2KN}{\beta}n^2\biggr)\,,
\end{aligned}
\label{ZU1n}
\end{equation}
where, as in Section~\ref{sec:thermo},
\begin{equation}
\tilde{\Omega}(\beta^{-1},\mu)
=N\biggl(\fz\Bigl(\mu+\frac{2\pi}{\beta}\calE\Bigr)
-\gz(\calE)\beta^{-1}\biggr)\,,\qquad
\gz'(\calE)=2\pi\fz'\Bigl(\mu+\frac{2\pi}{\beta}\calE\Bigr)=-2\pi\calQ\,.
\end{equation}

The sum over $n$ in \eqref{ZU1n} is evaluated using the Poisson summation formula:
\begin{equation}
\begin{aligned}
Z_{\UU(1)}(\beta^{-1},\mu)
&\sim N^{-1/2}e^{-\beta\tilde{\Omega}(\beta^{-1},\mu)}
\sum_{Q=-\infty}^{\infty} \exp\biggl(-\frac{(Q-\calQ N)^2}{2KN}\biggr)\\
&\approx \sum_{Q=-\infty}^{\infty}
N^{-1/2} \exp\Bigl(-\beta\bigl(\tilde{F}(\beta^{-1},Q)-\mu Q\bigr)\Bigr)\,,
\end{aligned}
\label{ZU1Q}
\end{equation}
where
\begin{equation}
\tilde{F}(\beta^{-1},\calQ N)
=\tilde{\Omega}(\beta^{-1},\mu)+\mu\calQ N
=N\bigl(\calF_{0}(\calQ)-\sz(\calQ)\beta^{-1}\bigr)\,.
\end{equation}
An important feature of the second line in \eqref{ZU1Q} is that the argument of $\tilde{F}$ is the integer charge $Q$ being summed over, and not the mean charge $\calQ N$; consequently the entropic prefactor of each term in the sum is $e^{N \calS (Q/N)}$ and not $e^{N\calS (\calQ)}$. (Since $d\sz/d\calQ=2\pi\calE$, the ratio of such factors in the $Q+1$ and $Q$ sectors is $e^{2\pi\calE}$.)
So, when we multiply the second line in \eqref{ZU1Q} by $N^{-3/2}Z_{\Sch}(\alpha_SN\beta^{-1})$, we obtain an expression for $Z(\beta^{-1},\mu)$ that is equivalent to \eqref{lnZQ}. Finally, this yields (\ref{DQE}), where the density of states at charge $Q$ has a prefactor, $e^{N \mathcal{S}(Q/N)}$, with entropy evaluated at the same charge $Q$. 

On the other hand, if $N$ is very large, the sum over $n$ in \eqref{ZU1n} is reduced to the $n=0$ term. Multiplying it by the same factor, we get
\begin{equation}
\ln Z(\beta^{-1},\mu)
\approx -\beta\tilde{\Omega}(\beta^{-1},\mu)
+\frac{2\pi^2\alpha_SN}{\beta}-2\ln\beta\qquad
\text{for } N\gg\beta\gg1\,.
\end{equation}
The absence of a $\ln N$ term is consistent with $1/N$ expansion.

\section{Renormalization theory}
\label{secRG}

In this section, we describe the physics at intermediate time scales, $1\ll \tau \ll \beta$, generalizing the ideas in Ref.~\cite{KS17-soft} section 3. More exactly, we will study the renormalization of both symmetric and anti-symmetric perturbations to the Green function and the self energy.

\subsection{General idea}
The $(G,\tilde{\Sigma})$ action \eqref{GSigma tilde action} is suited for the perturbative study near conformal point $(G_c,\tilde{\Sigma}_c)$, which is an exact saddle point for $\sigma=0$. We will work at zero temperature, i.e.\ $G_c=G_{\beta=\infty}$,\, $\tilde{\Sigma}_c=\tilde{\Sigma}_{\beta=\infty}$, see \eqref{zeroT Green}. The actual UV source $\sigma$, consisting of a delta function and its derivative, is strong in the UV (i.e.\ for $\tau:=\tau_1-\tau_2\sim 1$), and therefore, is hard to study without numerics. However, it is possible to introduce a weaker perturbation in a slightly extended UV region such that its effect at $\tau\gg 1$ reproduces the actual correction $(\delta G,\delta\tilde{\Sigma})$ to the conformal solution. This method has been applied to the Majorana SYK model in Ref.~\cite{KS17-soft} section 3, yielding a derivation of the Schwarzian action as well as the relation between its coefficient and the UV-sourced correction to the Green function. 

\begin{figure}[t]
\center
\includegraphics{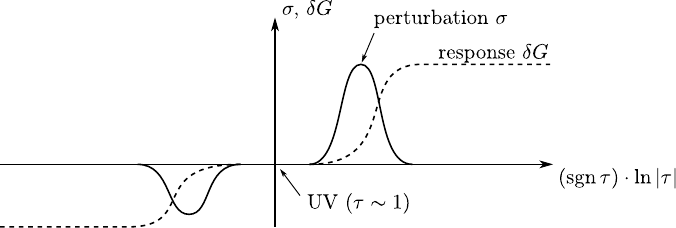}
\caption{RG flow of a perturbation $\sigma$ (solid line), generating the response $\delta G$ (dashed line) at larger time scales.}
\label{fig: responses}
\end{figure}

One useful property of the Majorana SYK model is anti-symmetry under time reflection. Namely, the perturbation source $\delta'(\tau)$ is an anti-symmetric function of time, and the ladder kernel that propagates the perturbation preserves this symmetry. As a consequence, the responses $\delta G(\tau)$, $\delta\tilde{\Sigma}(\tau)$ are also anti-symmetric in time. However, that is not the case for the complex SYK model, see Fig.~\ref{fig: responses} for an illustration. The actual UV source $\sigma(\tau_1,\tau_2)=\delta'(\tau_{12})-\mu \delta(\tau_{12})$ has both anti-symmetric and symmetric parts. More importantly, the ladder kernel (which will be studied later) mixes anti-symmetric and symmetric functions. The mixing effect will be characterized by a $2\times 2$ matrix that generalizes the number $k_{c}(h)$ of the Majorana SYK model~\cite{MS16-remarks,KS17-soft}. 

In general, renormalization theory determines how UV sources manifest themselves at intermediate scales, and thus, affect the IR physics. For instance, the interaction between the UV source and the IR deformation of the conformal solution due to reparameterization of time $\varphi(\tau)$ contributes to the local part of the effective action for the $\varphi$ field: it generates the Schwarzian term, which further determines such properties as specific heat. For the complex SYK model, the new ingredient is the perturbation due to chemical potential (or charge $\calQ$), sourcing the asymmetry of the Green function characterized by $\calE$ or $\theta$. The nontrivial relation \eqref{charge theta} between $\theta$ and $\calQ$ can be reproduced using renormalization, which further supports the statement that the charge is determined by the intermediate asymptotics of $G$.

To apply the renormalization theory for $\beta=\infty$, we write the charge as
\begin{equation}
\calQ=\int \sigma(\tau)\cdot (-\tau G(-\tau))\,d\tau
\label{defQ1}
\end{equation}
(cf.\ \eqref{defQ}), with $\sigma$ being small at each individual time scale but possibly spanning multiple scales. This way, $\sigma$ is regarded as a combination of infinitesimal perturbation sources. Focusing on a particular scale $\xi=\ln|\tau|$, we may characterize the cumulative effect of all sources at smaller scales by some value of $\calE$. The additional source at scale $\xi$ contributes both to $\calQ$ (via integral \eqref{defQ1}) and to $\calE$ (via renormalization). Thus, one can calculate $d\calE/d\calQ$, as elaborated in the following sections. The change in the asymmetry parameter $\calE$ is propagated by the RG flow and further augmented by any sources present at larger scales.

\subsection{Linear response to the perturbation $\sigma$}

\subsubsection{Quadratic expansion of the $(G,\tilde{\Sigma})$ action}
In this section, it will be convenient to treat bilocal functions $f(\tau_1,\tau_2)$ as operators (i.e.\ matrices indexed by $\tau_1$ and $\tau_2$) for which one can consider the product and the trace. A similar interpretation is also applicable to functions of four times. For example,
\begin{gather}
\Tr \bigl(f \cdot g\bigr)
=\int d\tau_1\,d\tau_2\, f(\tau_1,\tau_2)\, g(\tau_2,\tau_1) \,,\\[2pt]
f^{T}Ag = \int d\tau_1\,d\tau_2\,d\tau_3\,d\tau_4\,
f(\tau_2,\tau_1)\, A(\tau_1,\tau_2;\tau_3,\tau_4)\, g(\tau_3,\tau_4).
\end{gather}
With this in mind, we can express the $(G,\tilde{\Sigma})$ action \eqref{GSigma tilde action} as follows:
\begin{equation}
\frac{I[G,\tilde{\Sigma}]}{N}
= -\ln \det\bigl(-\tilde{\Sigma}\bigr)
- \frac{(-1)^{\frac{q}{2}}}{q}\,
\Tr\bigl(G^{\frac{q}{2}} \cdot G^{\frac{q}{2}}\bigr)
-\Tr\bigl(\tilde{\Sigma} \cdot  G\bigr) + 
\Tr\bigl(\sigma \cdot G\bigr)\,.
\end{equation}
Here the power $G^{q/2}$ is taken element-wise. Next, we expand the action $I=I[G,\tilde{\Sigma}]$ to the second order in the variations around the conformal point, $\delta G=G-G_c$,\, $\delta\tilde{\Sigma}=\tilde{\Sigma}-\tilde{\Sigma}_c$, ignoring the constant term $I[G_c,\tilde{\Sigma}_c]$:
\begin{equation}
    \frac{I_2}{N} = \frac{1}{2} \begin{pmatrix}
    \delta \tilde{\Sigma}^T &  \delta G^T 
    \end{pmatrix} 
    \begin{pmatrix}
    W_{\Sigma} & -\one \\
    -\one & W_{G}
    \end{pmatrix} 
    \begin{pmatrix}
    \delta \tilde{\Sigma} \\
    \delta G 
    \end{pmatrix} + \Tr \left( \sigma \cdot \delta G \right),\quad   
    W_{\Sigma} = \frac{\delta^2 I}{(\delta \tilde{\Sigma})^2} \,, \quad W_{G} =  \frac{\delta^2 I}{(\delta G)^2} \,.
    \label{eq: quadratic action} 
\end{equation}
The matrices $W_{\Sigma}$ and $W_{G}$ can also be expressed as follows:
\begin{equation}
W_{\Sigma} = \frac{\delta G}{\delta \tilde{\Sigma}} \,, \qquad
W_{G} = \frac{\delta\Sigma}{\delta G} \,,
\end{equation}
where the functional dependences of $G$ on $\tilde{\Sigma}$ and of $\Sigma$ on $G$ are given by the Schwinger-Dyson equations:
\begin{equation}
G = -\tilde{\Sigma}^{-1}, \qquad
\Sigma(\tau_1,\tau_2) = G(\tau_1,\tau_2)^{\frac{q}{2}} \left( - G(\tau_2,\tau_1)\right)^{\frac{q}{2}-1}\,. 
\end{equation}
(The equation $\tilde{\Sigma}=\Sigma+\sigma$ is not used.) These are the explicit formulas and diagrammatic representations of those matrices:
\begin{align}
&W_{\Sigma}(\tau_1,\tau_2;\tau_3,\tau_4) = G_c(\tau_1,\tau_3) G_c(\tau_4,\tau_2) = 
\begin{tikzpicture}[baseline={([yshift=-4pt]current bounding box.center)}]
\draw[thick, mid arrow] (40pt,12pt)--(0pt,12pt);
\draw[thick, mid arrow] (0pt,-12pt)--(40pt,-12pt);
\node at (-5pt,12pt) {\scriptsize $1$};
\node at (-5pt,-12pt) {\scriptsize $2$};
\node at (48pt,12pt) {\scriptsize $3$};
\node at (48pt,-12pt) {\scriptsize $4$};
\end{tikzpicture}\,,
\label{WSigma}\\[5pt]
&\begin{aligned}
W_{G}(\tau_1,\tau_2;\tau_3,\tau_4) &= \left( -1 \right)^{\frac{q}{2}-1}
\left[
\frac{q}{2} G_c(\tau_1,\tau_2)^{\frac{q}{2}-1} G_c(\tau_2,\tau_1)^{\frac{q}{2}-1} \delta (\tau_1,\tau_3) \delta (\tau_2,\tau_4)  \right. \\
&\hspace{50pt} \left.+ \left( \frac{q}{2}-1 \right) G_c(\tau_1,\tau_2)^{\frac{q}{2}} 
G_c(\tau_2,\tau_1)^{\frac{q}{2}-2} \delta (\tau_1,\tau_4) \delta (\tau_2,\tau_3) 
\right]\\
&= (-1)^{\frac{q}{2}-1} \left(
\frac{q}{2}
\begin{tikzpicture}[baseline={([yshift=-4pt]current bounding box.center)}]
\draw[thick, densely dotted] (10pt,15pt)--(0pt,15pt);
\draw[thick, densely dotted] (0pt,-15pt)--(10pt,-15pt);
\draw[thick, mid arrow] (0pt,15pt)..controls (5pt,7pt) and (5pt,-7pt)..(0pt,-15pt);
\draw[thick, mid arrow] (0pt,-15pt)..controls (-5pt,-7pt) and (-5pt,7pt)..(0pt,15pt);
\node at (-5pt,15pt) {\scriptsize $1$};
\node at (-5pt,-15pt) {\scriptsize $2$};
\node at (18pt,15pt) {\scriptsize $3$};
\node at (18pt,-15pt) {\scriptsize $4$};
\end{tikzpicture}
+
\left(\frac{q}{2}-1 \right) 
\begin{tikzpicture}[baseline={([yshift=-4pt]current bounding box.center)}]
\draw[thick, densely dotted] (0pt,15pt)--(20pt,-15pt);
\draw[thick, densely dotted] (20pt,15pt)--(0pt,-15pt);
\draw[thick, mid arrow] (0pt,-15pt)..controls (5pt,-7pt) and (5pt,7pt)..(0pt,15pt);
\draw[thick, mid arrow] (0pt,-15pt)..controls (-5pt,-7pt) and (-5pt,7pt)..(0pt,15pt);
\node at (-5pt,15pt) {\scriptsize $1$};
\node at (-5pt,-15pt) {\scriptsize $2$};
\node at (28pt,15pt) {\scriptsize $3$};
\node at (28pt,-15pt) {\scriptsize $4$};
\end{tikzpicture}
\right)\,.
\end{aligned}
\label{WG}
\end{align}
(An arrow pointing from $\tau'$ to $\tau$ denotes $G_c(\tau,\tau')$.)

\subsubsection{Ladder kernels} 
To calculate the effects of the perturbation source $\sigma$, we may first eliminate $\delta\tilde{\Sigma}$ from the quadratic action by evaluating it at the saddle point, $\delta\tilde{\Sigma}=W_{\Sigma}^{-1} \delta G$:
\begin{equation}
    \frac{I_2[\delta G]}{N} =  \frac{1}{2}  \delta G^T \left(W_{G} - W_{\Sigma}^{-1}  \right) \delta G + \Tr \left( \sigma \cdot  \delta G\right)\,.
\end{equation}
We further take the saddle point with respect to $\delta G$ and find its equilibrium value,
\begin{equation}
    \delta G = - \left(W_{G} - W_{\Sigma}^{-1} \right)^{-1} \sigma \,,
\end{equation}
which may be interpreted as the sum of ladder diagrams. The corresponding $\delta\tilde{\Sigma}$ is expressed in a similar way:
\begin{equation}
    \begin{aligned}
    \delta G &=  (1-\underbrace{W_{\Sigma} {W_{G}}}_{=:K_G})^{-1} W_{\Sigma}\sigma =\left(1+ K_G + K^2_G + \ldots \right) W_{\Sigma}\sigma\,, \\
    \delta \tilde{\Sigma} &= W_{\Sigma}^{-1} \delta G = (1- \underbrace{W_{G} W_{\Sigma}}_{=:K_{\Sigma}})^{-1} \sigma = (1+K_{\Sigma}+K_{\Sigma}^2+\ldots) \sigma \,.
    \end{aligned}
    \label{response}
\end{equation}
The ladder kernels $K_G$, $K_{\Sigma}$ are products of $W_{\Sigma}$ and $W_G$ in different orders, and thus, have the same spectrum (excluding $0$). Let us give their diagrammatic representations:
\begin{equation}
\begin{aligned}
K_{G}(\tau_1,\tau_2;\tau_3,\tau_4) &=  \left( -1 \right)^{\frac{q}{2}-1} 
\left(
\frac{q}{2}
\begin{tikzpicture}[baseline={([yshift=-4pt]current bounding box.center)}]
\draw[thick, mid arrow] (40pt,15pt)--(0pt,15pt);
\draw[thick, mid arrow] (0pt,-15pt)--(40pt,-15pt);
\draw[thick, mid arrow] (40pt,15pt)..controls (45pt,7pt) and (45pt,-7pt)..(40pt,-15pt);
\draw[thick, mid arrow] (40pt,-15pt)..controls (35pt,-7pt) and (35pt,7pt)..(40pt,15pt);
\node at (-5pt,15pt) {\scriptsize $1$};
\node at (-5pt,-15pt) {\scriptsize $2$};
\node at (48pt,15pt) {\scriptsize $3$};
\node at (48pt,-15pt) {\scriptsize $4$};
\end{tikzpicture}
+\left(\frac{q}{2}-1 \right) 
\begin{tikzpicture}[baseline={([yshift=-4pt]current bounding box.center)}]
\draw[thick, far arrow] (40pt,-15pt)--(0pt,15pt);
\draw[thick, near arrow] (0pt,-15pt)--(40pt,15pt);
\draw[thick, mid arrow] (40pt,15pt)..controls (45pt,7pt) and (45pt,-7pt)..(40pt,-15pt);
\draw[thick, mid arrow] (40pt,15pt)..controls (35pt,7pt) and (35pt,-7pt)..(40pt,-15pt);
\node at (-5pt,15pt) {\scriptsize $1$};
\node at (-5pt,-15pt) {\scriptsize $2$};
\node at (48pt,15pt) {\scriptsize $3$};
\node at (48pt,-15pt) {\scriptsize $4$};
\end{tikzpicture}
\right)
\,,\\[5pt]
K_{\Sigma}(\tau_1,\tau_2;\tau_3,\tau_4) &=  \left( -1 \right)^{\frac{q}{2}-1} 
\left(
\frac{q}{2}
\begin{tikzpicture}[baseline={([yshift=-4pt]current bounding box.center)}]
\draw[thick, mid arrow] (40pt,15pt)--(0pt,15pt);
\draw[thick, mid arrow] (0pt,-15pt)--(40pt,-15pt);
\draw[thick, mid arrow] (0pt,15pt)..controls (5pt,7pt) and (5pt,-7pt)..(0pt,-15pt);
\draw[thick, mid arrow] (0pt,-15pt)..controls (-5pt,-7pt) and (-5pt,7pt)..(0pt,15pt);
\node at (-5pt,15pt) {\scriptsize $1$};
\node at (-5pt,-15pt) {\scriptsize $2$};
\node at (48pt,15pt) {\scriptsize $3$};
\node at (48pt,-15pt) {\scriptsize $4$};
\end{tikzpicture}
+\left(\frac{q}{2}-1 \right) 
\begin{tikzpicture}[baseline={([yshift=-4pt]current bounding box.center)}]
\draw[thick, far arrow] (0pt,15pt)--(40pt,-15pt);
\draw[thick, near arrow] (40pt,15pt)--(0pt,-15pt);
\draw[thick, mid arrow] (0pt,-15pt)..controls (5pt,-7pt) and (5pt,7pt)..(0pt,15pt);
\draw[thick, mid arrow] (0pt,-15pt)..controls (-5pt,-7pt) and (-5pt,7pt)..(0pt,15pt);
\node at (-5pt,15pt) {\scriptsize $1$};
\node at (-5pt,-15pt) {\scriptsize $2$};
\node at (48pt,15pt) {\scriptsize $3$};
\node at (48pt,-15pt) {\scriptsize $4$};
\end{tikzpicture}
\right) \,.
\end{aligned}
\label{KGSigma}
\end{equation}

\subsubsection{Calculation of $K_G(h)$ and $K_{\Sigma}(h)$}
Due to $\SL(2,\RR)$ symmetry, $K_G$ and $K_\Sigma$ preserve power-law functions such as $\sigma(\tau) \sim \tau^{2\Delta-1-h}$. More exactly, we will consider perturbation sources of the form
\begin{equation}
\sigma(\tau) = 
\begin{pmatrix}
c_+\\
c_-
\end{pmatrix} |\tau|^{1-h}
\tilde{\Sigma}_c(\tau)
:= \begin{cases}
c_+ |\tau|^{1-h}\tilde{\Sigma}_c(\tau), & \tau>0 \\
c_- |\tau|^{1-h}\tilde{\Sigma}_c(\tau), & \tau<0
\end{cases}\,,
\end{equation}
which generate the responses
\begin{equation}
\delta G(\tau) = 
\begin{pmatrix}
\delta G_+\\
\delta G_-
\end{pmatrix} |\tau|^{1-h} G_c (\tau) \,, \qquad 
\delta\tilde{\Sigma}(\tau) = 
\begin{pmatrix}
\delta \tilde{\Sigma}_+\\
\delta \tilde{\Sigma}_-
\end{pmatrix} |\tau|^{1-h}\tilde{\Sigma}_c(\tau)\,.
\label{perturbations}
\end{equation} 
The goal is to find the $2\times 2$ matrices $W_\Sigma(h)$, $W_G(h)$ relating such coefficient vectors, excluding the $\tau$-dependent factors. For example, since $W_\Sigma=\delta G/\delta\tilde{\Sigma}$, we have $\left(\begin{smallmatrix} \delta G+\\ \delta G_- \end{smallmatrix}\right) = W_\Sigma(h) \left(\begin{smallmatrix} \delta\tilde{\Sigma}_+\\ \delta\tilde{\Sigma}_- \end{smallmatrix}\right)$. To calculate $W_{\Sigma}(h)$, we use the equation $\delta G(i\omega) = G(i\omega)^2 \delta \tilde{\Sigma}(i\omega)$, where $G(i\omega)$ is given by \eqref{zeroT Green frequency}. It should be combined with the Fourier transform
\begin{equation}
\bigintssss
\begin{pmatrix}
a_+ |\tau|^{-\alpha}, \quad \tau>0 \\
a_- |\tau|^{-\alpha} , \quad \tau<0
\end{pmatrix} e^{i\omega \tau} d\tau = 
\begin{pmatrix}
a'_+ |\omega|^{-1+\alpha}, \quad \omega>0 \\
a'_- |\omega|^{-1+\alpha} , \quad \omega<0
\end{pmatrix}\,,
\quad \begin{pmatrix}
a_+'\\
a_-'
\end{pmatrix}
= M(\alpha)
\begin{pmatrix}
a_+\\
a_-
\end{pmatrix}\,,
\end{equation}
where
\begin{equation}
M(\alpha) = \Gamma(1-\alpha) \begin{pmatrix}
i^{1-\alpha} & i^{-1+\alpha} \\
i^{-1+\alpha} & i^{1-\alpha}
\end{pmatrix}\,, \quad 
M(\alpha)^{-1} = \frac{\Gamma(\alpha)}{2\pi} 
\begin{pmatrix}
i^{-\alpha} & i^\alpha \\
i^\alpha & i^{-\alpha}
\end{pmatrix}\,.
\end{equation}
The relevant values of $\alpha$ are $2\Delta-1+h$ for $\delta G$ and $1-2\Delta+h$ for $\delta \tilde{\Sigma}$. Thus,
\begin{equation}
\begin{aligned}
W_{\Sigma}(h) ={}& - \frac{\Gamma(2-2\Delta)}{\Gamma(2\Delta)} 
\begin{pmatrix}
-e^{-\pi \calE} & 0 \\
0 & e^{\pi \calE}
\end{pmatrix} M(2\Delta-1+h)^{-1} 
\begin{pmatrix}
e^{-2i \theta} & 0 \\
0 & e^{2i \theta}
\end{pmatrix}
\\
&\cdot M(1-2\Delta+h) 
\begin{pmatrix}
-e^{\pi \calE} & 0 \\[3pt]
0 & e^{-\pi \calE}
\end{pmatrix} \\[3pt]
{}={}& \frac{\Gamma(2\Delta-1+h) \Gamma(2\Delta-h) }{\Gamma(2\Delta) \Gamma(2\Delta-1) \sin (2\pi \Delta) }
\begin{pmatrix}
\sin (\pi h + 2\theta) & - \sin (2\pi \Delta) + \sin (2\theta) \\
-\sin (2\pi \Delta) - \sin (2\theta) & \sin (\pi h -2\theta)
\end{pmatrix}
\end{aligned}
\end{equation}
 The matrix $W_G(h)$ is obtained from \eqref{WG}; it is in fact independent of $h$:
\begin{equation}
W_G(h) = \begin{pmatrix}
\frac{q}{2} & \frac{q}{2}-1 \\
\frac{q}{2}-1  & \frac{q}{2}
 \end{pmatrix}\,.
\end{equation}
Finally, 
\begin{equation}
  K_G(h)=W_{\Sigma}(h)W_G(h)\,, \quad  K_{\Sigma}(h) = W_G(h) W_{\Sigma}(h)\,.
\end{equation}
Note that
\begin{equation}
K_G(1-h) =
\begin{pmatrix}
0 & 1\\
1 & 0
\end{pmatrix}
K_{\Sigma}(h)^T 
\begin{pmatrix}
0 & 1 \\
1 & 0
\end{pmatrix} \,;
\label{eqn: kernel symmetry}
\end{equation}
this equation is related to the fact that  $K_G(\tau_1,\tau_2;\tau_3,\tau_4)=K_{\Sigma}(\tau_4,\tau_3;\tau_2,\tau_1)$.  Therefore, $K_G(h)$, $K_\Sigma(h)$, $K_G(1-h)$, $K_\Sigma(1-h)$ share the same eigenvalues.

\subsubsection{Resonant response}

Resonances occur at special values of $h$ such that $\det (1-K_\Sigma(h))=0$. In particular, $h=-1,0,1,2$ are solutions of this equation, also see Appendix~\ref{sec: operator spectrum} for discussions on the other solutions. Among them, the $h=2$ and $h=1$ perturbation sources determine the coefficient $\alpha_S$ in the effective action and the parameters $\calE,\calQ$, respectively. The dual values, $1-h=-1,0$, correspond to IR degrees of freedom, namely, the fluctuating fields $\varphi(\tau)$ and $\lambda(\tau)$.

Let $h=h_I$ be some resonance. The power-law source $\sigma(\tau)\sim \tau^{2\Delta-1-h_I}$ results in a divergent response, and therefore, has to be regulated. For this purpose, we multiply the source by a window function $u(\ln|\tau|)$:
\begin{equation}
\sigma_I(\tau) = \begin{pmatrix}
c^I_+ \\
c^I_-
\end{pmatrix} |\tau|^{1-h_I}\tilde{\Sigma}_c (\tau)\,
u\left(\ln |\tau|\right)\,, \qquad \int_{-\infty}^{+\infty} u(\xi) d\xi = 1\,.
\label{source2}
\end{equation}
Assuming that $u$ has finite support, $\sigma_I$ vanishes in the IR so that any response at sufficiently large scales is due to RG flow. On the other hand, the window should be sufficiently wide and $u(\xi)$ vary slowly with $\xi$, such that $\sigma_I (\tau)$ can be decomposed into power-law sources with $h$ close to $h_I$.

Following the argument in  Ref.~\cite{KS17-soft} section 3.1 we conclude that at sufficiently large $\tau$, 
\begin{equation}
\frac{\delta\tilde{\Sigma}(\tau)}{\tilde{\Sigma}_c (\tau)}= \begin{pmatrix}
\delta \tilde{\Sigma}_+\\
\delta \tilde{\Sigma}_-
\end{pmatrix}|\tau|^{1-h_I}\,, \quad \text{where} \quad 
\begin{pmatrix}
\delta \tilde{\Sigma}_+ \\
\delta \tilde{\Sigma}_- 
\end{pmatrix}
=\Res_{h=h_I}\bigl(K_{\Sigma}(h)-1\bigr)^{-1}
\begin{pmatrix}
c^I_+ \\
c^I_-
\end{pmatrix}\,.
\label{compare 1}
\end{equation}
A similar formula can be obtained for $\delta G$. Note that this result is independent of the details of window function $u$.

\subsubsection{The $h=1$ resonance}

As already mentioned, this resonance is related to the parameter $\calE$ and $\calQ$. So, let us find the residue of $(K_{\Sigma}(h)-1)^{-1}$ at $h=1$.

First, we compute $W_{\Sigma}(1)$ and $W'_{\Sigma}(1)$:
\begin{equation}
\begin{aligned}
W_{\Sigma}(1) &= \begin{pmatrix}
0 & -1 \\
-1 & 0
\end{pmatrix}
+ \frac{\sin (2 \theta)}{\sin (2\pi \Delta)} \begin{pmatrix}
-1 & 1 \\
-1 & 1
\end{pmatrix}\,, \\[3pt]
W'_{\Sigma}(1) & = -\frac{1}{1-2\Delta} W_{\Sigma}(1) - \pi \frac{\cos (2\theta)}{\sin (2\pi \Delta)} \begin{pmatrix}
1 & 0 \\
0 & 1
\end{pmatrix}\,.
\end{aligned}
\end{equation}
Thus,
\begin{equation}
\begin{aligned}
K_{\Sigma}(1) &= W_G W_{\Sigma}(1)=\begin{pmatrix}
1- \frac{q}{2} & -\frac{q}{2} \\
- \frac{q}{2} & 1-\frac{q}{2}
\end{pmatrix}
+ \frac{\sin (2 \theta)}{\sin (2\pi \Delta)} (q-1) \begin{pmatrix}
-1 & 1 \\
-1 & 1
\end{pmatrix}\,, \\[3pt]
K_{\Sigma}'(1) & =W_G W'_{\Sigma}(1)= -\frac{1}{1-2\Delta} K_{\Sigma}(1) - \pi \frac{\cos (2\theta)}{\sin (2\pi \Delta)} \begin{pmatrix}
\frac{q}{2} & \frac{q}{2}-1 \\
\frac{q}{2}-1 & \frac{q}{2}
\end{pmatrix}\,.
\end{aligned}
\end{equation}
The matrix $K_{\Sigma}(1)$ has eigenvalues $-(q-1)$ and $1$ as expected. The left and right eigenvectors associated with the eigenvalue $1$ are:
\begin{equation}
v^L= \begin{pmatrix}
1 & -1
\end{pmatrix}\,, \quad
v^R = \frac{1}{2} \begin{pmatrix}
1 \\
-1
\end{pmatrix}
- (1-\Delta) \frac{\sin (2\theta)}{\sin  (2\pi \Delta)} 
\begin{pmatrix}
1  \\
1
\end{pmatrix} \,, \quad v^L v^R =1\,.
\end{equation}
By abuse of notation, we introduce the number
\begin{equation}
k'(1):= v^L K_{\Sigma}' (1) v^R = -\frac{1}{1-2\Delta} - \pi \frac{\cos (2\theta)}{\sin (2\pi \Delta)}\,
\label{eqn: kprime(1)}
\end{equation}
not actually defining $k(h)$. Thus,
\begin{equation}
\Res_{h=1} \bigl(K_{\Sigma}(h)-1\bigr)^{-1} = \frac{1}{k'(1)} v^R v^L \,.
\label{Res_h=1}
\end{equation}

\subsection{Calculation of $d\calQ/d\calE$}

As part of the renormalization scheme for $\calQ$ and $\calE$, we calculate the variations of these parameters due to a perturbation source at a particular scale. More specifically, we consider the source \eqref{source2} with $h_I=1$:
\begin{equation}
\delta\sigma (\tau) = 
\begin{pmatrix}
c_+ \\
c_-
\end{pmatrix} \tilde{\Sigma}_c(\tau)\,
u\left(\ln |\tau|\right)\,, \qquad \int_{-\infty}^{+\infty} u(\xi) d\xi = 1\,.
\end{equation}
To find $\delta\calQ$, we integrate $\delta\sigma(\tau)$ against $-\tau G_c(-\tau)$, see \eqref{defQ1}. The functions $G_c=G_{\beta=\infty}$ and $\tilde{\Sigma}_c=\tilde{\Sigma}_{\beta=\infty}$ are given by \eqref{zeroT Green}; notice that $\tilde{\Sigma}_c(\tau)\cdot(-\tau G_c(-\tau)) =b\tau^{-1}$. Hence,
\begin{equation}
\delta\calQ = b \left( c_+ -c_-\right) = b v^L \begin{pmatrix}
c_+ \\
c_-
\end{pmatrix} \,.
\label{compare 2}
\end{equation}
The source also determines $\delta \tilde{\Sigma}$ through equations \eqref{compare 1} and \eqref{Res_h=1}:  
\begin{equation}
\frac{\delta \tilde{ \Sigma} (\tau)}{\tilde{\Sigma}(\tau)}  = \delta\calQ\cdot \frac{1}{b k'(1)} v^R \,.
\label{dSigma1}
\end{equation}
This result may be interpreted as a change of the asymmetry parameter $\calE$. Indeed, it follows from Eq.~\eqref{zeroT Green} that
\begin{equation}
b^{-1} \frac{d b}{d\calE} = - 2\pi \frac{\sinh (2\pi \calE)}{\cos (2\pi \Delta)+ \cosh (2\pi \calE)} = - 2\pi \frac{\sin (2\theta)}{\sin (2\pi \Delta)} \,,
\end{equation}
and hence,
\begin{equation}
\frac{\delta \tilde{\Sigma}(\tau)}{\tilde{\Sigma}(\tau)} = \delta \calE 
\begin{pmatrix}
\pi + (1-\Delta) b^{-1} \frac{d b}{d\calE} \\[2pt]
-\pi + (1-\Delta) b^{-1} \frac{d b}{d\calE} 
\end{pmatrix} = \delta\calE\cdot 2\pi v^R \,.
\label{dSigma2}
\end{equation}
Comparing \eqref{dSigma1} with \eqref{dSigma2}, we get:
\begin{equation}
\frac{d\calQ}{d\calE} = 2\pi b k'(1)= - \frac{\sin (2\pi \Delta)}{\cos (2\pi \Delta)+ \cosh (2\pi \calE)} - \pi (1-2\Delta)\frac{1+ \cos (2\pi \Delta) \cosh (2\pi \calE)}{( \cos (2\pi \Delta)+ \cosh (2\pi \calE) )^2}\,.
\label{dQdE}
\end{equation}
This formula can be written more compactly using the $\theta$ variable, 
\begin{equation}
\frac{d\calQ}{d\theta} = 2\pi b k'(1) \frac{d\calE}{d\theta}=- \frac{1}{\pi} - (1-2\Delta) \frac{\cos (2\theta)}{\sin (2\pi \Delta)}\,, \label{dQdtheta}
\end{equation}
which is
consistent with Eq.~(\ref{charge theta}) and the results in Refs.~\cite{GPS01,SS15,Davison17}.

\section{Computation of the compressibility} 
\label{sec:compress}

This section will present three different numerical approaches to computing the charge compressibility $K$ of the complex SYK model. We will limit these computations to the particle-hole symmetric case, where $Q=0$ and $\mu=0$. These computations will involve determination of the response of the particle-hole symmetric solution to small non-zero $Q$ or $\mu$:
\begin{enumerate}
\item In Section~\ref{sec:exact}, we will compute the compressibility by an exact diagonalization, evaluating  the ground state energy $E_0$ as a function of small $Q$. 

The value of $E_0$ is determined by the UV structure of the model, and we therefore expect $K$ to also be sensitive to the UV structure. This is as in Fermi liquid theory, where the compressibility involves a new Landau parameter, $F_0^s$, and is not determined by just the quasiparticle effective mass $m^\ast$. In contrast, in both Fermi liquid theory and the SYK models, the $T$-linear coefficient of the specific heat is determined by low energy physics: in Fermi liquid theory by $m^\ast$, and in the SYK model by the leading low energy deviation of the conformal solution \cite{Kit.KITP.2,MS16-remarks}.
\item In Section~\ref{numschwindys} we will numerically compute $K$ by an alternative method: full numerical solution of the Schwinger-Dyson equations of the SYK model. 

\item Finally, numerical approach in Section~\ref{kerneldiag} employs diagonalization of the two-particle kernel. 
\end{enumerate}
The values of $K$ obtained in these subsections are in excellent agreement. 
Throughout  the whole section we will recover $J$.

\subsection{Exact diagonalization}
\label{sec:exact}
In this subsection we perform exact diagonalization of the complex SYK Hamiltonian for $q=4$. 
The Hamiltonian \eqref{phsymham} 
commutes with the charge operator (\ref{chargeoper}): $[\hat{H},\hat{Q}]=0$. Therefore we can diagonalize $\hat{H}$ in each charge sector and find the ground state energy $E_{0}(Q)$.

\begin{figure}[t]
\center
\subfloat[$E_0(Q)$ vs. $Q$]
{\includegraphics[width=0.45\textwidth]{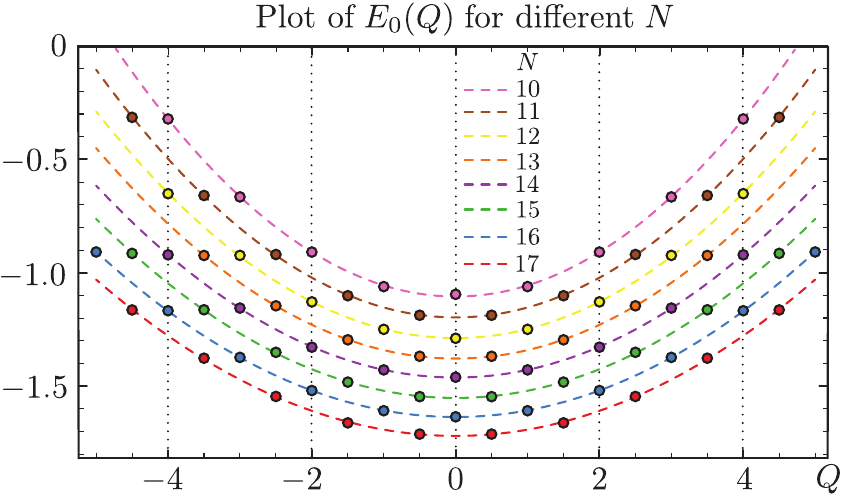}}
\qquad
\subfloat[ $E_0(0)$ vs. $N$]
{\includegraphics[width=0.42\textwidth]{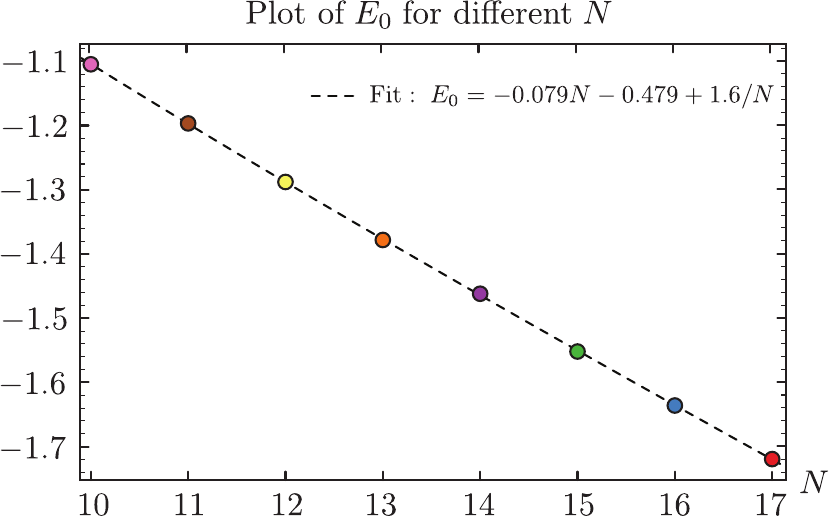}}
\caption{(a) The ground energy $E_{0}(Q)$ as a function of charge $Q$ in units $J$ ($q=4$). The number of samples for each charge are 250000 ($N=10$), 120000 ($N=11$), 50000 ($N=12$), 10000  ($N=13$), 2000  ($N=14$), 1000 ($N=15$), 500  ($N=16$), 200  ($N=17$). Dashed lines are fit by $E_{0}(Q)=E_{0}+a Q^{2}$.
(b) Plot for the ground energy $E_{0}$ at zero charge. The dashed line is a  fit by $E_{0}=- 0.079 N-0.479 + 1.6/N$. The leading large $N$ term to be compared with $2\epsilon_{0}$, where $\epsilon_{0}\approx -0.0406$ \cite{Cotler:2016fpe} 
}
\label{fig: ED}
\end{figure}

Our numerical results for the ground state energy $E_{0} (Q)$ are summarized in Fig.~\ref{fig: ED}(a).
\begin{enumerate}
\item 
Fitting these results to the form $E (Q) = E_0 + a Q^2$, we obtain the values of $E_0$ and $a$ as shown in the following table
\begin{center}
\begin{tabular}{@{} c|c|c|c|c|c|c|c|c @{}}
    \toprule
    $N$ & 10 & 11 & 12 & 13 & 14 & 15 & 16 & 17 \\ 
\hline
    $E_{0}$ & $-1.105$ & $-1.197$ & $-1.288$ & $-1.378$ & $-1.462$ & $-1.552$ & $-1.636$ & $-1.719$ \\ 
\hline
    $a$ & 0.0489 & 0.0437  & 0.0400  & 0.0371  & 0.0338 & 0.0316 &0.0292  & 0.0276  \\ 
  \end{tabular}
\end{center}

\item 
Finally, using $a = 1/(2 N K)$, we obtain the values of $K$ in the following table 
\begin{center}  
  \begin{tabular}{@{} c|c|c|c|c|c|c|c|c @{}}
    \toprule
    $N$ & 10 & 11 & 12 & 13 & 14 & 15 & 16 & 17 \\ 
\hline
    $K$ & $1.023$ & $1.041$ & $1.043$ & $1.037$ & $1.055$ & $1.056$ & $1.072$ & $1.067$ \\ 
  \end{tabular}
\end{center}
\end{enumerate}
Note that there is little dependence of $K$ on $N$.
We also show the $N$ dependence of $E_0$ in Fig.~\ref{fig: ED}(b).

\subsection{Schwinger-Dyson equation}
\label{numschwindys}
Here we briefly review the numerical solution of the Schwinger-Dyson equations for the complex SYK model. This has already been discussed in many papers, see particularly 
Refs.~\cite{MS16-remarks, Davison17}. Our main purpose is to show that this method gives compressibility $K$ very close to the result obtained in Section~\ref{sec:exact} from exact diagonalization. 

\begin{figure}[t]
\center
\subfloat[$\calQ/\mu$ vs. $T$]{
\includegraphics[width=0.44\textwidth]{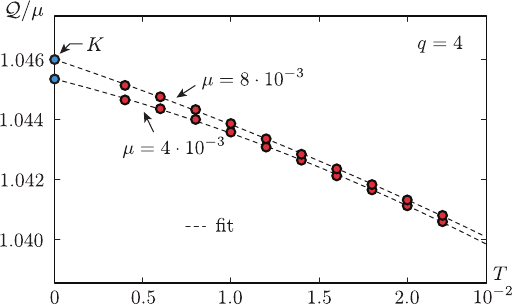}}\qquad
\subfloat[$K$ vs. $T$]{
\includegraphics[width=0.44\textwidth]{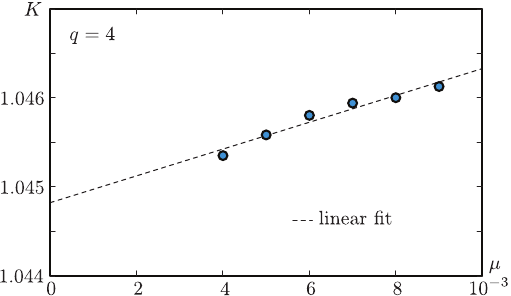} }
\caption{\label{Kcomp} (a) Plot of  ${\calQ}/ \mu$ for different temperatures $T$ and chemical potentials $\mu$ for $q=4$. (b) Plot of $K$ for different $\mu$. }
\end{figure}

We solve Schwinger-Dyson equation numerically using the well-known method of weighted iterations
\begin{equation}
    \Sigma_{j}(\tau)=J^{2}G_{j}^{q/2}(\tau)G_{j}^{q/2-1}(\beta-\tau)\,, \quad
G_{j+1}(i\omega_{n}) =(1-w) G_{j}(i\omega_{n})+\frac{w}{i \omega_{n}+\mu-\Sigma_{j}(i\omega_{n})} \,.
\end{equation}
For non-zero chemical potential it is convenient to start iterations with the conformal answer,  regulated at the boundaries $\tau=0^{+}$ and $\tau=\beta^{-}$, and  with the $\theta$ parameter corresponding  to specific charge $\calQ$ close to expected numerical value. This prevents iterations from falling into exponentially decaying solution.   
\begin{figure}[t]
\center
\includegraphics[width=0.55\textwidth]{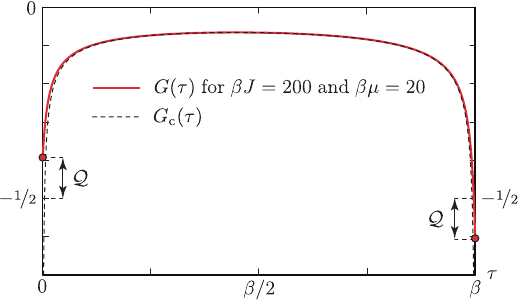}
\caption{\label{plotGmu} Plot of numerical solution for $G(\tau)$ at $q=4$, $\beta J =200$ and $\beta \mu=20$. The dashed line is conformal solution (\ref{Gconf})
with $\theta$ found from numerical $\calQ$ using the formula (\ref{charge theta}).}
\end{figure}
We find $\calQ$ numerically using the formula 
\begin{equation}
\calQ= \frac{1}{2}(G(0^{+})-G(\beta^{-}))\,. \label{numQ}
\end{equation}
For large $\beta J$ we can use equation (\ref{charge theta}) to find parameter $\theta$ in conformal solution (\ref{Gconf}). We plot an example of exact numerical $G(\tau)$ and its conformal fit $G_{\textrm{c}}(\tau)$ in Fig.~\ref{plotGmu}.
The grand potential can be computed  from the expression  \cite{Davison17} 
\begin{equation}
-\beta \frac{\Omega}{N} = \log \left(2 \cosh \frac{\beta \mu}{2}\right) +2 \textrm{Re} \sum_{n=0}^{\infty}\log\left(1-\frac{\Sigma(i\omega_{n})}{i\omega_{n}+\mu}\right) +\frac{q-1}{q} \sum_{n=-\infty}^{+\infty}\Sigma(i\omega_{n})G(i\omega_{n})\,,
\end{equation}
from which one can obtain the entropy as
\begin{equation}
S = -\beta \frac{\Omega}{N} - \beta \mu \calQ +\frac{2}{q}\sum_{n=-\infty}^{+\infty}\Sigma(i\omega_{n})G(i\omega_{n})\,,
\end{equation}
where $Q$ is computed numerically using (\ref{numQ}).
Finally compressibility in units $J$ can be obtained numerically by using the formula 
\begin{equation}
K =    \lim_{\mu\to 0} \frac{\calQ}{\mu} =  \frac{1}{N} \lim_{\mu\to 0} \frac{1}{2\mu} (G(0^{+})-G(\beta^{-}))\,.
\end{equation}
Numerically we fix $J=1$ and compute the ratio $\calQ/\mu$ for small $T$ and small $\mu$. We first approximate the result to the zero temperature to obtain $K$ as a function of small $\mu$, as shown in Fig.~\ref{Kcomp}, left. Then we approximate such $K(\mu)$ to $\mu=0$ (Fig.~\ref{Kcomp}.b, right).  
We did computations for $q=4$ and used $10^8$ grid points for the two-point function. The value of $K$ we found is 
\begin{equation}
K \approx 1.045/J\,. \label{KSDres}
\end{equation}
This result agrees quite well with the exact diagonalization result in the previous section and with the value of $K$ reported in \cite{Davison17}.

\subsection{Kernel diagonalization}
\label{kerneldiag}

This type of numerics was first done in  Ref.~\cite{MS16-remarks} for the antisymmetric kernel\footnote{We thank D. Stanford for sharing his code with us.}. In Appendix~\ref{app:eff} we discuss analytical approach for kernel diagonalization.  (also see Ref.~\cite{Davison17} Appendix F). 
The fluctuation analysis here is complementary to that in Section~\ref{secRG} in the sense that here we expand the fluctuations around the exact saddle while in the Section~\ref{secRG} we expand around the conformal saddle. 

We remind that we are working on the saddle with $\calQ=0$, where the general expressions for the kernel \eqref{KGSigma} have additional symmetry, i.e. they commute with the operator that switches two times, and thus we may analyze the kernel on the subspace of antisymmetric and symmetric functions separately. For this purpose, 
let us consider the symmetrized antisymmetric and symmetric kernels\footnote{Comparing to the general expression \eqref{KGSigma}, we ``average'' $K_G$ and $K_\Sigma$ in the sense that we separate $(q-2)$ rungs from one side to two sides, such that the final expression is hermitian. The superscript $\rA/\rS$ indicate the subspaces of the antisymmetric/symmetric functions of two time the kernels act on. We also need to replace the conformal $G_c$ by the exact Green function since we are expanding w.r.t. the exact saddle in this section.}
 \begin{equation}
K^{\rA/\rS}(\theta_{1},\theta_{2};\theta_{3},\theta_{4})=-\Big(\frac{q}{2}\pm(\frac{q}{2}-1)\Big)J^{2} |G(\theta_{12})|^{\frac{q-2}{2}}G(\theta_{13})G(\theta_{24})|G(\theta_{34})|^{\frac{q-2}{2}}\,,
\label{eqn: KAS}
 \end{equation}
where we fix $\beta=2\pi$ so all angles take values in the interval $[0,2\pi]$.
Since these kernels are invariant under
the translation of all four times, i.e. they 
commute with the operator $D = i(\partial_{\theta_{1}}+\partial_{\theta_{2}})$, one can look for the eigenfunctions 
of the kernels $\Psi_{h,n}^{\rA/\rS}(\theta_{1},\theta_{2})$, which are simultaneously  eigenfunctions of the operator $D$:
 \begin{equation}
\Psi_{h,n}^{\rA/\rS}(\theta_{1},\theta_{2}) = e^{i n \frac{\theta_{1}+\theta_{2}}{2}}\phi^{\rA/\rS}_{h,n}(\theta_{12})\,.
\end{equation}
Let us also define variants of the kernels with the parameter $n$ accordingly, 
\begin{equation}
K_n^{\rA/\rS} (\theta,\theta')= \int_{0}^{2\pi} K^{\rA/\rS} \left( s + \frac{\theta}{2}, s- \frac{\theta}{2} ; \frac{\theta'}{2} , - \frac{\theta'}{2}   \right) e^{-ins} ds\,, 
\end{equation}
such that $\phi^{\rA/\rS}_{h,n}(\theta)$ are the eigenfunctions with eigenvalue $k^{\rA/\rS}(h,n)$, more explicitly, 
\begin{equation}
\int_{0}^{2\pi}  K_n^{\rA/\rS} (\theta,\theta') \phi^{\rA/\rS}_{h,n}(\theta') \, d\theta' =k^{\rA/\rS}(h,n) \phi^{\rA/\rS}_{h,n}(\theta)\,.
\end{equation}
To numerically diagonalize kernels $K_n^{\rA/\rS}(\theta,\theta')$ in the space of antisymmetric/symmetric functions (on the discretized coordinates $\theta$, $\theta'$), it is more convenient to impose the symmetry explicitly, namely, we use $(K^{\rA}_n(\theta,\theta')-K^{\rA}_n(\theta,-\theta'))/2$ and $(K^{\rS}_n(\theta,\theta')+K^{\rS}_n(\theta,-\theta'))/2$ in the actual calculation. 

\begin{figure}[t]
\center
\subfloat[$k^{\rS}(1,1)$ vs. $\beta\calJ$]{
\includegraphics[width=0.45\textwidth]{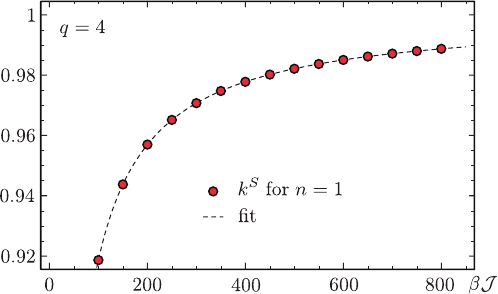} }
\qquad
\subfloat[ $n \alpha_K^{\rS}(n)$ vs. $n$]{
\includegraphics[width=0.44\textwidth]{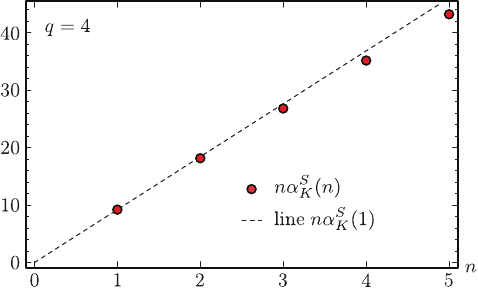} }
\caption{\label{kSn1} (a) Plot of numerical   $k^{\rS}(1,1)$ for $q=4$ and $n=1$. The dashed line is the fit (\ref{ksfit}). (b) Plot for $n\alpha_{K}^{\rS}(n)$ for $q=4$. One can see that within computational accuracy $\alpha_{K}^{\rS}(n)$ almost does not depend on $n$, confirming the expectation (\ref{ksexpect}).  We use $10^{8}$ grid points for numerical computation of $G(\theta)$
and  $10^{5}$  grid points for the kernel discretization in $\theta$ and $\theta'$ directions, so the kernel becomes a $10^{5}\times 10^{5}$ matrix. 
}
\end{figure}

We expect to find the highest eigenvalue of the  kernels $K^{\rA}_{n}(\theta,\theta')$ and $K^{\rS}_{n}(\theta,\theta')$ for large $\beta \mathcal{J}$, where $\mathcal{J}=\sqrt{q}J/2^{\frac{q-1}{2}}$ in the form
 \begin{equation}
k^{\rA}(2,n) = 1 -\frac{\alpha_{K}^{\rA}}{\beta \mathcal{J}}|n|+O\Bigl(\frac{1}{(\beta \mathcal{J})^{2}}\Bigr)\,, \quad k^{\rS}(1,n) = 1 -\frac{\alpha_{K}^{\rS}}{\beta \mathcal{J}}|n|+O\Bigl(\frac{1}{(\beta \mathcal{J})^{2}}\Bigr)\,. \label{ksexpect}
\end{equation}
These eigenvalues correspond to $h=2$ and $h=1$ modes. 
The Schwarzian coupling $\alpha_{\textrm{S}}$ and compressibility $K$ is related to $\alpha_{K}^{\rA}$ and $\alpha_{K}^{\rS}$ through the formulas \footnote{We notice that we have additional factor of 2 for $\alpha_S$ in comparison to  \cite{MS16-remarks} because in our case $N$ is the number of complex fermions.} 
 \begin{equation}
\alpha_{{S}} = \frac{\alpha_{K}^{\rA}}{3\alpha_{0}q^{2}}\frac{1}{\mathcal{J}} , \quad K = \frac{\alpha_{K}^{\rS}}{\alpha_{0}(q-1)} \frac{1}{\mathcal{J}}\,, \label{KrelalS}
\end{equation}
where $\alpha_{0}=2\pi q  \cot({\pi}/{q})/((q-1)(q-2))$. We compute numerically $k^{\rS}$ for $q=4$ and different values of $\beta \mathcal{J}$ and $n$. The plot of $k^{\rS}$ for $q=4$ and $n=1$ is represented in Fig.~\ref{kSn1}(a). By fitting the data points by polynomial in $1/\beta \mathcal{J}$
we obtain 

\begin{equation}
k^{\rS}(1,1) =     1-\frac{9.2}{\beta \mathcal{J}}+\frac{130.5}{(\beta \mathcal{J})^{2}}-\frac{2377}{(\beta \mathcal{J})^{3}}\,. \label{ksfit}
\end{equation}
From this fit we find that $\alpha_{K}^{\rS}=9.2$ and from (\ref{KrelalS}) we obtain for $q=4$ 
\begin{equation}
K \approx 1.04/J\,. \label{Knumres}
\end{equation}
This agrees very well with $K$ obtained from the Schwinger-Dyson equation (\ref{KSDres}) and exact diagonalization.  We also plotted $n \alpha_{K}^{\rS}(n)$ in Fig.~\ref{kSn1}(b), where $\alpha_{K}^{\rS}(n)$ obtained from fitting $k^{\rS}$ for different $n$ and $\alpha_{K}^{\rS}(1)=9.2$. One can see that within computational accuracy $\alpha_{K}^{\rS}$ does not depend on $n$ in agreement with expectation (\ref{ksexpect}).

\begin{figure}[t]
\center
\subfloat[Eigenfunctions $\phi_{2,n}^{\rA}$]{
\includegraphics[width=0.45\textwidth]{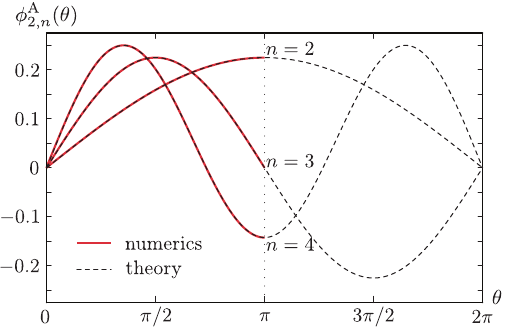}}
\qquad
\subfloat[Eigenfunctions $\phi_{1,n}^{\rS}$]{
\includegraphics[width=0.45\textwidth]{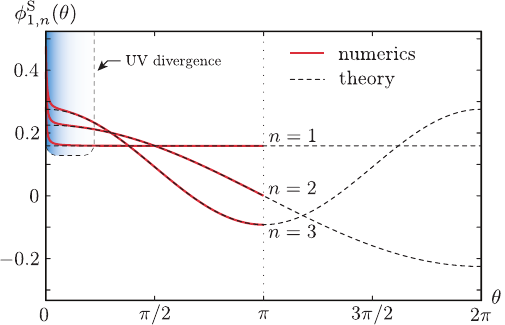}}
\caption{\label{phi21nAS} (a) numerical eigenfunctions $\phi_{2,n}^{\rA}$ for the antisymmetric kernel; note the perfect agreement between numerics and the analytic solution. (b) numerical eigenfunctions $\phi_{1,n}^{\rS}$ for the symmetric kernel. In this case one can see UV divergence near $\theta=0$ where the numerics disagrees with the theoretical conformal perturbation theory.
}
\end{figure}

Following the discussions in Ref.~\cite{MS16-remarks}, one might expect that the numerical result of $\alpha_K^{\rS}$ can be related to the deviation of the exact Green function from the conformal one  similar to the case for $\alpha_K^{\rA}$ (cf. Ref.~\cite{MS16-remarks} Eq.~(3.88)). We present a calculation following this procedure in Appendix \ref{app:GrishaK}. The result  does not agree with the numerical value of $K$ but agrees with the $K_{\text{linear}}$ (see \eqref{Kgamma relation})
\begin{equation}
K_{\textrm{linear}}  = -\left.\frac{2^{\frac{q+1}{2}}\alpha_{G}}{J\sqrt{q} \alpha_{0}}{k_{c}^{\rA}}'(2)\right|_{q=4} \approx 0.48/J\,.
\end{equation}
On the other hand, the numerical result for $\alpha_{S}$ from the anti-symmetric kernel agrees perfectly with the theoretical computation \cite{MS16-remarks}. The reason for the disagreement for $K$ presumably related to the fact that $K$ is a UV sensitive 
quantity, 
and the naive perturbation 
theory for the symmetric kernel  in $1/\beta \mathcal{J}$ series does not work well, e.g. the integrals obtained from higher corrections to the Green function have uncompensated power-law divergences which then contribute to the first order $1/\beta \mathcal{J}$ term, changing the final result. 
One sign of such a breakdown of perturbation theory is visible in our numerical results for eigenvectors of the symmetric kernel. They agree with the conformal kernel eigenfunctions   
everywhere except UV region, whereas for the antisymmetric eigenfunctions the agreement is perfect everywhere; see Fig.~\ref{phi21nAS}.  
The conformal kernel eigenfunctions, which are simultaneously  eigenfunctions of the Casimir with eigenvalues $h=2$ (anti-symmetric) and $h=1$ (symmetric) read
\begin{equation}
\begin{aligned}
&\phi_{2,n}^{\rA}(\theta) =  \frac{\gamma_{n}}{2\sin\frac{\theta}{2}}\Big(\frac{\sin \frac{n \theta}{2}}{\tan \frac{\theta}{2}}- n \cos \frac{n \theta}{2}\Big)\,,\quad 
\phi_{1,n}^{\rS}(\theta) = \frac{1}{2\pi|n|^{1/2}} \frac{\sin \frac{n \theta}{2}}{\sin \frac{\theta}{2}}\,,  \label{confeigfun} \\
& \hspace{80pt} \text{where} \quad \gamma_{n}^{2} =\frac{3}{\pi^{2}|n|(n^2-1)}  \,.
\end{aligned}
\end{equation}
The divergence of the eigenfunctions of the symmetric kernel in UV region is captured in the large $q$ limit (see Appendix \ref{largeqsymkern}).

\section{Bulk picture and zero-temperature entropy}
\label{sec:bulk}

In this section, we find the zero-temperature entropy $\sz$ of the complex SYK model by considering a massive Dirac fermion in AdS$_2$. The actual calculation is done in the Euclidean case, that is, on the hyperbolic plane. The asymmetry of the Green function \eqref{Gconf} may be interpreted as a phase factor with an imaginary phase, $2\pi i\calE$, suggestive of an imaginary $\UU(1)$ field acting on the Dirac fermion. (It corresponds to a real electric field in AdS$_2$.) The partition function in the presence of such a field yields the dependence of $\sz$ on $\calE$, and hence, on $\calQ$ via Eq.~(\ref{dQdE}). We will find that the $\sz$ so obtained is exactly equal to that obtained from direct computations for the complex SYK model~\cite{GPS01,SS15,Davison17}.

Our computation of $\sz$ should be contrasted with that for higher-dimensional charged black holes
\cite{Myers99,Faulkner09,SS15,Gaikwad:2018dfc,Nayak:2018qej,Moitra:2018jqs,Chaturvedi:2018uov,Sachdev19,Moitra:2019bub,Anninos:2019oka}, summarized in Appendix~\ref{app:em}. In the latter case, the value of $\sz$ in Eq.~(\ref{S0bh}) is determined by the horizon area and has no direct connection to the parameters of the SYK model. The present section interprets $\sz$ as the contribution of fermionic fields; such matter fields~\cite{Faulkner09} only make a subdominant contribution to thermodynamics in the conventional higher dimensional AdS/CFT correspondence.

\subsection{General idea}
For illustrative purposes, we will use the Majorana SYK model,
\begin{equation}\label{H_SYK}
\hat{H}_{\text{Majorana}}
= \frac{i^{q/2}}{q!}\sum_{j_1,\dots,j_q}J_{j_1\cdots j_q}\,
\hat{\chi}_{j_1}\dots\hat{\chi}_{j_q}\,,
\qquad
\overline{J_{j_1\cdots j_q}^2}=\frac{(q-1)!J^2}{N^{q-1}}\,.
\end{equation}
Among many methods of calculating its zero-temperature entropy $\sz=\sz_{\text{Majorana}}$, the formula
\begin{equation}
\sz_{\text{Majorana}}
= \int_0^{\frac{1}{2}-\Delta} \frac{\pi x }{\tan (\pi x)} dx
\label{SMaj}
\end{equation}
can be derived by evaluating $\frac{1}{2}\ln\det(-\tilde{\Sigma})$ with proper regularization\cite{Kit.KITP.1, Kit.KITP.2} (see also appendix~\ref{subsec: ln det}). Indeed, $\sz$ is defined as the zeroth order term in the $1/\beta$ expansion $\frac{\ln Z}{N} = -\frac{E_0}{N}\beta +\sz +O(\beta^{-1})$, where  $\ln Z$ may be approximated by minus the $(G,\tilde{\Sigma})$ action at the saddle point. As explained in appendix~\ref{subsec: ln det}, the double integral part of the action has $\beta$ and $O(\beta^{-1})$ terms but no constant term.

For the complex SYK model, $Z$ should be understood as the grand partition function, and $\sz$ should be replaced by its Legendre transform, $\gz(\calE)= \sz(\calQ)-2\pi\calE\calQ$. We will derive a formula similar to \eqref{SMaj} by considering $\ln Z$ in the $\beta\to\infty$ limit and extracting the constant term:\footnote{One may have noticed that the integrand in \eqref{gz_int} has a form similar to the Plancherel measure for the universal cover of $\SL(2,\RR)$. This analogy will be elucidated in Section~\ref{sec:Plancherel}.}
\begin{equation}
\gz(\calE) = \int_{0}^{\frac{1}{2}-\Delta}  \frac{2 \pi x \sin (2\pi x)  }{\cosh (2\pi \calE) - \cos(2\pi x) } dx\,.
\label{gz_int}
\end{equation}
For $\tilde{\Sigma}$ asymmetric in time and frequency, the direct calculation of $\det(-\tilde{\Sigma})$ is fraught with regularization difficulties. This is where the bulk picture offers a crucial advantage, replacing the tricky UV regularization with a simple subtraction of a boundary contribution.

In an abstract sense, the bulk is an artificial system that mimics most important properties of the real one. It may also be regarded as a heat bath for a small subset of sites~\cite{KS17-soft}. The following argument seems to apply to all large $N$ systems, but we will focus on the Majorana SYK model for simplicity. Consider adding an extra site to $N$ existing ones and modifying the couplings $J_{j_1,\dots,j_q}$ accordingly, multiplying them by $\bigl(\frac{N}{N+1}\bigr)^{\frac{q-1}{2}} \approx 1-\frac{q-1}{2N}$. In the thermodynamic limit, the logarithm of the partition function is proportional to $N$, and its change by the stated procedure is just $\frac{\ln Z}{N}$. Calling the original $N$ sites a ``bath'', we get:
\begin{equation}
\frac{\ln Z}{N}=\ln Z_{\text{full}}-\ln Z_{\text{bath}}
-\frac{q-1}{2}\,\frac{\partial\ln Z}{N\partial\ln J}\,,
\end{equation}
where ``full'' refers to the bath and the extra site together, but with the couplings unchanged. In the $\beta\to\infty$ limit, $\frac{\partial\ln Z}{N\partial\ln J}=-\frac{E_0}{N}\beta+O(\beta^{-1})$; hence, the last term in the above equation may be neglected.

To calculate $\ln Z_{\text{full}}-\ln Z_{\text{bath}}$, we may write the Hamiltonian as $\hat{H}_{\text{full}} =\hat{H}_{\text{bath}} +i\hat{\chi}\hat{\xi}$, where $\hat{\chi}$ represents the extra site and $\hat{\xi}$ is a certain operator acting on the other sites. When $N$ is large, $\hat{\xi}$ is Gaussian, meaning that the bath is completely characterized by the correlation function $\langle{\rm T} \hat{\xi}(\tau_1)\hat{\xi}(\tau_2)\rangle=-\Sigma(\tau_1,\tau_2)$ while higher correlators are obtained by Wick's theorem. This suggests the replacement of the real system by a collection of Grassmann variables $\Psi_j$ with a quadratic action $I=-\frac{1}{2}\sum_{j,k}B_{jk}\Psi_j\Psi_k$, where the indices take values on the time circle (for the extra site) and some abstract locations (for the bath). The full matrix $B$ has this structure:
\begin{equation}
B_{\text{full}}
=\begin{pmatrix} -\sigma & Y\\ -Y^T & B_{\text{bath}} \end{pmatrix}\,,\qquad
\sigma=\partial_\tau\,,\quad YB_{\text{bath}}^{-1}Y^{T}=-\Sigma\,,
\end{equation}
with $\sigma$ and $B_{\text{bath}}$ being square and skew-symmetric, and $Y$ rectangular. Using this artificial model, we get
\begin{equation}
\ln Z_{\text{full}}-\ln Z_{\text{bath}}
=\frac{1}{2}\ln\frac{\det B_{\text{full}}}{\det B_{\text{bath}}}
=\frac{1}{2}\ln\det(-\sigma-\Sigma)\,,
\end{equation}
where we have used the identity $\det\left(\begin{smallmatrix}A & B\\C & D \end{smallmatrix}\right)= \det D\cdot \det(A-BD^{-1}C)$.

While the previous description leaves many possibilities for choosing $B_{\text{bath}}$, the nicest one is a Majorana fermion with mass $M=\frac{1}{2}-\Delta$ on the hyperbolic plane. All its properties follow from those of the Dirac fermion, studied in the next subsection and appendix~\ref{sec:app_Dirac}. In this preliminary discussion, we use the Poincare half-plane model with the metric $ds^2=({d\tau^2+dy^2})/{y^2}$\, ($y>0$). A Majorana spinor $\psi$ has two components, $\psi_{\downarrow}$ and $\psi_{\uparrow}$. Solutions of the equation of motion have this asymptotic form:
\begin{equation}
\psi(\tau,y)=
\psi_+(\tau)\, y^{\Delta_+}\!\begin{pmatrix}
1\\ 1
\end{pmatrix}
+\psi_-(\tau)\, y^{\Delta_-}\!\begin{pmatrix}
1\\ -1
\end{pmatrix}\,\text{ for }y\to 0\,,\quad\:
\Delta_+=1-\Delta\,,\quad \Delta_-=\Delta\,.
\label{near-b_Poincare}
\end{equation}
The boundary condition $\psi_-(\tau)=0$ is chosen, which prescribes a sufficiently fast decay near the boundary. We will refer to it as the ``Dirichlet b.c.'' and to the condition $\psi_+(\tau)=0$ as the ``Neumann b.c.''.

Assuming that only the first term in \eqref{near-b_Poincare} is present, we can promote the asymptotic coefficient $\psi_+(\tau)$ to a field and identify it with the field $\xi(\tau)$ characterizing the bath. This is reasonable because the correlator
\begin{equation}
\langle\psi_{\pm}(\tau_1)\psi_{\pm}(\tau_2)\rangle
\sim \sgn(\tau_1-\tau_2)\,|\tau_1-\tau_2|^{-2\Delta_{\pm}}\qquad
(\text{``$+$'' for Dirichlet, }\, \text{``$-$'' for Neumann})
\end{equation}
matches $\langle\xi(\tau_1)\xi(\tau_2)\rangle=-\Sigma(\tau_1,\tau_2)$ if the ``$+$'' sign is chosen. The part of the action involving the boundary fermion $\chi(\tau)$ is
\begin{equation}
I_{\text{boundary}}=\int \biggl(\frac{1}{2}\chi\partial_{\tau}\chi+i\psi_{+}\chi\biggr)\,
d\tau\,.
\end{equation}
Since we are interested in low temperature properties, or large time scales, the $\chi\partial_{\tau}\chi$ term may be neglected. Thus, $\chi$ becomes a Lagrange multiplier field, forcing $\psi_{+}$ to vanish. This indicates a change from the Dirichlet to Neumann boundary condition. The corresponding asymptotic coefficient $\psi_-(\tau)$ may be identified with $\chi(\tau)$, whose correlator is $-G(\tau_1,\tau_2)$.

To summarize, the zero-temperature entropy of the Majorana SYK model is
\begin{equation}
\sz=\bigl[\ln Z_{\text{full}}-\ln Z_{\text{bath}}\bigr]_{\text{reg}}\,,
\end{equation}
where $[\cdots]_{\text{reg}}$ denotes the constant term in the $1/\beta$ expansion. The partition functions $Z_{\text{bath}}$ and $Z_{\text{full}}$ correspond to a Majorana fermion on the hyperbolic plane with the Dirichlet and Neumann boundary conditions, respectively. For the complex SYK model, one should consider $\gz(\calE)$ instead of $\sz$ and use a Dirac fermion. The calculation will follow. We note that this procedure is similar to that used to compute the influence of double trace operators on the free energy in the AdS/CFT correspondence~\cite{gubser2003double,Gubser:2002vv,Diaz:2007an,Giombi:2013yva,Giombi:2014xxa,Giombi:2016pvg}.

\subsection{Dirac fermion on the hyperbolic plane}

Now we describe a realization of the auxiliary ``bath'' system for the complex SYK model. The abstract action $I_{\text{bath}}=-\Psi^{\dag}B_{\text{bath}}\Psi$ is chosen in the form
\begin{equation}
I_{\text{Dirac}}
=\int i \bar{\psi}\left(\gamma^c\nabla_c + M\right)\psi\,\sqrt{g}\,d^2x\,,\qquad
\psi = \begin{pmatrix}
\psi_{\downarrow} \\ \psi_{\uparrow}
\end{pmatrix}\,,\quad
\bar{\psi} = \begin{pmatrix}
-\psi_{\uparrow}^* & \psi_{\downarrow}^*
\end{pmatrix}\,,
\end{equation}
where
\begin{equation}
\nabla_{\alpha}\psi
=\left(\partial_{\alpha} + \frac{1}{2} \omega_{\alpha bc} \Sigma^{bc}
-iA_\alpha\right) \psi\,.
\end{equation}
Specific to two dimensions, the spin connection factors into a scalar and a constant matrix:
\begin{equation}
\begin{pmatrix}
\omega_{\alpha 11} & \omega_{\alpha 12} \\
\omega_{\alpha 21} & \omega_{\alpha 22} 
\end{pmatrix} = \omega_{\alpha} \begin{pmatrix}
0 & -1 \\
1 & 0
\end{pmatrix} \,,\qquad
\partial_\alpha\omega_\beta-\partial_\beta\omega_\alpha
=-\frac{R}{2}\epsilon_{\alpha\beta}\,.
\end{equation}
(Further details, such as the expressions for the Dirac matrices $\gamma^1,\gamma^2$ and the spin matrices $\Sigma^{ab}= \frac{1}{4} \left[ \gamma^a, \gamma^b\right]$, can be found in appendix~\ref{sec:app_Dirac}.) The Majorana case differs in that $\psi_{\downarrow}$, $\psi_{\uparrow}$ are real, the $\UU(1)$ gauge field $A$ is absent, and the action has an overall factor $\frac{1}{2}$.

We use the Poincare disk model of the hyperbolic plane ${\rm H^2}$:
\begin{equation}
ds^2=4 \frac{dr^2+r^2 d\varphi^2}{(1-r^2)^2}\,.
\end{equation}
The $\UU(1)$ gauge field $A$ is imaginary (but becomes real upon the analytic continuation from the hyperbolic plane to the global anti-de Sitter space sharing a diameter of the Poincare disk). More specifically,
\begin{equation}
A_\alpha=-i\calE\omega_\alpha,\quad
\partial_{\alpha}A_{\beta}-\partial_{\beta}A_{\alpha}
=-i\calE\epsilon_{\alpha\beta}\,.
\end{equation}
Thus, the model is characterized by the Dirac mass $M$ and the field strength $\calE$. We also need to specify a boundary condition. To this end, we note that a general solution of the Dirac equation $(\gamma^c\nabla_c+M)\psi=0$ has this asymptotic form near the boundary:
\begin{equation}
\psi(r,\varphi) \approx\psi_{+}(\varphi)\,\eta_{+}(r,\varphi)
+\psi_{-}(\varphi)\,\eta_{-}(r,\varphi)\quad \text{for } r\to 1\,,
\end{equation}
where
\begin{equation}
\eta_{\pm} (r,\varphi)= \bigl(1-r^2\bigr)^{\Delta_{\pm}}
\begin{pmatrix}
e^{i \frac{(\varphi\pm\gamma)}{2}} \\
\pm e^{-i \frac{(\varphi\pm\gamma)}{2}}
\end{pmatrix} \,,\qquad
\Delta_{\pm}=\frac{1}{2}\pm\sqrt{M^2-\calE^2}\,,\quad 
\gamma=\arcsin\frac{\calE}{M}\,.
\label{eqn: etaD}
\end{equation}
\begin{figure}[t]
\centerline{\begin{tikzpicture}[scale=0.8, baseline={([yshift=-4pt]current bounding box.center)}]
    \draw[thick] (0pt,0pt) circle (40pt);
       \draw[thick, ->,>=stealth] (0pt,0pt) -- (15pt,0pt);
       \node[right] at (13pt,0pt) {$\Vec{e}_1$};
       \draw[thick, ->,>=stealth] (0pt,0pt) -- (0pt,15pt); 
       \node[above] at (0pt,13pt) {$\Vec{e}_2$};
       \draw[->,>=stealth] (15pt,15pt) -- ++(8pt,0pt);
       \draw[->,>=stealth] (15pt,15pt) -- ++(0pt,8pt);
       \draw[->,>=stealth] (-15pt,-15pt) -- ++(8pt,0pt);
       \draw[->,>=stealth] (-15pt,-15pt) -- ++(0pt,8pt);
       \draw[->,>=stealth] (15pt,-15pt) -- ++(8pt,0pt);
       \draw[->,>=stealth] (15pt,-15pt) -- ++(0pt,8pt);
       \draw[->,>=stealth] (-15pt,15pt) -- ++(8pt,0pt);
       \draw[->,>=stealth] (-15pt,15pt) -- ++(0pt,8pt);
       \draw[->,>=stealth, blue,thick] (28.28pt, 28.28pt) -- ++ (-20pt,20pt) node[right]{$\partial_\varphi$};
\end{tikzpicture}}
\caption{Local frame $(\Vec{e}_1,\Vec{e}_2)$ relative to which the Dirac spinor is defined.}\label{fig:frame}
\end{figure}
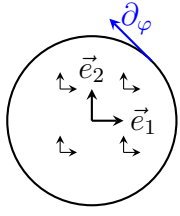
The dependence on the polar angle $\varphi$ in Eq.~\eqref{eqn: etaD} is a consequence of gauge choice: we use the local frame (vielbein) shown in Fig.~\ref{fig:frame}, whose orientation relative to the tangent vector $\partial_\varphi$ depends on $\varphi$. For the bath model, we postulate the Dirichlet boundary condition, $\psi_{-}(\varphi)=0$. But when the bulk fermion is coupled to a boundary fermion, the correct condition is Neumann, $\psi_{+}(\varphi)=0$.

The Euclidean propagator for each boundary condition,
\begin{equation}
C_{\pm}=-i(\gamma^c\nabla_c+M)^{-1}_{\pm},\qquad
C_{\pm}(x_1,x_0)=\bigl\langle\psi(x_1)\bar{\psi}(x_0)\bigr\rangle_{\pm}
\end{equation}
with the matrix structure
\begin{equation}
C_{\pm}=\begin{pmatrix}
-C^{\downarrow \uparrow}_{\pm} &  C^{\downarrow \downarrow}_{\pm}
\vspace{2pt}\\
-C^{\uparrow\uparrow}_{\pm} & C^{\uparrow \downarrow}_{\pm} 
\end{pmatrix}\,, \quad C^{jk}_{\pm} = \langle \psi_j \psi_k^* \rangle_{\pm}\,,
\end{equation}
is calculated in appendix~\ref{sec:app_Dirac}, see Eq.~\eqref{eqn: propagator Z1Z0}. In particular, when both $x_1=(r_1,\varphi_1)$ and $x_0=(r_0,\varphi_0)$ approach the boundary, the propagator becomes
\begin{equation}
C_{\pm}(r_1,\varphi_1;r_0,\varphi_0)
\approx\bigl\langle\psi_{\pm}(\varphi_1)
\bar{\psi}_{\pm}(\varphi_0)\bigr\rangle\,
\eta(r_1,\varphi_1)\bar{\eta}(r_0,\varphi_0)\quad \text{for } r_1,r_0\to 1\,,
\end{equation}
where for $0<\varphi_1-\varphi_0<2\pi$, we have
\begin{equation}
\bigl\langle\psi_{\pm}(\varphi_1)\psi_{\pm}(\varphi_0)\bigr\rangle
=\frac{\Gamma\bigl(\Delta_{\pm}+\frac{1}{2}+i\calE\bigr)\,
\Gamma\bigl(\Delta_{\pm}+\frac{1}{2}-i\calE\bigr)}{4\pi\,\Gamma(2\Delta_\pm)}\,
e^{\calE (\pi-\varphi_1+\varphi_0) }
\biggl( 2\sin \frac{\varphi_1-\varphi_0}{2} \biggr)^{-2\Delta_{\pm}}.
\end{equation}
Thus, $\langle\psi_{+}(\varphi_1)\psi_{+}(\varphi_0)\rangle \sim -\tilde{\Sigma}(\varphi_1,\varphi_0)$ and $\langle\psi_{-}(\varphi_1)\psi_{-}(\varphi_0)\rangle \sim -G(\varphi_1,\varphi_0)$ (up to some constant factors), where $\tilde{\Sigma}$ and $G$ are defined for the complex SYK model with
\begin{equation}
\Delta=\frac{1}{2}-\sqrt{M^2-\calE^2}\,.
\label{eqn:DM}
\end{equation}

\subsection{Subtraction of infinities and the ``spooky propagator''}

We are now in a position to evaluate the thermodynamic quantity
\begin{equation}
\gz(\Delta,\calE)=\bigl[\ln Z_{\text{full}}-\ln Z_{\text{bath}}\bigr]_{\text{reg}}
=\bigl[\ln\det(\gamma^c\nabla_c+M)_{-}
-\ln\det(\gamma^c\nabla_c+M)_{+}\bigr]_{\text{reg}}\,.
\end{equation}
Each of the two terms in the square brackets suffers from a UV divergence and the divergence due to infinite volume. The former is canceled due to the subtraction of the terms and the latter due to the regularization $[\cdots]_{\text{reg}}$, which amounts to the subtraction of a boundary contribution. The two terms exactly cancel each other if $M=|\calE|$. For $M>|\calE|$, it is convenient to take the derivative with respect to $M$ using the relation \eqref{eqn:DM} between $M$ and $\Delta$:
\begin{equation}
\frac{M}{\Delta-1/2}\,\frac{\partial\gz(\Delta,\calE)}{\partial\Delta}
=\bigl[\Tr(\gamma^c\nabla_c+M)_{-}^{-1}
-\Tr(\gamma^c\nabla_c+M)_{+}^{-1}\bigr]_{\text{reg}}
=i\bigl[\Tr(C_{-}-C_{+})\bigr]_{\text{reg}}\,.
\label{eqn:dertr1}
\end{equation}

In the last expression, $-C_{+}$ may be regarded as a propagator of a ghost particle. For this reason, we call the difference $C_{\spooky}=C_{-}-C_{+}$ the ``spooky propagator''. The function $C_{\spooky}(x_1,x_0)$ has no singularity at $x_1=x_0$ and may be interpreted as the bulk fermion propagating from point $x_0$ to the boundary, where it mixes with the boundary fermion, and then moving to point $x_1$.\,\footnote{More exactly, $C_{\spooky} \sim (\text{boundary to bulk})_+ \cdot G \cdot (\text{bulk to boundary})_+$. The boundary-to-bulk and bulk-to-boundary propagators are, actually, $\tilde{\SL}(2,\RR)$ intertwiners.} This is an explicit formula:
\begin{equation}
C_{\spooky}(r,\varphi;0)=\frac{M\sin(2\pi\Delta)}
{4i\cos(\pi(\Delta-i\calE))\cos(\pi(\Delta+i\calE))}
\begin{pmatrix}
-A_{\Delta,\frac{1}{2}+i\calE,-\frac{1}{2}-i\calE}(r^2) &
e^{i\varphi}A_{\Delta,\frac{1}{2}+i\calE,\frac{1}{2}-i\calE}(r^2) \\
e^{-i\varphi}A_{\Delta,\frac{1}{2}+i\calE,\frac{1}{2}-i\calE}(r^2) &
-A_{\Delta,-\frac{1}{2}+i\calE,\frac{1}{2}-i\calE}(r^2)
\end{pmatrix}\,,
\end{equation}
where
\begin{equation}
A_{\lambda,l,r}(u)=u^{(l+r)/2}(1-u)^{\lambda}
\textbf{F}(\lambda+l,\lambda+r,1+l+r;u)\,,\quad
\textbf{F}(a,b,c;u)=\frac{{}_2{\rm F}_1(a,b,c;u)}{\Gamma(c)}\,.
\end{equation}

Let us complete the calculation of $\gz(\Delta,\calE)$ using Eq.~\eqref{eqn:dertr1}. We have
\begin{equation}
\Tr C_{\spooky}=\int_{\rm H^2} \Tr C_{\spooky}(x,x)\,\sqrt{g(x)}\,d^2x
= \text{Area}({\rm H^2})\cdot\Tr C_{\spooky}(0,0)\,.
\end{equation}
The area of the hyperbolic plane is obviously infinite, but it can be made finite by regularization. Indeed, consider the disk $D_r$ of radius $r$ centered at the origin. It has the following area and boundary length:
\begin{equation}
\text{Area}(D_r)=4\pi\int_{0}^{r^2}\frac{dx}{(1-x)^2}
=\frac{4\pi r^2}{1-r^2}\,,\qquad
\text{Length}(\partial D_r)=\frac{4\pi r}{1-r^2}\,,
\end{equation}
so that
\begin{equation}
\lim_{r\to 1}\bigl(\text{Area}(D_r)-\text{Length}(\partial D_r)\bigr)=-2\pi\,.
\end{equation}
Hence, $[\Tr C_{\spooky}]_{\text{reg}}=-2\pi \Tr C_{\spooky}(0,0)$. Plugging this in \eqref{eqn:dertr1}, we get:
\begin{equation}
\frac{\partial\gz(\Delta,\calE)}{\partial\Delta}
=\frac{i\pi(1-2\Delta)}{M}\Tr C_{\spooky}(0,0)
=-\frac{\pi(1-2\Delta)\sin(2\pi\Delta)}
{2\cos(\pi(\Delta+i\calE))\cos(\pi(\Delta-i\calE))}\,.
\label{eqn:dgz/dD}
\end{equation}
(This is also equal to $-2\pi^2b$, where $b$ is defined in \eqref{zeroT Green}.) Thus,
\begin{equation}
\gz(\Delta,\calE)=\int_{\Delta}^{1/2}
\frac{\pi(1-2x)\sin(2\pi x)}{\cosh(2\pi\calE)+\cos(2\pi x)}\,dx\,.
\end{equation}

In conclusion, we rewrite Eq.~\eqref{eqn:dgz/dD} as follows,
\begin{equation}
\frac{\partial\gz(\Delta,\calE)}{\partial\Delta}
=-\pi\biggl(\frac{1}{2}-\Delta\biggr)
\Bigl(\tan\pi(\Delta+i\calE)+\tan\pi(\Delta-i\calE)\Bigr)\,,
\end{equation}
and note that it is consistent with the combination of \eqref{g and Q} and \eqref{charge theta}:
\begin{equation}
\frac{\partial\gz(\Delta,\calE)}{\partial\calE}=-2\pi\calQ
=2\theta - i\pi\biggl(\frac{1}{2}-\Delta\biggr)
\Bigl(\tan\pi(\Delta+i\calE)-\tan\pi(\Delta-i\calE)\Bigr)\,.
\end{equation}
Indeed, both equations give the same result for the mixed derivative if we use the fact that $\partial(2\theta)/\partial\Delta =-i\pi \bigl(\tan\pi(\Delta+i\calE)-\tan\pi(\Delta-i\calE)\bigr)$.

\subsection{Relation to the Plancherel factor}
\label{sec:Plancherel}

For readers who are familiar with the Plancherel measure for $\tilde{\SL}(2,\RR)$~\cite{Pukanszky64,Kitaev:2017hnr}, it may be tempting to relate the key ingredient in the entropy formula,
\begin{equation}
\Tr C_{\spooky}(0,0) =  \frac{i M \sin (2\pi \Delta)}{2 \cos (\pi (\Delta+i\calE))  \cos (\pi (\Delta-i\calE)) }\,,
\label{eqn: 5.4 goal}
\end{equation}
to the Plancherel factor. The latter also appears in the decomposition of the unit operator $\one^\nu$ acting on $\nu$-spinors (for an arbitrary real $\nu$) on the hyperbolic plane~\cite{Kitaev:2017hnr}: 
\begin{equation}
\one^\nu = \frac{1}{2\pi} \Biggl(
\int_{0}^{+\infty} ds\, \frac{s \sinh (2\pi s)}{\cosh (2\pi s)+\cos (2\pi \nu)}\, \Pi^{\nu}_{1/2+is} + \sum_{\substack{\lambda = |\nu|-p > 1/2\\
p=0,1,2,\ldots}} \left( \lambda - \frac{1}{2} \right) \Pi_{\lambda}^{\nu}
\Biggr) \,,
\label{eqn: decomposition of 1}
\end{equation}
where $\Pi^{\nu}_{\lambda}$ is the projector onto the eigenspace of the $\tilde{\SL}(2,\RR)$ Casimir operator with the eigenvalue $\lambda(1-\lambda)$. The operators $\Pi^{\nu}_{\lambda}$ are defined by integral kernels that depend on pairs of points $x_1,x_0 \in {\rm H^2}$; the normalization is such that $\Pi^{\nu}_{\lambda}(x,x)=1$.

We will make the connection to the Plancherel factor explicit by deriving \eqref{eqn: 5.4 goal} from \eqref{eqn: decomposition of 1}, bypassing the full calculation of the Dirac propagator. As explained in appendix~\ref{sec:app_Dirac}, the components of a Dirac spinor have different effective spins $\nu$, equal to ${\downarrow}=-\frac{1}{2}-i\calE$ and ${\uparrow}=\frac{1}{2}-i\calE$. The Dirac operator is represented by the matrix
\begin{equation}
\gamma^c\nabla_c+M =
\begin{pmatrix}
M & 2\nabla_- \\ 
2 \nabla_+ & M
\end{pmatrix}\,,
\label{eqn: block D}
\end{equation}
where $\nabla_+$ and $\nabla_-$ are certain differential operators changing the value of $\nu$ by $1$ and $-1$, respectively. (Here the subscripts ``$\pm$'' have nothing to do with boundary conditions.) The Casimir operator is expressed in terms of $\nabla_\pm$ by Eq.~\eqref{eqn: Casimir}, so both $4\nabla_-\nabla_+$ for $\nu={\downarrow}$ and $4\nabla_+\nabla_-$ for $\nu={\uparrow}$ are equal to $\frac{1}{4}+\calE^2-Q$. Using this and the formula $\Delta=\frac{1}{2}-\sqrt{M^2-\calE^2}$, we obtain the following expression for the propagator:
\begin{equation}
C =
\begin{pmatrix}
  -C^{\downarrow\uparrow} &  C^{\downarrow\downarrow}
  \vspace{2pt}\\
  -C^{\uparrow\uparrow} & C^{\uparrow\downarrow} 
\end{pmatrix}
=-i(\gamma^c\nabla_c+M)^{-1}
= -i\bigl(Q-\Delta(1-\Delta)\bigr)^{-1}
\begin{pmatrix}
  M & -2\nabla_- \\
  -2\nabla_+ & M
\end{pmatrix}.
\label{eqn: Cgen}
\end{equation}
Let us first calculate the matrix element involving $\nu=\frac{1}{2}-i\calE$ spinors with Dirichlet boundary condition (indicated by the subscript ``$+$''),
\begin{equation}
C^{\uparrow\downarrow}_{+}
= -iM\bigl(Q-\Delta(1-\Delta)\bigr)^{-1}_{+}\,.
\end{equation}
The general idea is to use the Casimir eigendecomposition \eqref{eqn: decomposition of 1}; the role of boundary condition will become clear later.

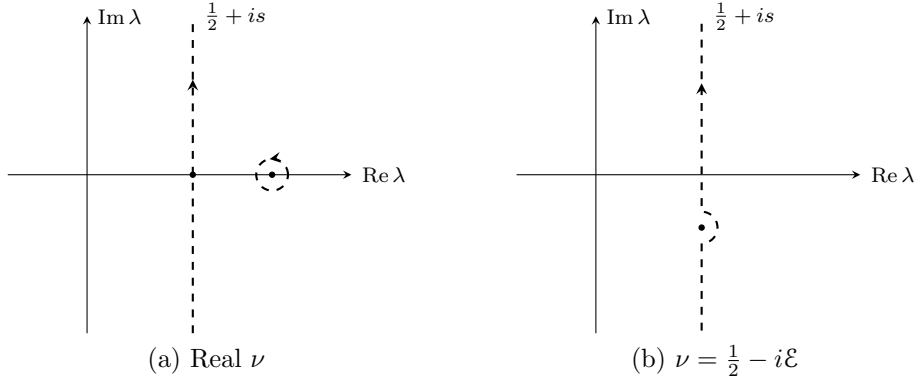
\begin{figure}[t]\centering
\subfloat[Real $\nu$]{
\begin{tikzpicture}[baseline={(current bounding box.center)}]
\draw [->,>=stealth] (-30pt,0pt) -- (100pt,0pt) node[right]{\scriptsize $\Re \lambda$};
\draw [->,>=stealth] (0pt,-60pt) -- (0pt,60pt) node[right]{\scriptsize $\Im \lambda$};
\draw[far arrow, dashed,thick] (40pt,-60pt)--(40pt,60pt) node[right]{\scriptsize $\frac{1}{2}+is$};
\filldraw (40pt,0pt) circle (1pt);
\filldraw (70pt,0pt) circle (1pt) ;
\draw[thick, dashed, near arrow] (70pt,0pt) circle (6pt);
\end{tikzpicture}}
\hspace{30pt}
\subfloat[$\nu=\frac{1}{2}-i\calE$]{
\begin{tikzpicture}[baseline={(current bounding box.center)}]
\draw [->,>=stealth] (-30pt,0pt) -- (100pt,0pt) node[right]{\scriptsize $\Re \lambda$};
\draw [->,>=stealth] (0pt,-60pt) -- (0pt,60pt)
node[right]{\scriptsize $\Im \lambda$};
\draw[far arrow, dashed, thick, dash phase=-1pt]
  (40pt,-60pt) -- (40pt,-26pt)
  arc (-90:90:6pt)
  -- (40pt,60pt)
  node[right]{\scriptsize $\frac{1}{2}+is$};
\filldraw (40pt,-20pt) circle (1pt) ;
\end{tikzpicture}}
\caption{Contour $\Gamma$ in Eq.~\eqref{eqn: new one} includes the vertical line $\Re\lambda=\frac{1}{2}$ and also encircles the points $\lambda=\nu+n$ (for integer $n$) between that line and the line $\Re(\lambda-\nu)=\frac{1}{2}$.}
\label{fig: new contour}
\end{figure}

For the task at hand, it is convenient to transform Eq.~\eqref{eqn: decomposition of 1} to a different form, which generalizes to complex values of $\nu$:
\begin{equation}
\one^\nu = \frac{i}{4\pi } \int_{\Gamma}  d\lambda\,
\Bigl(\lambda-\frac{1}{2}\Bigr)
\tan\biggl(\pi\Bigl(\lambda-\frac{1}{2}-\nu\Bigr)\biggr)\,
\Pi^{\nu}_{\lambda}\,,
\label{eqn: new one}
\end{equation}
where  the contour $\Gamma$ is illustrated in Fig.~\ref{fig: new contour}(a). It is obtained by a deformation of the vertical line $\Re(\lambda-\nu)=\frac{1}{2}$ and consists of the line from $\frac{1}{2}-i\infty$ to $\frac{1}{2}+i\infty$ and circles surrounding the poles of $\tan\bigl(\pi\bigl(\lambda-\frac{1}{2} -\nu\bigr)\bigr)$ in the strip ${\frac{1}{2}< \Re\lambda < \frac{1}{2}+\Re\nu}$ or ${\frac{1}{2}+\Re\nu < \Re\lambda < \frac{1}{2}}$ (depending on the sign of $\Re\nu$). The rewriting is based on this representation of the Plancherel factor,
\begin{equation}
\frac{s \sinh (2\pi s)}{\cosh (2\pi s)+\cos (2\pi \nu)}=-\frac{is}{2}\Bigl(  \tan \bigl(\pi (is-\nu)\bigr) -\tan \bigl(\pi (-is-\nu)\bigr) \Bigr)\,,
\end{equation}
and the symmetry $\Pi^\nu_{\lambda}=\Pi^\nu_{1-\lambda}$, which allows one to extend the integral in \eqref{eqn: decomposition of 1} from a half-line to a full line. More explicitly,
\begin{equation}
\int_{0}^{+\infty} ds\, \frac{s \sinh (2\pi s)}{\cosh (2\pi s)+\cos (2\pi \nu)} \Pi^{\nu}_{1/2+is}\,
= \frac{i}{2} \int_{\frac{1}{2}-i\infty}^{\frac{1}{2}+i\infty}
d\lambda\,
\Bigl(\lambda-\frac{1}{2}\Bigr)
\tan\biggl(\pi\Bigl(\lambda-\frac{1}{2}-\nu\Bigr)\biggr)\,
\Pi^{\nu}_{\lambda}\,.
\end{equation}
The discrete series contribution (i.e.\ the second term in \eqref{eqn: decomposition of 1}) can be treated as residues of the same integrand, which leads to the expression \eqref{eqn: new one}. Note that when $\lambda$ and $\nu$ are arbitrary complex numbers, $\Pi^\nu_{\lambda}$ is no longer an orthogonal projector. Formally, it is just a function of $x_1,x_0 \in {\rm H^2}$, and $\one^\nu$ should likewise be interpreted as a (generalized) function, namely, $g(x_0)^{-1/2}\delta(x_1-x_0)$, where $g$ is the determinant of the metric tensor.

Given this caveat, we will proceed with caution. It is true that
\begin{equation}
Q\Pi^\nu_{\lambda}=\Pi^\nu_{\lambda}Q=\lambda(1-\lambda)\Pi^\nu_{\lambda}\,.
\end{equation}
However, the following corollary holds only for the Dirichlet boundary condition and is qualified by a restriction on $\lambda$:
\begin{equation}
C^{\uparrow\downarrow}_{+}\Pi^{1/2-i\calE}_{\lambda}
= -iM\bigl(\lambda(1-\lambda)-\Delta(1-\Delta)\bigr)^{-1}
\Pi^{1/2-i\calE}_{\lambda}\quad\:
\text{for}\quad \Delta<\Re\lambda<1-\Delta\,.
\label{eqn: PiC}
\end{equation}
Indeed, we should require that the left-hand side of the above equation be well-defined, meaning the absolute convergence of the corresponding integral:
\begin{equation}
\bigl(C^{\uparrow\downarrow}_{+}\Pi^{1/2-i\calE}_{\lambda}\bigr)(x_1,x_0)
=\int C^{\uparrow\downarrow}_{+}(x_1,x)
\Pi^{1/2-i\calE}_{\lambda}(x,x_0)\,\sqrt{g(x)}\,d^2x\,.
\end{equation}
To check this condition, let us use polar coordinates, $x=(r,\varphi)$. As $r$ tends to $1$, the propagator $C^{\uparrow\downarrow}_{+}(x_1,x)$ scales as $(1-r)^{1-\Delta}$, whereas $\Pi^\nu_{\lambda}(x,x_0)$ has terms proportional to $(1-r)^{\lambda}$ and $(1-r)^{1-\lambda}$. Since $\sqrt{g(x)}\sim(1-r)^{-2}$, the convergence condition is exactly as indicated in Eq.~\eqref{eqn: PiC}.

\begin{figure}[t]
\center
\subfloat[Contour $\Gamma$ for $C_+$]{
\begin{tikzpicture}[baseline={(current bounding box.center)}]
\draw [->,>=stealth] (-10pt,0pt) -- (90pt,0pt) node[right]{\scriptsize $\Re \lambda$};
\draw [->,>=stealth] (0pt,-60pt) -- (00pt,60pt) node[right]{\scriptsize $\Im \lambda$};
\draw[far arrow, dashed, thick]
  (40pt,-60pt) -- (40pt,-11pt) arc (-90:90:5pt) -- (40pt,60pt)
  node[right]{\scriptsize $\frac{1}{2}+is$};
\filldraw (40pt,-6pt) circle (1pt) ;
\filldraw (60pt,0pt) circle (1pt); 
\node[below] at (63pt,-3pt) {\scriptsize $\Delta_+$};
\filldraw (20pt,0pt) circle (1pt);
\node[below] at (23pt,-3pt) {\scriptsize $\Delta_-$};
\end{tikzpicture}}
\hspace{15pt}
\subfloat[Contour $\tilde{\Gamma}$ for $C_-$]{
\begin{tikzpicture}[baseline={(current bounding box.center)}]
\draw [->,>=stealth] (-10pt,0pt) -- (90pt,0pt) node[right]{\scriptsize $\Re \lambda$};
\draw [->,>=stealth] (0pt,-60pt) -- (00pt,60pt) node[right]{\scriptsize $\Im \lambda$};
\draw[dashed,thick,blue, mid arrow] (40pt,-60pt)--(40pt,-25pt); 
\draw[dashed,thick,blue]
  (40pt,-15pt) -- (40pt,-11pt) arc (-90:90:5pt) -- (40pt,15pt);
\filldraw (40pt,-6pt) circle (1pt);
\draw[dashed,thick,blue, mid arrow] (40pt,25pt)--(40pt,60pt); 
\node[right] at (40pt,60pt) {\scriptsize $\frac{1}{2}+is$};
\filldraw (60pt,0pt) circle (1pt); 
\node[below] at (75pt,-3pt) {\scriptsize $\Delta_+$};
\filldraw (20pt,0pt) circle (1pt); 
\node[below] at (10pt,-3pt) {\scriptsize $\Delta_-$};
\draw[thick,densely dashed,blue] (15pt,0pt) arc (-180:0:5pt);
\draw[thick,densely dashed,blue]  (25pt,0pt) arc (180:90:15pt);
\draw[thick,densely dashed,blue]  (15pt,0pt) arc (180:90:25pt);

\draw[thick,densely dashed,blue] (65pt,0pt) arc (0:180:5pt);
\draw[thick,densely dashed,blue]  (55pt,0pt) arc (0:-90:15pt);
\draw[thick,densely dashed,blue]  (65pt,0pt) arc (0:-90:25pt);
\draw[red,thick, <-,>=stealth] (21.21pt,6.84pt) arc (160:35:20pt);
\draw[red,thick, <-,>=stealth] (58.79pt,-6.84pt) arc (-20:-145:20pt);
\end{tikzpicture}}
\hspace{15pt}
\subfloat[Contour $\Gamma_{\spooky}\sim\tilde{\Gamma}-\Gamma$ for $C_{\spooky}$]{
\begin{tikzpicture}[baseline={(current bounding box.center)}]
\draw [->,>=stealth] (-10pt,0pt) -- (90pt,0pt) node[right]{\scriptsize $\Re \lambda$};
\draw [->,>=stealth] (0pt,-60pt) -- (00pt,60pt) node[right]{\scriptsize $\Im \lambda$};
\filldraw (60pt,0pt) circle (1pt); 
\node[below] at (63pt,-8pt) {\scriptsize $\Delta_+$};
\filldraw (20pt,0pt) circle (1pt); 
\node[below] at (23pt,-8pt) {\scriptsize $\Delta_-$};
\draw[thick,densely dashed,blue, near arrow, xscale=-1] (-20pt,0pt) circle (8pt);
\draw[thick,densely dashed,blue, near arrow] (60pt,0pt) circle (8pt);
\filldraw (40pt,-6pt) circle (1pt);
\end{tikzpicture}\hspace{8pt}}
\caption{Switching the boundary condition from Dirichlet to Neumann amounts to exchanging the poles $\Delta_-$ and $\Delta_+$. The procedure should be accompanied by a deformation of the integration contour as shown in (b). The difference contour $\tilde{\Gamma}-\Gamma$ is homologous (in the complement of singularities) to the one shown in (c).}
\label{fig: spooky}
\end{figure}
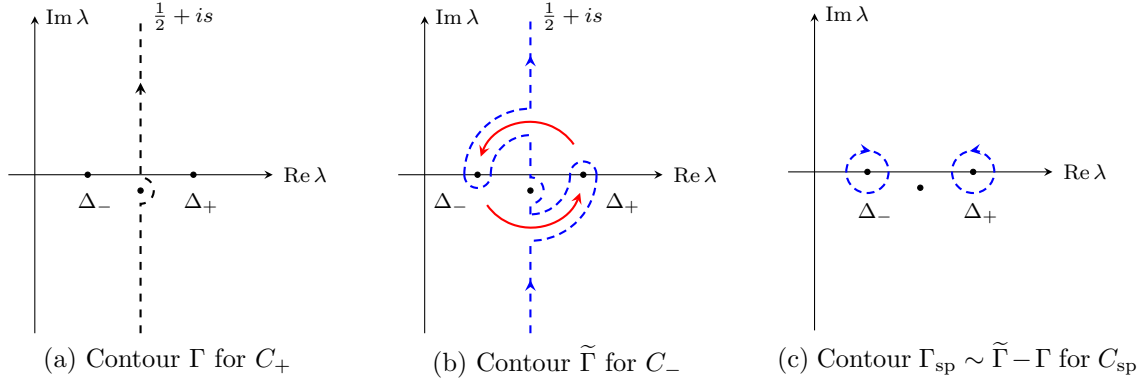
 
We now apply the decomposition of identity \eqref{eqn: new one} together with Eq.~\eqref{eqn: PiC}:
\begin{equation}
C^{\uparrow\downarrow}_{+}
= C^{\uparrow\downarrow}_{+} \cdot \one^{1/2-i\calE}
=\frac{M}{4\pi } \int_{\Gamma}  d\lambda\,
\frac{\bigl(\lambda-\frac{1}{2}\bigr)
\tan\bigl(\pi(\lambda+i\calE)\bigr)}
{(\lambda-\Delta)(1-\lambda-\Delta)}\,
\Pi^{1/2-i\calE}_{\lambda} \,.
\label{eqn: C_+ integral}
\end{equation}
 Note that the contour $\Gamma$ passes between the poles of the integrand at $\Delta_-=\Delta$ and $\Delta_+=1-\Delta$. The propagator $C^{\uparrow\downarrow}_{-}$ with Neumann boundary condition cannot be obtained in the same way, but we can use analytic continuation in $M$. Suppose that $M>|\calE|$ initially. As $M$ changes to $-M$ avoiding the branch cut between $\calE$ and $-\calE$, the numbers $\Delta_+$ and $\Delta_-$ are swapped, and the propagator $C^{\uparrow\downarrow}_{+}$ turns into $-C^{\uparrow\downarrow}_{-}$ for the original value of $M$. On the right-hand side of \eqref{eqn: C_+ integral}, the analytic continuation should involve a deformation of the integration contour such that it avoids the moving poles, see Fig.~\ref{fig: spooky}. Thus,
\begin{equation}
C^{\uparrow\downarrow}_{-}
=\frac{M}{4\pi } \int_{\tilde{\Gamma}}  d\lambda\,
\frac{\bigl(\lambda-\frac{1}{2}\bigr)
\tan\bigl(\pi(\lambda+i\calE)\bigr)}
{(\lambda-\Delta)(1-\lambda-\Delta)}\,
\Pi^{1/2-i\calE}_{\lambda} \,.
\label{eqn: C_- integral}
\end{equation}
The ``spooky propagator'' $C^{\uparrow\downarrow}_{\spooky} = C^{\uparrow\downarrow}_{-} - C^{\uparrow\downarrow}_{+}$ is given by the integral of the same function along the difference contour $\Gamma_{\spooky}\sim\tilde{\Gamma}-\Gamma$, which consists of circles wrapping the points $\lambda=\Delta_-$ (clockwise) and $\lambda=\Delta_+$ (counterclockwise) as shown in Fig.~\ref{fig: spooky}(c). Hence, the spooky propagator is determined by the residues of the integrand at $\Delta_-$ and  $\Delta_+$:
\begin{equation}
C^{\uparrow\downarrow}_{\spooky}
=  \frac{i M}{4} \Bigl(\tan \bigl( \pi (\Delta+i\calE)   \bigr) + \tan \bigl( \pi (\Delta-i\calE)\bigr) \Bigr)    \Pi^{1/2-i\calE}_{\Delta} \,.
\end{equation}

The calculation of the other diagonal element of the propagator matrix, $-C^{\downarrow\uparrow}$ (in all three variants) is completely analogous; we just need to use $\nu=-\frac{1}{2}-i\calE$. Restricting to coincident points and using the normalization condition $\Pi^{\nu}_{\lambda}(x,x)=1$, we obtain the final result: 
\begin{equation}
\Tr C_{\spooky}(0,0) = \frac{i M}{2} \Bigl(\tan \bigl( \pi \bigl( \Delta+i\calE \bigr)   \bigr) + \tan \bigl( \pi \bigl( \Delta-i\calE \bigr)   \bigr) \Bigr)  \,,
\end{equation}
which is equivalent to Eq.~\eqref{eqn: 5.4 goal}.

\section{Discussion}
\label{sec:conc}

One of the main new physical consequences of our computations on the complex SYK model is the many-body density of states in Eq.~(\ref{DQE}). For each total charge $Q$, the energy dependence of the density of states is the same as in the Schwarzian theory, with a ground state energy $E_0 (Q)$, and a zero temperature entropy $\calS (Q/N)$ determined by the value of $Q$. Although this result is natural from the physical point of view, we derived it from the effective action \eqref{Seff}, which describes an ensemble with fluctuating $Q$. The presence of the particle-hole asymmetry parameter $\calE$ in the action was essential for the consistency of that calculation.

The other parameters in the effective action in Eq.~(\ref{Seff}) are the charge compressibility $K$, and $\gamma$, the coefficient of the $T$-linear specific heat at fixed $Q$. While the value of $\gamma$ was determined by a low-energy analysis using conformal perturbation theory \cite{MS16-remarks,KS17-soft}, we have shown here that a similar procedure does not apply for $K$. This is highlighted by the UV divergence in the eigenmodes of the symmetric sector of the two-particle kernel shown in Fig.~\ref{phi21nAS}. 
It is necessary to account for high energy contributions to obtain the correct value of $K$, and we presented three such computations in Sections~\ref{sec:exact}, \ref{numschwindys}, and \ref{kerneldiag}; the numerical values so obtained were consistent with each other. These distinct behaviors of $\gamma$ and $K$ are analogous to those in the Fermi liquid theory: the quasiparticle effective mass $m^\ast$ determines the specific heat, but an additional Landau parameter, $F_0^s$, is needed for the compressibility.

We presented a new computation of the zero temperature entropy $\sz$ of the complex SYK model in Section~\ref{sec:bulk}. The entropy was shown to be equal to the difference in the logarithm of the partition function of a massive Dirac fermion on ${\rm H^2}$ between Neumann and Dirichlet boundary conditions, in a manner similar to the influence of double-trace operators in the usual AdS/CFT correspondence \cite{gubser2003double,Gubser:2002vv,Diaz:2007an,Giombi:2013yva,Giombi:2014xxa,Giombi:2016pvg}. This bulk approach correctly reproduced the $\calQ$ and $\calE$ dependence of $\sz$ in the SYK model. 

The above computation of the entropy should be contrasted with that in higher dimensional black holes whose near-horizon geometry has an AdS$_2$ factor (reviewed in Appendix~\ref{app:em}), where the entropy is given by the horizon area in the higher-dimensional space, and arises from degrees of freedom unrelated to the fermions. While all entropies mentioned here obey Eq.~(\ref{dSdQ}), the functional form of $\sz (\calQ)$ is different for the higher-dimensional black holes \cite{SS15}. Probe Dirac fermions can be added to such higher-dimensional black holes \cite{Faulkner09,Iqbal:2011ae}, 
and their Green function agrees with those of the SYK model \cite{SS10,SS15}; however such fermions
only contribute $\mathcal{O}(1)$ entropy in the distinct large-$N$ limit of the usual AdS/CFT correspondence.

\section*{Acknowledgments}

We thank
Wenbo Fu, Antoine Georges, 
Simone Giombi, Luca Iliesiu, 
Igor Klebanov, Sung-Sik Lee, 
Juan Maldacena, 
Max Metlitski, 
Olivier Parcollet, 
Xiao-Liang Qi, 
Shinsei Ryu, 
Wei Song, Douglas Stanford and
Cenke Xu
for useful discussions.
Y.G.\ is supported by the Gordon and Betty Moore Foundation EPiQS Initiative through Grant (GBMF-4306) and DOE grant, DE-SC0019030. A.K.\ is supported by the Simons Foundation under grant~376205 and through the ``It from Qubit'' program, as well as by the Institute of Quantum Information and Matter, a NSF Frontier center funded in part by the Gordon and Betty Moore Foundation. 
S.S.\ and G.T.\ are supported by DOE grant, DE-SC0019030. 
G.T.\  acknowledges support from the MURI grant W911NF-14-1-0003 from ARO and by DOE grant DE-SC0007870.
This work was performed in part at KITP, University of California, Santa Barbara supported by the NSF under grant PHY-1748958.

\appendix

\section{Luttinger-Ward analysis and the anomalous contribution to charge}
\label{app:GPS}

In this section, we will discuss frequency domain derivations of the charge formula \eqref{charge theta} for general $q$ following the strategy in  Ref.~\cite{GPS01} (GPS) appendix A. 
Here we aim to provide an alternative route to the discussions in
Section~\ref{section: charge} that may be more transparent to the readers familiar with Luttinger's theorem and Luttinger-Ward functional. We also draw attention to the comparison with perturbative anomalies in quantum field theory.

\subsection{IR divergence and anomaly}
Instead of Feynman propagator used in Ref.~\cite{GPS01}, we will work with the imaginary time Green function for convenience and express the charge as the following integral
\begin{equation}
\calQ=G(0^+) + \frac{1}{2} = \int_{-\infty}^{\infty} \frac{d\omega}{2\pi} G(i\omega) e^{i\omega 0^-} + \frac{1}{2} \,.
\label{QGint}
\end{equation}
We proceed by the standard Luttinger-Ward procedure (see Ref.~\cite{luttinger1960ground}), i.e. inserting the identity $1=\partial_z \left( G(z)^{-1} + \Sigma(z) \right)$, 
which leads to the expression
\begin{equation}
\calQ- \frac{1}{2} =\int_{-i \infty}^{i \infty} \frac{dz}{2\pi i} G(z)\left( \partial_z G^{-1}(z) + \partial_z \Sigma(z) \right)  e^{z 0^-} \,.
\end{equation}
This manipulation is similar to the manipulations done in Eq.~\eqref{charge formula}. However, instead of further anti-symmetrizing the integrand, we split the two terms in braces with an explicit cut-off:
\begin{equation}
\text{r.h.s.}=\underbrace{{\rm P} \int_{-i \infty}^{i \infty} \frac{dz}{2\pi i} G(z) \partial_z G^{-1}(z)e^{z 0^-}  }_{I_1} + \underbrace{{\rm P} \int_{-i \infty}^{i \infty} \frac{dz}{2\pi i} G(z)  \partial_z \Sigma(z) e^{z0^-} }_{I_2}
\,,
\end{equation}
where the principal value is implemented by a symmetric cut-off in frequency-domain:
\begin{equation}
{\rm P} \int_{-i\infty}^{+i\infty} :=\lim_{\eta \rightarrow 0} \left( \int_{-i\infty}^{-i\eta} + \int_{i\eta}^{+i\infty} \right).
\label{principal value}
\end{equation}
We emphasis that the regularization is crucial in the discussion here: the integral $I_1$ and $I_2$ are logarithmically divergent, so their value depends on the regularization-scheme. More explicitly, the logarithmic divergence arises from the IR asymptotics of the Green function (non-Fermi liquid behavior) $G(z) \sim z^{2\Delta- 1}$. On the contrary, the analogous integrals in the standard Luttinger-Ward analysis for the Fermi liquid are well defined (i.e. with no divergence) and one can further prove the second integral actually vanishes due to the existence of the Luttinger-Ward functional \cite{luttinger1960ground}. 

The situation here is very similar to the perturbative anomalies in quantum field theory (e.g. see the discussions in \cite{peskin2018introduction} chapter 19, a similar comment was also made in Ref.~\cite{altshuler1998luttinger}). A particularly simple example is the two dimensional massless QED, where the Feynman diagrams formally satisfy the Ward identity both for vector and axial current. However, the regularizations will make it impossible to have both gauge invariance and the axial current conservation. 

In this section, we choose to use a regulator \eqref{principal value} (following GPS) that let $I_1$ term inherit the physical meaning as in the Fermi liquid, while let $I_2$ term carries the anomalous contribution. As a comparison, the time domain symmetric regulator used in Section~\ref{section: charge} will set $I_2=0$ but shift the value of $I_1$ integral. In any case, the sum $I_1+I_2$ is regularization-scheme independent and determines the physical charge. 

\subsection{Calculation of $I_1$ integral}
Let us first evaluate the $I_1$ integral explicitly
\begin{equation}
I_1 = - {\rm P} \int_{-i \infty}^{i \infty} \frac{dz}{2\pi i} \partial_z \log G(z) e^{z 0^-} = -\left( \int_{0+i\eta}^{\infty+i\eta} - \int_{0-i\eta}^{\infty-i\eta }  \right)  \frac{dz}{2\pi i} \partial_z \log G(z) e^{z0^-}\,.
\end{equation}
In the second step we bend the contour close to the real axis so that we can proceed using the analytic properties of Green function, thus
\begin{equation}
    I_1= - \int_{0^+}^{+\infty} \frac{dz}{2\pi i} \partial_z \log	 \frac{G(z+i\eta)}{G(z-i\eta)} e^{z0^-}  = - \frac{1}{\pi} \lim_{\eta \rightarrow 0}  \left( 
\arg G(\infty+i\eta) -\arg G(i\eta)
 \right)\,.
\end{equation}
We conclude that $I_1$ is determined by the phase difference between UV and IR asymptotics of Green function as in the usual Luttinger-Ward analysis for Fermi liquid:
\begin{equation}
I_1 = - \frac{1}{2} - \frac{\theta}{\pi}\,.
\end{equation}

\subsection{Anomalous Luttinger-Ward term $I_2$ at $q=4$} 
Now, we calculate $I_2$ integral in the present regularization-scheme.   
Before moving to the evaluation of $I_2$ for general $q$, let us first review/simplify and remark on the detailed calculations performed in Ref.~\cite{GPS01} appendix A for $q=4$ model. 

In the reference aforementioned, $I_2$ is expressed using spectral function: 
\begin{equation}
I_2 = {\rm P} \int_{-i \infty}^{i \infty} \frac{dze^{z0^-}}{2\pi i} \int_{-\infty}^{+\infty} d\omega_0 \int_{\{\omega\}} d^3 \omega \frac{\rho (\omega_0) \rho(\omega_1) \rho(\omega_2) \rho(\omega_3) }{(z-\omega_0)(z-(\omega_1+\omega_2-\omega_3))^2} \,,
\label{I2 integral}
\end{equation}
where the domain $\{\omega \}$ is defined as $ \{\omega \}: =\{ \omega_1,\omega_2 >0, \omega_3<0 \} \cup \{ \omega_1,\omega_2 <0, \omega_3>0 \}$\footnote{The sign here is due to the sign structure of the time arguments in self energy  $\Sigma(\tau)=G(\tau)^{q/{2}}(-G(-\tau))^{q/{2}-1}$.}. 
The spectral function $\rho(\omega)=-\frac{1}{\pi} \Im G^R(\omega)$ 
has the following IR asymptotics
\begin{equation}
\rho(\pm \omega) =\frac{1}{\pi}  \underbrace{\sin \left( \Delta \pi \pm \theta \right) \sqrt{\frac{\Gamma(2-2\Delta)}{\Gamma(2\Delta)}} b^{\Delta-\frac{1}{2}}}_{s_{\pm}} \omega^{2\Delta-1} \quad \text{for} \quad 0 < \omega \ll 1 \,.
\label{eqn: spectral}
\end{equation}
The UV behavior of $\rho(\omega)$ has to be determined by numerics. However, $I_2$ integral only depends on the IR asymptotics and therefore universal. 
A simple argument is as follows. Without the cutoff, the $z$ integration in Eq.~\eqref{I2 integral} will run into a logarithmic divergence at small frequency, which can be seen by power counting of the IR asymptotics of $\rho(\omega)$. 
Now assume we consider a variation of the spectral function $\delta \rho(\omega)$ that does not change the IR asymptotics of $\rho(\omega)$, e.g. $\delta \rho(\omega) \sim \omega^{-1/2+s}$ with $s>0$ at $\omega \rightarrow 0$.
Then the corresponding variation of the integral $\delta I_2$
is free of IR divergence as the
$\omega$ integrations contribute a term asymptotic to $z^{-1+s}$ at small $z$ and the $z$ integration is IR finite now. Therefore, for the variation $\delta I_2$, there is no obstruction to take the $\eta\rightarrow 0$ limit first, namely replacing the principal value by an integration along imaginary axis. Then we can integrate $z$ first by deforming the contour to right half plane and picking up the residues:
\begin{equation}
\int^{i \infty }_{-i \infty}  \frac{e^{z0^-} dz}{2\pi i(z-x)(z-y)^2} =
\begin{cases}
\sgn(x) (x-y)^{-2} &  xy<0 \\
0 & xy>0
\end{cases} \,,
\end{equation}
where
$x=\omega_0$ and $y=\omega_1+\omega_2-\omega_3$ in $I_2$. Next, we finish the $d^4 \omega$ integration and have
\begin{equation}
\delta I_2 =  \int_0^{+\infty} d^4 \omega \frac{\delta (\rho(\omega_0)\rho(\omega_3) \rho(-\omega_1)\rho(-\omega_2) -\rho(-\omega_0)\rho(-\omega_3) \rho(\omega_1)\rho(\omega_2)   )}{(\omega_0+\omega_3-\omega_1-\omega_2)^2} =0\,.
\end{equation}
That is to say, for a variation $\delta \rho(\omega)$ that does not change the IR asymptotics of spectral function, the integral $I_2$ is unchanged. 
This conclusion also allows us to substitute the exact $\rho(\omega)$ by its IR form while calculating $I_2$. Thus, 
\begin{equation}
I_2 = {\rm P}
\int_{-i\infty}^{i\infty} \frac{dz e^{z0^-}}{2\pi i} \int_0^{+\infty} \frac{d^4\omega}{\pi^4} \frac{1}{\sqrt{\omega_0 \omega_1 \omega_2 \omega_3}} \frac{s_+^3s_--s_-^3s_+}{(z-\omega_0)(z-(\omega_1+\omega_2+\omega_3))^2}\,,
\label{eqn: explicit I2}
\end{equation} 
where $s_{\pm}$ are defined in Eq.~\eqref{eqn: spectral} and characterize the spectral asymmetry. Finally, we evaluate the explicit integrals, first for $d^4 \omega$ and then $dz$:
\begin{equation}
    I_2 = {\rm P}
\int_{-i\infty}^{i\infty} \frac{dz e^{z0^-}}{2\pi i z} \frac{s_+^3s_--s_-^3s_+}{\pi} =-  \frac{s_+^3 s_- -s_-^3 s_+}{2\pi} = - \frac{\sin 2\theta}{4} \,.
\end{equation}
We may call $I_2$ the ``anomalous'' Luttinger-Ward term as it arises from a formally vanishing integral. As we mentioned before, its counterpart in Fermi liquid is well-defined and indeed vanishes. The anomaly discussed here also shares some similarity with the perturbative anomaly in quantum field theory. 

\subsection{Dimensional Regularization}
It may be useful to also present a ``dimensional regularization'' version of the calculation for the anomalous term $I_2$. More explicitly, we use the following form for Green function 
\begin{equation}
G_{\eta}(i\omega) = \begin{cases}
G(i\omega) & \text{for} \quad |\omega| \gtrsim 1 \\
G(i\omega) |\omega|^{2\eta} & \text{for} \quad |\omega| \ll 1
\end{cases}\,,
\end{equation}
where $\eta$ is a small positive number that will be taken to zero in the end. 
Note that the regulator here is symmetric in $\omega$.\footnote{In contrast to the one used in Section~\ref{section: charge} which is symmetric in $\tau$.} In other words, the present regulator has the same symmetry as the hard cut-off used above and they should give the same value for $I_2$. 

The small shift in the scaling of Green function will induce a small shift in both the scaling and the prefactor in the spectral function:
\begin{equation}
\rho_{\eta}(\pm \omega) =\frac{1}{\pi} \omega^{2\eta} \underbrace{ \sin \left(  (\Delta+\eta) \pi \pm \theta \right) \sqrt{\frac{\Gamma(2-2\Delta)}{\Gamma(2\Delta)}} b^{\Delta-\frac{1}{2}}}_{s_{\eta,\pm}} \omega^{2\Delta-1} \quad \text{for} \quad 0 < \omega \ll 1 \,.
\end{equation}
The shift in the prefactor follows from the analyticity of Green function $G_{\eta}(z)$ on the upper half plane. On the other hand, the shift in scaling  saves the $I_2$ integral from IR logarithmic divergence and allows us to do the $dz$ integration first. Therefore, 
\begin{equation}
 I_2 =  \int_0^{+\infty} d^4 \omega\frac{ (\rho_\eta (\omega_0)\rho(\omega_3) \rho(-\omega_1)\rho(-\omega_2) -\rho_\eta(-\omega_0)\rho(-\omega_3) \rho(\omega_1)\rho(\omega_2)   )}{(\omega_0+\omega_3-\omega_1-\omega_2)^2}  \,.
  \end{equation}
  We can proceed by inserting the explicit expressions for $\rho$ and $\rho_{\eta}$ with an intermediate transition scale $\omega_\Lambda$ above which the integrand identically cancels,
  \begin{equation}
      I_2= (s_{\eta,+} s_- - s_{\eta,-} s_+) s_+ s_- \int_0^{\omega_\Lambda} \frac{d\omega_0}{\pi} \int_0^{\infty} \frac{d\omega^3}{\pi^3} \frac{\omega_0^{2\eta}}{\sqrt{\omega_0 \omega_ 1\omega_2 \omega_3}} \frac{1}{(\omega_0+\omega_3-\omega_1-\omega_2)^2}\,.
  \end{equation}
The actual value of $\omega_\Lambda$ is not important and will not enter the final result. Now the $d\omega$ integration are straightforward to perform. After
 taking the limit $\eta\rightarrow 0$, we have
 \begin{equation}
I_2
= - \frac{(s_{\eta,+} s_- - s_{\eta,-} s_+) s_+ s_-}{2\pi^2 \eta}  = -\frac{\sin 2\theta}{4} \,.
\end{equation}

\subsection{Generalization to $q>4$} 
It is straightforward to generalize the explicit calculations to $q>4$ using the same regularization-scheme. 
First of all, the argument that the $I_2$ term only relies on the IR asymptotics of the spectral function \eqref{eqn: spectral} still applies.  
Therefore, we end up with an explicit integral that generalizes Eq.~\eqref{eqn: explicit I2}: 
\begin{equation}
I_2 = {\rm P}
\int_{-i\infty}^{i\infty} \frac{dz e^{z0^-}}{2\pi i} \int_0^{+\infty} \frac{d^q\omega}{\pi^q} \frac{ (s_+^{q/2+1}s_-^{q/2-1}-s_-^{q/2+1}s_+^{q/2-1})}{\left(\omega_0 \omega_1 \omega_2 \ldots \omega_{q-1}\right)^{1-2\Delta}(z-\omega_0)(z-(\omega_1+\omega_2+\ldots +\omega_{q-1}))^2} \,.
\end{equation} 
It is useful to note that the $d^q\omega$ integration is the ``multivariate Beta function''. Or more explicitly, we can use the following formula
\begin{equation}
\int_0^{+\infty} \frac{d^{n}x}{(x_1 x_2 \ldots x_n)^{\alpha}} \delta (y-x_1-x_2-\ldots - x_n) = y^{n-1 - n\alpha} \frac{\Gamma(1-\alpha)^n}{\Gamma(n-n\alpha)},\quad 0< \alpha < 1 \,.
\end{equation}
Thus, the integral simplifies and we insert the expressions for $s_{\pm}$ to get
\begin{equation}
\begin{aligned}
I_2 &=  {\rm P}
\int_{-i\infty}^{i\infty} \frac{dz e^{z0^-}}{2\pi i} \int_0^{+\infty} \frac{d x dy y^{1-2\Delta}}{\pi^q x^{1-2\Delta} }  \frac{ (s_+^{q/2+1}s_-^{q/2-1}-s_-^{q/2+1}s_+^{q/2-1})}{ (z-x)(z-y)^2} \frac{\Gamma(2\Delta)^{q-1}}{\Gamma(2-2\Delta)}  \\
&= - \left( \frac{1}{2} -\Delta \right) \frac{\sin (2\theta)}{\sin (2\pi \Delta)}\,.
\end{aligned}
\end{equation}
Putting together with $I_1$, we reproduce the charge formula
\begin{equation}
    \calQ = - \frac{\theta}{\pi} -  \left( \frac{1}{2} -\Delta \right) \frac{\sin (2\theta)}{\sin (2\pi \Delta)} \,.
    \label{eqn: temp1}
\end{equation}

\subsection{$q \rightarrow 2$ limit and the semicircle}
  
\begin{figure}[t]
\center
\begin{tikzpicture}[scale=1.5, baseline={(current bounding box.center)}]
\filldraw (0pt,0pt) circle (1pt) node[above right]{$O$};
\filldraw (-10pt,0pt) circle (1pt) node[below]{$-\mu$};
\filldraw (-50pt,0pt) circle (1pt) node[below]{$-2J-\mu$};
\filldraw (30pt,0pt) circle (1pt) node[below]{$2J-\mu$};
\filldraw (-10pt,0pt) circle (1pt) node[above left]{$C$};
\filldraw (-10pt,40pt) circle (1pt) node[above]{A};
\filldraw (0pt,38.7pt) circle (1pt) node[above right]{$B$};
\draw[->,>=stealth] (-60pt,0pt)-- (60pt,0pt) node[right]{$\omega$};
\draw[->,>=stealth] (0pt,0pt) -- (0pt,60pt) node[right]{$\rho(\omega)$};
\draw[thick] (30pt,0pt) arc (0:180:40pt);
\draw[thick] (-10pt,0pt) -- (-10pt,40pt);
\draw[densely dashed] (-10pt,0pt) -- (0pt,38.7pt);
\end{tikzpicture}
\caption{Spectral function $\rho(\omega)$ for the $q=2$ model has the shape of semicircle as expected (after rescaling the $x$ or $y$ axis). }
\label{fig: semicircle}
\end{figure}
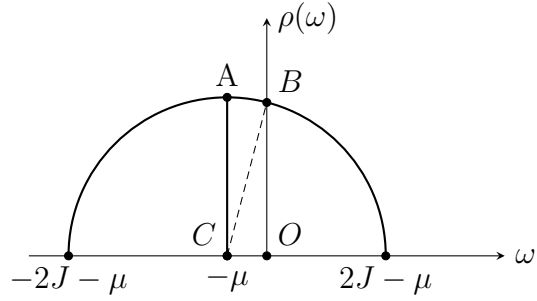

 We would like to check if the charge formula \eqref{eqn: temp1} 
  is valid at $q\rightarrow 2$ limit. Namely, whether 
  \begin{equation}
  \calQ = -\frac{\theta}{\pi}- \frac{\sin (2\theta)}{2\pi}
  \end{equation}
  is consistent with the result obtained by directly diagonalizing the quadratic random Hamiltonian. 
For $q=2$, we can solve the 
Schwinger Dyson equation 
\begin{equation}
G(z) = \frac{1}{z+\mu-\Sigma(z)} \,, \quad \Sigma(z)= J^2 G(z) \,,
\end{equation}
exactly with the UV term. We work in zero temperature and have recovered the $J$ dependence in the equation. The solution is given by 
\begin{equation}
G(z)= \frac{(z+\mu)\mp \sqrt{(z+\mu)^2-4J^2}}{2J^2}\,, \quad \text{take ``$\mp$'' sign for $\Im z \gtrless 0$}
\end{equation}
where the sign is determined by the requirement that the spectral function is non-negative and the Green function in the lower half plane is the complex conjugate of the upper half plane. 
The spectral function is nonzero on an interval $-2J-\mu<\omega<2J-\mu$ and has the semicircle form
\begin{equation}
\rho(\omega)= - \frac{1}{\pi} \Im G(\omega+i\epsilon) = \frac{1}{2\pi J^2} \sqrt{4J^2- (\omega+\mu)^2} \,, \quad -2J-\mu < \omega < 2J-\mu
\end{equation}
with area $1$ as expected and shown in Fig.~\ref{fig: semicircle}. The charge is determined by the following simple formula
\begin{equation}
\calQ= \int_{-\infty}^{0} \rho(\omega) d\omega - \frac{1}{2}
\label{eqn: semicircle charge}
\end{equation}
where the $-1/2$ is the shift required by the definition. Pictorially, $\calQ$ is equal to the area of $ABOC$ in Fig.~\ref{fig: semicircle}.

Next we determine phase factor $\theta$ by matching the IR asymptotics
\begin{equation}
z\rightarrow 0^+ i\,, \quad
G(z)= \frac{(z+\mu) -  \sqrt{(z+\mu)^2-4J^2}}{2J^2} \rightarrow \frac{\mu - i \sqrt{4J^2-\mu^2}}{2J^2} 
\end{equation}
with the definition
\begin{equation}
G(\pm i\omega) \rightarrow \mp i e^{\mp i \theta} \cdot |G(\pm i\omega)|  \,, \quad \omega \rightarrow 0\,,
\end{equation}
from which we have
\begin{equation}
\theta =- \arcsin \frac{\mu}{2J} = - \angle ACB \quad (\text{in Fig.~\ref{fig: semicircle}})\,.
\end{equation} 
Now we are ready to relate the charge $\calQ$ 
 to $\theta$ using the semicircle geometry shown in Fig.~\ref{fig: semicircle}
\begin{equation}
\begin{aligned}
\calQ &= \text{area of } ABOC=\frac{2}{\pi} \Bigg( \underbrace{ \frac{\angle ACB }{2} }_{\text{circular sector ABC}}  + \underbrace{\frac{\sin \angle BCO  \cos \angle BCO }{2}}_{\text{triangle BCO}} \Bigg)  \\
&
=- \frac{\theta}{\pi} - \frac{\sin (2\theta)}{2\pi} \quad (\text{where we have used }  \angle ACB = \angle BCO = -\theta )\,,
\end{aligned}
\end{equation} 
which is consistent with the charge formula \eqref{eqn: temp1}.

\section{Operator spectrum}
\label{sec: operator spectrum}

The solutions of the equation $\det (1-K_G(h))=0$ contain important information about the OPE of two fermions $\hat{\psi}_j^\dagger(\tau) \hat{\psi}_j(0)$.  For instance, at $\theta=0$, the matrix 
\begin{equation}
W_{\Sigma}(h) = 
\frac{\Gamma(2\Delta-1+h) \Gamma(2\Delta-h) }{\Gamma(2\Delta) \Gamma(2\Delta-1) \sin (2\pi \Delta) }
\begin{pmatrix}
\sin (\pi h ) & - \sin (2\pi \Delta)  \\
-\sin (2\pi \Delta) & \sin (\pi h )
\end{pmatrix} 
\end{equation}
is symmetric, and therefore, the eigenvectors of $K_G(h)=W_{\Sigma}(h)W_G(h)$ are $v_1=\begin{pmatrix}
1 & 1
\end{pmatrix}^T$ and $v_2=\begin{pmatrix}
1 & -1
\end{pmatrix}^T$ with eigenvalues denoted as $k^{\rA/\rS}$\cite{MS16-remarks, Klebanov:2016xxf,Bulycheva:2017uqj}
\begin{align}
K_G(h)v_1 = k^{\rA}(h)v_1 \,, \quad
k^{\rA}(h) &= \frac{\Gamma(2\Delta-h)\Gamma(2\Delta+h-1)}{\Gamma(2\Delta-2)\Gamma(2\Delta+1)} \left(1
- \frac{\sin (\pi h)}{\sin (2\pi \Delta)}
 \right) \,, \\
 K_G(h)v_2 = k^{\rS}(h)v_2 \,, \quad
 k^{\rS}(h) &= \frac{\Gamma(2\Delta-h)\Gamma(2\Delta+h-1)}{\Gamma(2\Delta-1)\Gamma(2\Delta)} \left(1
+ \frac{\sin (\pi h)}{\sin (2\pi \Delta)}
 \right) \,.
 \label{eqn: KSKA Q=0}
\end{align}
Using the reflection symmetry $h\leftrightarrow 1-h$ (cf. Eq.~\eqref{eqn: kernel symmetry}), we restrict our discussion to $h\geq 1/2$ and label them in ascending order in this section, i.e. $1/2\leq h^{\rA/\rS}_0 < h^{\rA/\rS}_1 < h^{\rA/\rS}_2\ldots $ are solutions of $k^{\rA/\rS}(h)=1$ respectively. In particular, the anti-symmetric sector, corresponding to the solutions of $k^{\rA}(h)=1$, reproduces the scaling dimensions of the operators appearing in $\hat{\chi}_j(\tau) \hat{\chi}_j(0)$ OPE of the Majorana SYK (which is determined by the equation $k_c(h)=1$ in the notation of Ref.~\cite{MS16-remarks,KS17-soft}). The leading one $h_0^{\rA}=2$ corresponds to the Schwarzian sector and responsible for the energy fluctuation. Analogously, the leading one in the symmetric sector $h_0^{\rS}=1$ is related to the $\UU(1)$ charge in the complex SYK model as we discussed in Section~\ref{secRG}. 

\begin{figure}[t]
\center
\subfloat[$k^{\rA}(h)$ with $\theta=\pi/8$]{
\includegraphics[scale=0.67]{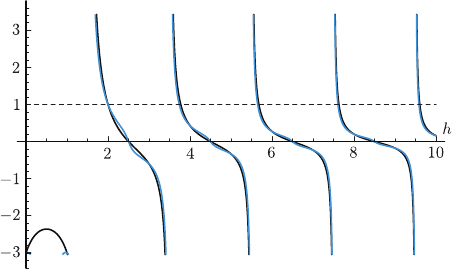}}\quad
\subfloat[$k^{\rA}(h)$ with $\theta=\pi/6$]{
\includegraphics[scale=0.67]{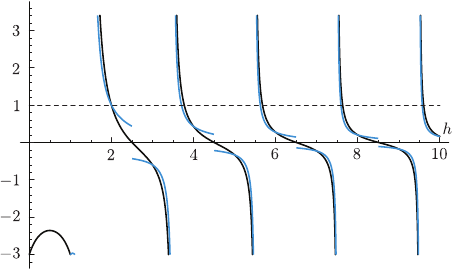}}\quad
\subfloat[$k^{\rA}(h)$ with $\theta=\pi/4$]{
\includegraphics[scale=0.67]{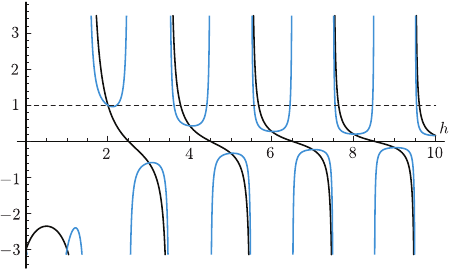}}\quad
\\
\subfloat[$k^{\rS}(h)$ with $\theta=\pi/8$]{
\includegraphics[scale=0.67]{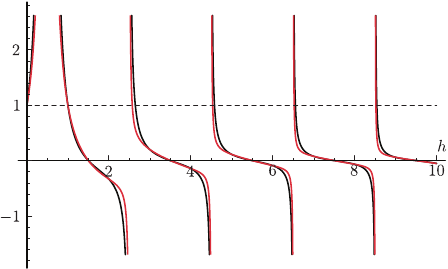}}\quad
\subfloat[$k^{\rS}(h)$ with $\theta=\pi/6$]{
\includegraphics[scale=0.67]{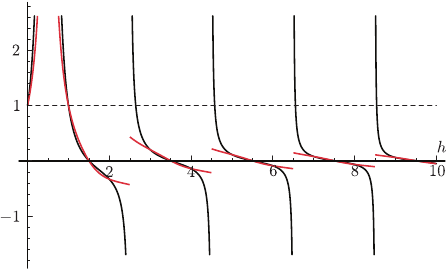}}\quad
\subfloat[$k^{\rS}(h)$ with $\theta=\pi/4$]{
\includegraphics[scale=0.67]{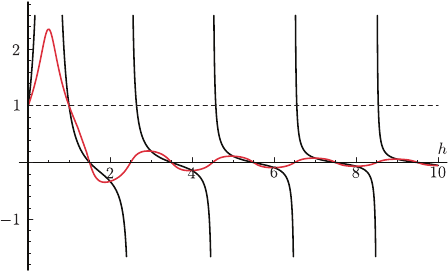}}
\caption{Plots of $k^{\rA}(h)$ and $k^{\rS}(h)$ for $\Delta=1/4$ and $\theta=\pi/8$, $\pi/6$ and $\pi/4$. Black lines are for reference, represent the value  of $k^{\rA/\rS}(h,\theta)$ for $\theta=0$. }
\label{fig: KS and KA}
\end{figure}

\begin{figure}[t]
\center
\subfloat[$h^{\rA/\rS}(\theta)$]{
\includegraphics[scale=1]{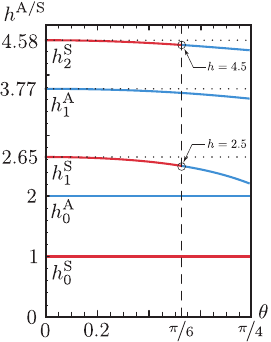}}
\quad
\subfloat[${k^{\rA/\rS}}'(h_0^{\rA/\rS},\theta)$]{
\includegraphics[scale=1]{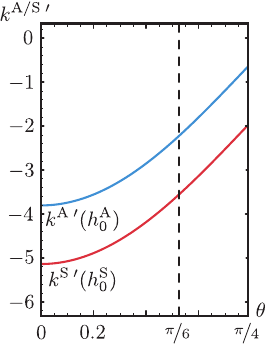}}
\quad
\subfloat[${k^{\rA/\rS}}'(h_1^{\rA/\rS},\theta)$]{
\includegraphics[scale=1]{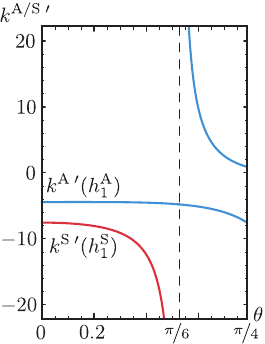}}
\caption{We set $\Delta=1/4$ for all plots in this figure. (a) Plots of $h^{\rA}(\theta)$ and $h^{\rS}(\theta)$ as functions of $\theta$. 
All dimensions $h_{k}^{\rS}$ for $k \geq  1$ disappear at $\theta=\pi/6$ and reappear in the anti-symmetric branch.
(b) Plots of ${k^\rA}'(2,\theta)$ and ${k^\rS}'(1,\theta)$ (derivative w.r.t.\ $h$, not $\theta$) as functions of $\theta$. Note that ${k^\rS}'(1,\theta)= -(1-2\Delta)^{-1} - \pi \cos (2\theta)/\cos (2\pi \Delta)$ (for an arbitrary $\Delta$) is denoted by $k'(1)$  in Eq.\eqref{eqn: kprime(1)}. 
(c) Plots of ${k^\rA}'(h_1^\rA,\theta)$ and ${k^\rS}'(h_1^\rS,\theta)$ as functions of $\theta$. Note ${k^\rS}'(h_1^\rS,\theta)\to-\infty$ as $\theta$ approaches the critical value $\pi/6$ from the left. For $\theta$ slightly above $\pi/6$, the solution reappears in the anti-symmetric branch with a divergent ($+\infty$) derivative ${k^\rA}'$ . 
}
\label{fig: h1 vs theta}
\end{figure}

For general $\theta$, the matrix $W_{\Sigma}(h)$ has no symmetry and the symmetric/anti-symmetric sectors generally mix via  $2\times 2$ ladder kernel $K_G$ (or $K_{\Sigma}$). Let us denote the two eigenvalues by $k^{\rA}(h,\theta)$ (anti-symmetric branch) and $k^{\rS}(h,\theta)$ (symmetric branch) as a generalization of the notation $k^{\rA/\rS}(h)$. Their explicit formulas are as follows, 
\begin{equation}
\begin{aligned}
k^{\rA}(h,\theta) &= \frac{\Gamma(2\Delta-h)\Gamma(2\Delta+h-1)}{\Gamma(2\Delta+1)\Gamma(2\Delta-1)} 
\cdot
\Bigg( 2\Delta-1 + \frac{\cos (2\theta) \sin (\pi h)}{\sin (2\pi \Delta)} \\
&
-   \sqrt{
\sin (2\theta)^2 \Bigl(  1- \Bigl( \frac{\sin (\pi h)}{\sin (2\pi \Delta)} \Bigr)^2 \Bigr)+ \Bigl( \cos (2\theta) + (2\Delta-1) \frac{\sin (\pi h)}{\sin (2\pi \Delta)}  \Bigr)^2 
} 
\Bigg)
\end{aligned}\,,
\label{eqn: KA}
\end{equation}
\begin{equation}
\begin{aligned}
k^{\rS}(h,\theta) &= \frac{\Gamma(2\Delta-h)\Gamma(2\Delta+h-1)}{\Gamma(2\Delta+1)\Gamma(2\Delta-1)} 
\cdot
\Bigg( 2\Delta-1 + \frac{\cos (2\theta) \sin (\pi h)}{\sin (2\pi \Delta)} \\
&
+    \sqrt{
\sin (2\theta)^2 \Bigl(  1- \Bigl( \frac{\sin (\pi h)}{\sin (2\pi \Delta)} \Bigr)^2 \Bigr)+ \Bigl( \cos (2\theta) + (2\Delta-1) \frac{\sin (\pi h)}{\sin (2\pi \Delta)}  \Bigr)^2 
} 
\Bigg)\,.
\end{aligned}
\label{eqn: KS}
\end{equation}
To illustrate, we plot $k^{\rA}(h,\theta)$ (blue lines) and $k^{\rS}(h,\theta)$ (red lines) as functions of $h$ for $\Delta=1/4$ and $\theta=\pi/8$, $\pi/6$, $\pi/4$ together with $\theta=0$ (black lines, as a reference) in Fig.~\ref{fig: KS and KA}. 
Here are some comments:
\begin{enumerate}
\item For integer value $h=n \in \ZZ$, one can check that $k^{\rA/\rS}(h,\theta)$ is independent of $\theta$, i.e. $k^{\rA/\rS}(h,\theta)=k^{\rA/\rS}(h,0)$ for any $\theta$. 

An immediate corollary is that 
$k^{\rA}(2,\theta)=1$ and $k^{\rS}(1,\theta)=1$ for all $\theta$, i.e. the scaling dimensions of the energy and charge operator are protected (as well as their dual field with $h=-1$ and $0$ respectively).
\item General solutions of $k^{\rA/\rS}(h,\theta)=1$ depend on $\theta$. In Fig.~\ref{fig: h1 vs theta} we plot the value of $h_1^{\rA/\rS}(\theta)$ as functions of $\theta$ for $\Delta=1/4$. 

\item From Fig~\ref{fig: KS and KA}, we notice that 
for $\Delta=1/4$, there is a critical value $\theta_c=\pi/6$ which is determined by the following equation
\begin{equation}
\cos (2\theta_c) = 1-2\Delta \,.
\label{eqn: critical theta}
\end{equation}
Above the critical value, namely for $\theta>\theta_c$, the solutions $h_{n\geq 1}^{\rS}$ disappear. In other words, the only solutions for $k^{\rS}(h,\theta\geq \theta_c)=1$ are $h_0^{\rS}=1$ and its dual $0$.

To explain why Eq.~\eqref{eqn: critical theta} is relevant, let us analyze the pole structure of $k^{\rS}(h,\theta)$. 
Naively the expression \eqref{eqn: KS} has simple poles at $h=2\Delta+m$ with $m\in \ZZ^{\geq 0}$ due to the 
overall factor $\Gamma(2\Delta-h)$. However, the expression in big parentheses in \eqref{eqn: KS},
\begin{equation}
2\Delta-1 + (-1)^m \cos (2\theta) + \left|\cos (2\theta) + (2\Delta-1) (-1)^m )\right|\quad \text{at} ~~h=2\Delta+m\,. 
 \end{equation}
has zeros that cancel some of the poles. 
Indeed the poles at odd $m$, i.e. at $h=2\Delta+1$, $2\Delta+3$, $\ldots$ are canceled in $k^{\rS}(h,\theta)$. Furthermore, 
when $\cos (2\theta) < 1-2\Delta$, the poles at even $m$ i.e. at $h=2\Delta$, $2\Delta+2$, $\ldots$ are also canceled. At critical value $\cos (2\theta) = 1-2\Delta$, there is a discontinuity for $k^{\rS}(h,\theta_c)$ at $h=2\Delta+2k$ for $k\geq 1$. Explicit calculation yields (for $\Delta=1/4$)
 \begin{equation}
 \lim_{h\rightarrow (2\Delta+2k)^-} k^{\rS}(h,\theta_c) = -\frac{\sqrt{3}}{4k}\,, \qquad \lim_{h\rightarrow (2\Delta+2k)^+} k^{\rS}(h,\theta_c) = \frac{\sqrt{3}}{4k} \,.
 \end{equation}
 Parallel discussions apply to the anti-symmetric branch $k^{\rA}(h,\theta)$ 
  where an additional set of solutions to the equation $k^{\rA}(h,\theta)=1$ emerges when $\theta>\theta_c$, as shown in Fig.~(\ref{fig: KS and KA}.c). Technically, this is related to the additional set of poles at $h=2\Delta+2k$.  
  
  \item An immediate consequence of the ``branch switching'' phenomenon described above is as follows. If we write the scaling dimensions (i.e.\ the solutions of the equation $k^{\rA/\rS}(h)=1$) as $h=2\Delta+m+\delta h$, then $\delta h$ changes sign for those solutions that move from the symmetric to the anti-symmetric branch. 
This happens as the slope ${k^{\rA/\rS}}'(h)$ diverges (see Fig.~\ref{fig: h1 vs theta} (c)). The divergence of the slope seems to suggest the vanishing of the corresponding OPE coefficient (cf. Ref.~\cite{MS16-remarks} Eq.~(3.54), assuming the Plancherel factor part does not diverge at these points).
 \end{enumerate}

One final remark is that for general $\Delta\neq 1/4$, there is an additional subtlety that the eigenvalues $k^{\rA/\rS}$ can be generally complex numbers for certain range of $\theta$, but we will not discussed the details in this paper. 

\section{Low energy contribution to $K$}
\label{app:GrishaK}

Our computations of fluctuations in this appendix, and in Section~\ref{kerneldiag}, follow the methods of Ref.~\cite{MS16-remarks}; these methods are related to those in Section~\ref{secRG}.  
The key difference is that in Section~\ref{secRG} we expand the $(G,\Sigma)$ action w.r.t. the \emph{conformal} saddle, while in this appendix and Section~\ref{kerneldiag} we expand $(G,\Sigma)$ action w.r.t. the \emph{exact} saddle. We will  
use subscript ``$_c$'' and ``$_{\textrm{exact}}$'' to emphasis the contrast. 

As we discussed in Section~\ref{sec:compress}, we expect that all energy scales contribute to the numerical value of the compressibility $K$, and a low energy conformal perturbation approach (similar to that used successfully for the specific heat in Ref.~\cite{MS16-remarks}) does not yield the correct value of $K$, instead it reproduces $K_{\text{linear}}$ discussed in Section~\ref{sec: RG charge compressibility} (cf. Eq.~\eqref{Kgamma relation}). The UV divergence of the symmetric kernel eigenmodes in Section~\ref{kerneldiag} provides explicit evidence for this claim. Nevertheless, we will present the low energy analysis of the symmetric kernel here, as it could be useful for other investigations.

\subsection{Effective action for  fluctuations around the saddle point}
\label{app:eff}
In this section we consider the $(G,\Sigma)$ action for the complex SYK model with zero chemical potential and derive effective action for  quadratic fluctuations around the saddle point of the action. In this section we recover $J$.

The $(G,\Sigma)$ action for the complex SYK with zero chemical potential is (cf. Eq.~\eqref{Gsigma action})  
\begin{equation}
\frac{I}{N} = -\ln \det (-\partial_{\tau}-\Sigma) -\int d\tau_{1}d\tau_{2} \left[\Sigma(\tau_{1},\tau_{2})G(\tau_{2},\tau_{1})+\frac{J^{2}}{q}(-G(\tau_{1},\tau_{2})G(\tau_{2},\tau_{1}))^{\frac{q}{2}}\right]\,. \label{cSYKGSact}
\end{equation}
The crucial difference from the Majorana SYK model is that now the bilocal fields $G(\tau_{1},\tau_{2})$ and $\Sigma(\tau_{1},\tau_{2})$ are not necessarily antisymmetric under exchange of variables $\tau_{1}\leftrightarrow \tau_{2}$.  

The saddle point $G_{\textrm{exact}}, \Sigma_{\textrm{exact}}$ of the action (\ref{cSYKGSact}) is  the exact solution of the Schwinger-Dyson equations \eqref{SchwDyseq}. 
Now we consider small fluctuations around the \emph{exact} saddle point $G_{\textrm{exact}},\Sigma_{\textrm{exact}}$:
\begin{equation}
G(\tau_{1},\tau_{2}) = G_{\textrm{exact}}(\tau_{12}) +\delta G(\tau_{1},\tau_{2}), \quad \Sigma(\tau_{1},\tau_{2}) = \Sigma_{\textrm{exact}}(\tau_{12}) +\delta \Sigma(\tau_{1},\tau_{2})\,,
\end{equation}
and expand the $(G,\Sigma)$ action up to quadratic terms. Next we integrate out Gaussian fluctuations of the $\delta \Sigma$ field 
and obtain the Gaussian action for the fluctuations $\delta G$, which is convenient to parametrize as $
\delta G(\tau_{1},\tau_{2})=|G_{\textrm{exact}}(\tau_{12})|^{\frac{2-q}{2}} g(\tau_{1},\tau_{2})$
\begin{equation}
\begin{aligned}
\frac{I}{N} =& \frac{1}{2}J^{2}(q-1) \int_{0}^{\beta}  d^4\tau\, g^{\rA}(\tau_{1},\tau_{2})  \big((K^{\rA}_{\textrm{exact}}(\tau_{1},\tau_{2};\tau_{3},\tau_{4}))^{-1}-1\big)  g^{\rA}(\tau_{3},\tau_{4})\\
&-\frac{1}{2}J^{2}\int_{0}^{\beta}  d^4\tau\,g^{\rS}(\tau_{1},\tau_{2})  \big((K^{\rS}_{\textrm{exact}}(\tau_{1},\tau_{2};\tau_{3},\tau_{4}))^{-1}-1\big)  g^{\rS}(\tau_{3},\tau_{4})\,, \label{ASactfluct}
\end{aligned}
\end{equation}
where we also decomposed fluctuations $g(\tau_{1},\tau_{2})$ on symmetric and antisymmetric parts $
g(\tau_{1},\tau_{2})=g^{\rS}(\tau_{1},\tau_{2})+g^{\rA}(\tau_{1},\tau_{2})$, so $g^{\rS/\rA}(\tau_{2},\tau_{1})=\pm g^{\rS/\rA}(\tau_{1},\tau_{2})$ and introduced  the antisymmetric and symmetric kernels $K^{\rA/\rS}_{\textrm{exact}}$ whose explicit expressions have been shown in Eq.~\eqref{eqn: KAS}. We copy the formulas here with the emphasis of subscript ``$_{\textrm{exact}}$''
\begin{equation}
\begin{aligned}
&K^{\rA}_{\textrm{exact}}(\tau_{1},\tau_{2};\tau_{3},\tau_{4}) = -J^{2}(q-1)|G_{\textrm{exact}}(\tau_{12})|^{\frac{q-2}{2}}G_{\textrm{exact}}(\tau_{13})G_{\textrm{exact}}(\tau_{24})|G_{\textrm{exact}}(\tau_{34})|^{\frac{q-2}{2}} \,, \\
&K^{\rS}_{\textrm{exact}}(\tau_{1},\tau_{2};\tau_{3},\tau_{4}) = -J^{2}|G_{\textrm{exact}}(\tau_{12})|^{\frac{q-2}{2}}G_{\textrm{exact}}(\tau_{13})G_{\textrm{exact}}(\tau_{24})|G_{\textrm{exact}}(\tau_{34})|^{\frac{q-2}{2}}\,. \label{defofKAKS}
\end{aligned}
\end{equation}

In the  large $\beta J$ limit the exact Green function $G_{\textrm{exact}}$ in the kernels $K^{\rA}_{\textrm{exact}}$ and $K^{\rS}_{\textrm{exact}}$ can be approximated by  the conformal solution $G_{c}$, so one obtains conformal kernels $K^{\rA}_{c}$ and $K^{\rS}_{c}$. The spectrum of the conformal kernels can be computed exactly and was  given in \eqref{eqn: KSKA Q=0}
\begin{equation}
\begin{aligned}
k^{\rS}_{c}(h) &= \frac{\Gamma(2\Delta-h)\Gamma(2\Delta+h-1)}{\Gamma(2\Delta-2)\Gamma(2\Delta+1)} \left(1
- \frac{\sin (\pi h)}{\sin (2\pi \Delta)}
 \right) \,, \\
 k^{\rA}_{c}(h) &= \frac{\Gamma(2\Delta-h)\Gamma(2\Delta+h-1)}{\Gamma(2\Delta-1)\Gamma(2\Delta)} \left(1
+ \frac{\sin (\pi h)}{\sin (2\pi \Delta)}
 \right) \,,
\end{aligned}
\label{eqn: KSKA Q=0 v2}
\end{equation}
where $h$ labels the $\SL(2,\RR)$ representations: for the antisymmetric case $h=2,4,6,\dots$ and for the symmetric case $h=1,3,5,\dots$ and there is also a principal series $h=1/2+is$,  $s\in \RR^+$  for both cases. 
Note $k^{\rA}_{c}(h=2)=1$ and $k^{\rS}_{c}(h=1)=1$, i.e. if one uses conformal kernels in \eqref{ASactfluct} then the effective action for  these special modes is zero, which would also indicates  instability \cite{Kim:2019upg}. Actually, this problem appeared because we replaced the exact kernels by conformal ones. The exact eigenvalues $k^{\rA}$ and $k^{\rS}$ which correspond to $h=2$ and $h=1$ modes  differ from $1$ by $1/(\beta J)$ corrections. 
In order to find these corrections one uses $1/(\beta J)$ correction to the conformal Green function   \cite{MS16-remarks,KS17-soft}
\begin{equation}
G_{c}(\tau)\to G_{c}(\tau)\left(1-\frac{\alpha_{G}}{\beta \mathcal{J}}f_{0}(\tau)\right), \quad f_{0}(\tau) =2 +\frac{\pi - 2\pi |\tau|/\beta}{\tan \frac{\pi |\tau|}{\beta}} \,,
\end{equation}
where $\mathcal{J}=\sqrt{q}J/2^{\frac{q-1}{2}}$ and $\alpha_{G}$ is a UV dependent constant, which can be found numerically. 
Next using the corrected Green function in the kernels one finds corrections to their eigenvalues. 

We  find the $1/(\beta J)$ correction to the conformal kernels using perturbation theory \cite{MS16-remarks}. At the first order one  computes diagonal matrix elements of the perturbation. In our case the perturbation to the conformal kernels consists of two parts and reads 
\begin{equation}
V^{\rA/\rS}(\tau_{1},\tau_{2};\tau_{3},\tau_{4}) =-\frac{2\alpha_{G}}{\beta \mathcal{J}} K^{\rA/\rS}_{c}(\tau_{1},\tau_{2};\tau_{3},\tau_{4})\Big(\frac{q-2}{2}f_{0}(\tau_{12})+f_{0}(\tau_{13})\Big)\,.
\end{equation}
The part of $V$ which involves $f_{0}(\tau_{12})$ is called the rung term and the part with $f_{0}(\tau_{13})$ is called the rail term. The corrections to the eigenvalues are simply 
\begin{equation}
\delta k^{\rA}= \langle \Psi^{\rA} |V^{\rA}| \Psi^{\rA}\rangle, \quad 
\delta k^{\rS}= \langle \Psi^{\rS} |V^{\rS}| \Psi^{\rS}\rangle\,,
\end{equation}
where for $ |\Psi^{\rA}\rangle $ and $ |\Psi^{\rS}\rangle$ we take unperturbed conformal eigenfunctions, so $K^{\rA}_{c}|\Psi^{\rA}\rangle = k^{\rA}_{c}|\Psi^{\rA}\rangle$ and $K^{\rS}_{c}|\Psi^{\rS}\rangle = k^{\rS}_{c}|\Psi^{\rS}\rangle$.

For the antisymmetric case the correction for $h=2$ mode was already found in \cite{MS16-remarks,KS17-soft} and reads ($k^{\rA}=1+\delta k^{\rA}$)
\begin{equation}
\delta k^{\rA}(2,n) = -\frac{\alpha_{K}^{\rA}}{\beta \mathcal{J}}|n|, \quad \alpha_{K}^{\rA} =- \alpha_{G}q {k^{\rA}_{c}}'(2)\,. \label{cortokA}
\end{equation}
In the next subsection we are going to compute  correction to $h=1$ eigenvalue of the symmetric kernel. We expect to find a similar to (\ref{cortokA}) form, but with some other coefficient $\alpha_{K}^{\rS}$.
We will find that such obtained value  $\alpha_{K}^{\rS}$ does not agree with the numerical computation in the section \ref{kerneldiag}.  The reason for this disagreement is hidden in the 
fact that when we use conformal Green function in the kernels instead of the exact one,  we implicitly assume that the eigenvalues  $k^{\rA}(h=2)$
and $k^{\rS}(h=1)$ are not affected by the UV domain ($\tau<1/J$ or $\tau>\beta-1/J$), where the conformal Green function $G_{c}$ diverges, but the exact $G_{\textrm{exact}}$
goes to $1/2$. And in fact this turned out to be correct for the antisymmetric case. But it is not correct for the symmetric case. On a technical level the exact $h=1$ eigenfunctions $\Psi_{1,n}^{\textrm{exact}}(\tau_{1},\tau_{2})$ of the symmetric kernel do not approach completely conformal eigenfunctions $\Psi_{1,n}(\tau_{1},\tau_{2})$ at the $\beta J \to \infty$ limit. There is always a  discrepancy in the UV domain.  The exact eigenfunctions grow as $\beta J$ at the coincident points $\tau_{1}=\tau_{2}$, whereas the conformal ones approach a constant. 
This effect is nicely captured in the large $q$ limit, which we discuss in the appendix \ref{largeqsymkern}.

\subsection{Symmetric sector}
\label{app:symm}

The rung and rail integrals in the symmetric sector are (we omit factor $-\frac{2\alpha_{G}}{\beta \mathcal{J}}$ for brevity)
\begin{equation}
\begin{aligned}
&\delta k_{\textrm{rung}}^{\rS} =\frac{(q-2)}{2} \int_{-\pi}^{+\pi}d\theta_{1}d\theta_{2}d\theta_{3}d\theta_{4} \Psi_{1,n}^{S*}(\theta_{1},\theta_{2})K^{\rS}_{c}(\theta_{1},\theta_{2};\theta_{3},\theta_{4}) f_{0}(\theta_{12})\Psi_{1,n}^{\rS}(\theta_{3},\theta_{4})\,, \\
&\delta k_{\textrm{rail}}^{\rS} = \int_{-\pi}^{+\pi}d\theta_{1}d\theta_{2}d\theta_{3}d\theta_{4} \Psi_{1,n}^{S*}(\theta_{1},\theta_{2})K^{\rS}_{c}(\theta_{1},\theta_{2};\theta_{3},\theta_{4}) f_{0}(\theta_{13})\Psi_{1,n}^{\rS}(\theta_{3},\theta_{4})\,,
\end{aligned}
\end{equation}
where the conformal kernel, $h=1$ wave functions and the correction to the conformal propagator are
\begin{equation}
\begin{aligned}
&K^{\rS}_{c}(\theta_{1},\theta_{2};\theta_{3},\theta_{4}) = -\frac{1}{4(q-1)\alpha_{0} } \frac{\textrm{sgn}(\theta_{13})\textrm{sgn}(\theta_{24})}{|\sin \frac{\theta_{12}}{2}|^{1-2\Delta}| \sin \frac{\theta_{13}}{2}|^{2\Delta}| \sin \frac{\theta_{24}}{2}|^{2\Delta}| \sin \frac{\theta_{34}}{2}|^{1-2\Delta}}\,, \\
&\Psi_{1,n}^{\rS}(\theta_{1},\theta_{2}) =  \frac{e^{-in \frac{\theta_{1}+\theta_{2}}{2}}}{2 \pi |n|^{1/2}}\frac{\sin \frac{n  \theta_{12}}{2}}{\sin \frac{ \theta_{12}}{2}}\,, \quad f_{0}(\theta) = 2 +\frac{\pi -|\theta|}{\tan \frac{|\theta|}{2}}, \quad \alpha_{0} = \frac{2\pi q  \cot(\frac{\pi}{q})}{(q-1)(q-2)}, 
\end{aligned}
\end{equation}
where we fix $\beta=2\pi$ so all angles take values in the interval $[-\pi,\pi]$.
It is convenient to represent different integrals by Feynman diagrams.  Let us introduce Feynman rules. We denote propagators as
\begin{equation}
\begin{tikzpicture}[baseline={([yshift=-4pt]current bounding box.center)}]
\draw[thick](0pt,0pt)-- (30pt,0pt);
\node at (0pt,7pt) {\normalsize $\theta_{1} $};
\node at (32pt,7pt) {\normalsize $\theta_{2} $};
\node at (15pt,7pt) {\normalsize $\alpha $};
\end{tikzpicture} = \frac{1}{(\sin^{2}\frac{\theta_{12}}{2})^{\alpha}}, \qquad 
\begin{tikzpicture}[baseline={([yshift=-4pt]current bounding box.center)}]
\draw[thick, mid arrow](0pt,0pt)-- (30pt,0pt);
\node at (0pt,7pt) {\normalsize $\theta_{1} $};
\node at (32pt,7pt) {\normalsize $\theta_{2} $};
\node at (15pt,7pt) {\normalsize $\alpha $};
\end{tikzpicture} = \frac{\textrm{sgn}(\theta_{12})}{(\sin^{2}\frac{\theta_{12}}{2})^{\alpha}} \label{propsonunit}
\end{equation}
A very useful tool  for computation in a conformal theory is  the star-triangle identities \cite{Symanzik:1972wj}
\begin{equation}
\begin{aligned}
&\int_{-\pi}^{+\pi} \frac{d\theta_{0}}{(\sin^{2}\frac{\theta_{01}}{2})^{\alpha_{1}}(\sin^{2}\frac{\theta_{02}}{2})^{\alpha_{2}}(\sin^{2}\frac{\theta_{03}}{2})^{\alpha_{3}}} = 
\frac{b_{\alpha_{1},\alpha_{2}}}{(\sin^{2}\frac{\theta_{12}}{2})^{\frac{1}{2}-\alpha_{3}}(\sin^{2}\frac{\theta_{13}}{2})^{\frac{1}{2}-\alpha_{2}}(\sin^{2}\frac{\theta_{23}}{2})^{\frac{1}{2}-\alpha_{1}}}\,,  \\
&\int_{-\pi}^{+\pi} \frac{d\theta_{0} \textrm{sgn}(\theta_{01})\textrm{sgn}(\theta_{02})}{(\sin^{2}\frac{\theta_{01}}{2})^{\alpha_{1}}(\sin^{2}\frac{\theta_{02}}{2})^{\alpha_{2}}(\sin^{2}\frac{\theta_{03}}{2})^{\alpha_{3}}} = 
\frac{f_{\alpha_{1},\alpha_{2}}\textrm{sgn}(\theta_{31})\textrm{sgn}(\theta_{32})}{(\sin^{2}\frac{\theta_{12}}{2})^{\frac{1}{2}-\alpha_{3}}(\sin^{2}\frac{\theta_{13}}{2})^{\frac{1}{2}-\alpha_{2}}(\sin^{2}\frac{\theta_{23}}{2})^{\frac{1}{2}-\alpha_{1}}}\,,
\end{aligned}
\end{equation}
where $\alpha_{1}+\alpha_{2}+\alpha_{3}=1$ and  
\begin{align}
b_{\alpha_{1},\alpha_{2}} = 2\sqrt{\pi}\frac{\Gamma(\frac{1}{2}-\alpha_{1})\Gamma(\frac{1}{2}-\alpha_{2})\Gamma(\frac{1}{2}-\alpha_{3})}{\Gamma(\alpha_{1})\Gamma(\alpha_{2})\Gamma(\alpha_{3})}\,, \quad  f_{\alpha_{1},\alpha_{2}} = 2\sqrt{\pi}\frac{\Gamma(1-\alpha_{1})\Gamma(1-\alpha_{2})\Gamma(\frac{1}{2}-\alpha_{3})}{\Gamma(\frac{1}{2}+\alpha_{1})\Gamma(\frac{1}{2}+\alpha_{2})\Gamma(\alpha_{3})}\,.
\end{align}
Graphical representation of the star-triangle identities is 
\begin{equation}
\begin{tikzpicture}[baseline={([yshift=-4pt]current bounding box.center)}]
\draw[thick](0pt,0pt)-- (0pt,20pt);
\draw[thick](0pt,0pt)-- (17pt,-10pt);
\draw[thick](0pt,0pt)-- (-17pt,-10pt);
\node at (-6pt,10pt) {\normalsize $\alpha_{1}$};
\node at (13pt,-1pt) {\normalsize $\alpha_{2}$};
\node at (-14pt,-1pt) {\normalsize $\alpha_{3}$};
\node at (7pt,20pt) {\normalsize $\theta_{1} $};
\node at (23pt,-10pt) {\normalsize $\theta_{2} $};
\node at (-22pt,-9pt) {\normalsize $\theta_{3} $};
\end{tikzpicture} =b_{\alpha_{1},\alpha_{2}}
\begin{tikzpicture}[baseline={([yshift=-4pt]current bounding box.center)}]
\draw[thick](0pt,20pt)-- (17pt,-10pt);
\draw[thick](17pt,-10pt)-- (-17pt,-10pt);
\draw[thick](-17pt,-10pt)-- (0pt,20pt);
\node at (7pt,20pt) {\normalsize $\theta_{1} $};
\node at (23pt,-10pt) {\normalsize $\theta_{2} $};
\node at (-22pt,-9pt) {\normalsize $\theta_{3} $};
\node at (31pt,5pt) {\normalsize $\nicefrac{1}{2}-\alpha_{3} $};
\node at (-32pt,5pt) {\normalsize $\nicefrac{1}{2}-\alpha_{2} $};
\node at (0pt,-22pt) {\normalsize $\nicefrac{1}{2}-\alpha_{1} $};
\end{tikzpicture} \quad 
\begin{tikzpicture}[baseline={([yshift=-4pt]current bounding box.center)}]
\draw[thick, mid arrow](0pt,0pt)-- (0pt,20pt);
\draw[thick, mid arrow](0pt,0pt)-- (17pt,-10pt);
\draw[thick](0pt,0pt)-- (-17pt,-10pt);
\node at (-7pt,10pt) {\normalsize $\alpha_{1}$};
\node at (13pt,0pt) {\normalsize $\alpha_{2}$};
\node at (-13pt,0pt) {\normalsize $\alpha_{3}$};
\node at (7pt,20pt) {\normalsize $\theta_{1} $};
\node at (23pt,-10pt) {\normalsize $\theta_{2} $};
\node at (-22pt,-10pt) {\normalsize $\theta_{3} $};
\end{tikzpicture} =f_{\alpha_{1},\alpha_{2}}
\begin{tikzpicture}[baseline={([yshift=-4pt]current bounding box.center)}]
\draw[thick](0pt,20pt)-- (17pt,-10pt);
\draw[thick, mid arrow](-17pt,-10pt)-- (17pt,-10pt);
\draw[thick, mid arrow](-17pt,-10pt)-- (0pt,20pt);
\node at (7pt,20pt) {\normalsize $\theta_{1} $};
\node at (23pt,-10pt) {\normalsize $\theta_{2} $};
\node at (-22pt,-10pt) {\normalsize $\theta_{3} $};
\node at (31pt,5pt) {\normalsize $\nicefrac{1}{2}-\alpha_{3} $};
\node at (-32pt,5pt) {\normalsize $\nicefrac{1}{2}-\alpha_{2} $};
\node at (0pt,-22pt) {\normalsize $\nicefrac{1}{2}-\alpha_{1} $};
\end{tikzpicture}
\end{equation}
In addition to the star-triangle identities we also list a couple of useful integrals.  One is the integral, which gives the delta-function
\begin{equation}
\int_{-\pi}^{\pi} d\theta_{0} \frac{\textrm{sgn}(\theta_{10})\textrm{sgn}(\theta_{02})}{(\sin^{2}\frac{\theta_{10}}{2})^{1-\alpha} (\sin^{2}\frac{\theta_{02}}{2})^{\alpha}}=
\begin{tikzpicture}[baseline={([yshift=-4pt]current bounding box.center)}]
\draw[thick, mid arrow](0pt,0pt)-- (36pt,0pt);
\draw[thick, mid arrow](36pt,0pt)-- (72pt,0pt);
\node at (0pt,7pt) {\normalsize $\theta_{1} $};
\node at (40pt,7pt) {\normalsize $\theta_{0} $};
\node at (19pt,7pt) {\normalsize $1-\alpha $};
\node at (74pt,7pt) {\normalsize $\theta_{2} $};
\node at (54pt,7pt) {\normalsize $\alpha $};
\end{tikzpicture}=f_{1-\alpha}f_{\alpha}\delta(\theta_{12})\,,
 \label{delint}
\end{equation}
where $f_{\alpha}=i2^{2\alpha+1} \cos(\pi \alpha) \Gamma(1-2\alpha)$.
The other useful integral, where $\alpha_{1},\alpha_{2},\alpha_{3}$ are arbitrary real numbers is
\begin{equation}
\int_{-\pi}^{+\pi}\frac{d\theta_{1}d\theta_{2}d\theta_{3} }{(\sin^{2}\frac{\theta_{12}}{2})^{\alpha_{1}}(\sin^{2}\frac{\theta_{23}}{2})^{\alpha_{2}}(\sin^{2}\frac{\theta_{13}}{2})^{\alpha_{3}}} =
\int_{-\pi}^{\pi} d\theta_{1}d\theta_{2}d\theta_{3}
\begin{tikzpicture}[baseline={([yshift=-4pt]current bounding box.center)}]
\draw[thick](0pt,20pt)-- (17pt,-10pt);
\draw[thick](17pt,-10pt)-- (-17pt,-10pt);
\draw[thick](-17pt,-10pt)-- (0pt,20pt);
\node at (7pt,20pt) {\normalsize $\theta_{1} $};
\node at (23pt,-10pt) {\normalsize $\theta_{2} $};
\node at (-22pt,-10pt) {\normalsize $\theta_{3} $};
\node at (17pt,5pt) {\normalsize $\alpha_{1} $};
\node at (-17pt,5pt) {\normalsize $\alpha_{3} $};
\node at (0pt,-15pt) {\normalsize $\alpha_{2} $};
\end{tikzpicture}  = b_{\alpha_{1},\alpha_{2},\alpha_{3}}\,, \label{b3int}
\end{equation}
where 
\begin{align}
b_{\alpha_{1},\alpha_{2},\alpha_{3}}= 8\pi^{\frac{3}{2}} \frac{\Gamma(\frac{1}{2}-\alpha_{1})\Gamma(\frac{1}{2}-\alpha_{2})\Gamma(\frac{1}{2}-\alpha_{3})\Gamma(1-\alpha_{1}-\alpha_{2}-\alpha_{3})}{\Gamma(1-\alpha_{1}-\alpha_{2})\Gamma(1-\alpha_{1}-\alpha_{3})\Gamma(1-\alpha_{2}-\alpha_{3})} \,. \label{b3}
\end{align}
The integral (\ref{b3int}) can be computed by setting one angle to zero and projecting to a line, where one can use Feynman parameters.

Since the correction to the conformal propagator $f_{0}(\theta_{12})$ is obtained from  three point function of two fermions with the operator of dimension $h=-1$, it  can be represented as the integral
\begin{equation}
f_{0}(\theta_{12})=\int_{-\pi}^{+\pi}d\theta_{0}\frac{(\sin^{2}\frac{\theta_{01}}{2})^{\frac{1}{2}}(\sin^{2}\frac{\theta_{02}}{2})^{\frac{1}{2}}}{(\sin^{2}\frac{\theta_{12}}{2})^{\frac{1}{2}}}=\begin{tikzpicture}[baseline={([yshift=-4pt]current bounding box.center)}]
\draw[thick](0pt,10pt)-- (17pt,-10pt);
\draw[thick](17pt,-10pt)-- (-17pt,-10pt);
\draw[thick](-17pt,-10pt)-- (0pt,10pt);
\node at (2pt,15pt) {\normalsize $\theta_{0} $};
\node at (23pt,-10pt) {\normalsize $\theta_{2} $};
\node at (-22pt,-10pt) {\normalsize $\theta_{1} $};
\node at (19pt,3pt) {\normalsize $-\nicefrac{1}{2} $};
\node at (-21pt,3pt) {\normalsize $-\nicefrac{1}{2} $};
\node at (0pt,-17pt) {\normalsize $\nicefrac{1}{2} $};
\end{tikzpicture}\
\end{equation}
Since the integrals $\delta k_{\textrm{rung}}^{\rS}$ and $\delta k_{\textrm{rail}}^{\rS}$ are logarithmically divergent we introduce a soft cutoff $\eta \to 0$ by multiplying   $f_{0}(\theta)$ by $(\sin^{2}\frac{\theta}{2})^{\eta}$, so the new graphical representation for $f^{\eta}_{0}(\theta_{12})$ is 
\begin{equation}
f_{0}^{\eta}(\theta_{12})=\int_{-\pi}^{+\pi}d\theta_{0}\frac{(\sin^{2}\frac{\theta_{01}}{2})^{\frac{1}{2}}(\sin^{2}\frac{\theta_{02}}{2})^{\frac{1}{2}}}{(\sin^{2}\frac{\theta_{12}}{2})^{\frac{1}{2}-\eta}}=\begin{tikzpicture}[baseline={([yshift=-4pt]current bounding box.center)}]
\draw[thick](0pt,10pt)-- (17pt,-10pt);
\draw[thick](17pt,-10pt)-- (-17pt,-10pt);
\draw[thick](-17pt,-10pt)-- (0pt,10pt);
\node at (2pt,15pt) {\normalsize $\theta_{0} $};
\node at (23pt,-10pt) {\normalsize $\theta_{2} $};
\node at (-22pt,-10pt) {\normalsize $\theta_{1} $};
\node at (19pt,3pt) {\normalsize $-\nicefrac{1}{2} $};
\node at (-21pt,3pt) {\normalsize $-\nicefrac{1}{2} $};
\node at (0pt,-17pt) {\normalsize $\nicefrac{1}{2}-\eta $};
\end{tikzpicture}\
\end{equation}
Finally we notice that $f_{1-\Delta}f_{\Delta}=-4(q-1)\alpha_{0}$ and the conformal kernel can be depicted as
\begin{equation}
K_{c}^{\rS}(\theta_{1},\theta_{2};\theta_{3},\theta_{4})=\frac{1}{f_{1-\Delta}f_{\Delta}}\begin{tikzpicture}[baseline={([yshift=-4pt]current bounding box.center)}]
\draw[thick,  mid arrow](-15pt,-15pt)-- (15pt,-15pt);
\draw[thick,  mid arrow](-15pt,15pt)-- (15pt,15pt);
\draw[thick](-15pt,-15pt)-- (-15pt,15pt);
\draw[thick](15pt,-15pt)-- (15pt,15pt);
\node at (-21pt,15pt) {\normalsize $\theta_{1} $};
\node at (22pt,15pt) {\normalsize $\theta_{3} $};
\node at (-21pt,-15pt) {\normalsize $\theta_{2} $};
\node at (22pt,-15pt) {\normalsize $\theta_{4} $};
\node at (0pt,23pt) {\normalsize $\Delta$};
\node at (0pt,-22pt) {\normalsize $\Delta$};
\node at (-35pt,0pt) {\normalsize $\nicefrac{1}{2}-\Delta$};
\node at (35pt,0pt) {\normalsize $\nicefrac{1}{2}-\Delta$};
\end{tikzpicture}\
\end{equation}

\subsection{Computation of the rung integral}
In  case of the rung integral we can simply integrate over $\theta_{3}$ and $\theta_{4}$ to obtain 
\begin{align}
&\delta k_{\textrm{rung}}^{\rS} =\frac{(q-2)}{2} \int_{-\pi}^{+\pi}d\theta_{1}d\theta_{2} f_{0}^{\eta}(\theta_{12})|\Psi_{1,n}(\theta_{1},\theta_{2})|^{2}\,. \label{rungint1}
\end{align}
Next, it is convenient to use  decomposition
\begin{align}
\sin^{2}\frac{n\theta}{2}=\sum_{k=1}^{n}c_{k}\big(\sin^{2}\frac{\theta}{2}\big)^{k}, \quad c_{k}= \frac{2 n (-4)^{k-1}\Gamma(n+k)}{\Gamma(n-k+1)\Gamma(2k+1)}\,, 
\label{multangl}
\end{align}
which can be derived from multiple-angle formula and properties of the Chebyshev polynomials. Therefore the integral (\ref{rungint1}) takes the form (in what follows we assume that $n>0$, so $|n|=n$).
\begin{equation}
\delta k_{\textrm{rung}}^{\rS}=\frac{(q-2)}{8\pi^{2}n}\sum_{k=1}^{n}c_{k}\begin{tikzpicture}[baseline={([yshift=-4pt]current bounding box.center)}]
\draw[thick](0pt,10pt)-- (17pt,-10pt);
\draw[thick](17pt,-10pt)-- (-17pt,-10pt);
\draw[thick](-17pt,-10pt)-- (0pt,10pt);
\node at (2pt,15pt) {\normalsize $\theta_{0} $};
\node at (23pt,-10pt) {\normalsize $\theta_{2} $};
\node at (-22pt,-10pt) {\normalsize $\theta_{1} $};
\node at (19pt,3pt) {\normalsize $-\nicefrac{1}{2} $};
\node at (-21pt,3pt) {\normalsize $-\nicefrac{1}{2} $};
\node at (0pt,-20pt) {\scriptsize $\nicefrac{3}{2}-k-\eta $};
\end{tikzpicture} = \frac{(q-2)}{8\pi^{2}n}\sum_{k=1}^{n}c_{k} b_{-\frac{1}{2},-\frac{1}{2},\frac{3}{2}-k-\eta}\,.
\end{equation}
For $k=1$ we have $c_{1}=n^{2}$ and we find using (\ref{b3})
\begin{align}
b_{-\frac{1}{2},-\frac{1}{2}, \frac{1}{2}-\eta}= 8\pi^{2}\Big(\frac{1}{2\eta}-\log2 +1\Big)+O(\eta)\,.
\end{align}
For $k>1$ there is no divergence and we can set $\eta=0$ and compute 
\begin{align}
\sum_{k=2}^{n}c_{k}b_{-\frac{1}{2},-\frac{1}{2}, \frac{3}{2}-k} =\sum_{k=2}^{n}c_{k}\frac{8 \pi ^{3/2} \Gamma \left(k+\frac{1}{2}\right)}{(k-1) \Gamma (k)} = 8\pi^{2}n^{2}\Big(-H_{n}+\frac{1}{2n}+\frac{1}{2}\Big)\,,
\end{align}
where $H_{n}=\sum_{m=1}^{n}\frac{1}{m}$ is the Harmonic number.
Therefore we finally find 
 \begin{align}
\delta k_{\textrm{rung}}^{\rS} = n(q-2)\Big(\frac{1}{2\eta}-\log 2 -H_{n}+\frac{1}{2n}+\frac{3}{2}\Big)\,.
\end{align}

\subsection{Computation of the rail integral}
In this case we represent $h=1$ eigenmodes  $\Psi_{1,n}^{\rS}(\theta_{1},\theta_{2})$ in the form
\begin{align}
\Psi^{\rS}_{1,n}(\theta_{1},\theta_{2}) = \frac{i}{4\pi \sqrt{n}} (e^{-in \theta_{1}}-e^{-in \theta_{2}})\frac{\textrm{sgn}(\theta_{12})}{(\sin^{2}\frac{\theta_{12}}{2})^{\frac{1}{2}}}\,,
\end{align}
therefore we find 
\begin{align}
\Psi^{S*}_{1,n}(\theta_{1},\theta_{2})\Psi^{\rS}_{1,n}(\theta_{3},\theta_{4}) = \frac{1}{16\pi^{2} n}(e^{in\theta_{13}}-e^{in\theta_{14}}-e^{in\theta_{23}}+e^{in\theta_{24}})\frac{\textrm{sgn}(\theta_{12})\textrm{sgn}(\theta_{34})}{(\sin^{2}\frac{\theta_{12}}{2}\sin^{2}\frac{\theta_{34}}{2})^{\frac{1}{2}}}\,.
\end{align}
Since the integral $\delta k_{\textrm{rail}}^{\rS}$ is real we can take only real part in $\Psi^{S*}_{1,n}(\theta_{1},\theta_{2})\Psi^{\rS}_{1,n}(\theta_{3},\theta_{4})$, so we get
\begin{align}
\Psi^{S*}_{1,n}(\theta_{1},\theta_{2})\Psi^{\rS}_{1,n}(\theta_{3},\theta_{4}) \to  \frac{-1}{8\pi^{2} n}\frac{(\sin^{2}\frac{n\theta_{13}}{2}-\sin^{2}\frac{n\theta_{14}}{2}-\sin^{2}\frac{n\theta_{23}}{2}+\sin^{2}\frac{n\theta_{24}}{2})\textrm{sgn}(\theta_{12})\textrm{sgn}(\theta_{34})}{(\sin^{2}\frac{\theta_{12}}{2}\sin^{2}\frac{\theta_{34}}{2})^{\frac{1}{2}}}\,.
\end{align}
Thus finally we decompose  $\delta k_{\textrm{rail}}^{\rS}$ into a sum of four integrals 
\begin{equation}
\delta k_{\textrm{rail}}^{\rS}= \frac{-1}{8\pi^{2}n f_{1-\Delta}f_{\Delta}}\left(\begin{tikzpicture}[baseline={([yshift=-4pt]current bounding box.center)}]
\draw[thick,  mid arrow](-17pt,-17pt)-- (17pt,-17pt);
\draw[thick,  mid arrow](-17pt,17pt)-- (17pt,17pt);
\draw[thick, mid arrow](-17pt,17pt)--(-17pt,-17pt);
\draw[thick, mid arrow](17pt,17pt)--(17pt,-17pt);
\draw[thick](-17pt,17pt)--(0pt,30pt);
\draw[thick](17pt,17pt)--(0pt,30pt);
\draw[thick,dashed] (-17pt,17pt)..controls (-5pt,23pt) and (5pt,23pt)..(17pt,17pt);
\node at (2pt,35pt) {\normalsize $\theta_{0} $};
\node at (-23pt,17pt) {\normalsize $\theta_{1} $};
\node at (24pt,17pt) {\normalsize $\theta_{3} $};
\node at (-23pt,-17pt) {\normalsize $\theta_{2} $};
\node at (24pt,-16pt) {\normalsize $\theta_{4} $};
\node at (16pt,25pt) {\tiny $-\nicefrac{1}{2} $};
\node at (-17pt,25pt) {\tiny $-\nicefrac{1}{2} $};
\node at (0pt,12pt) {\tiny $\frac{1}{2}+\Delta-\eta$};
\node at (0pt,-22pt) {\tiny $\Delta$};
\node at (-29pt,0pt) {\tiny $1-\Delta$};
\node at (29pt,0pt) {\tiny $1-\Delta$};
\node at (0pt,-35pt) {\normalsize $(\textrm{A}) $};
\end{tikzpicture}+\begin{tikzpicture}[baseline={([yshift=-4pt]current bounding box.center)}]
\draw[thick,  mid arrow](-17pt,-17pt)-- (17pt,-17pt);
\draw[thick,  mid arrow](-17pt,17pt)-- (17pt,17pt);
\draw[thick, mid arrow](-17pt,17pt)--(-17pt,-17pt);
\draw[thick, mid arrow](17pt,17pt)--(17pt,-17pt);
\draw[thick](-17pt,17pt)--(0pt,30pt);
\draw[thick](17pt,17pt)--(0pt,30pt);
\draw[thick,dashed] (-17pt,-17pt)..controls (-5pt,-11pt) and (5pt,-11pt)..(17pt,-17pt);
\node at (2pt,35pt) {\normalsize $\theta_{0} $};
\node at (-23pt,17pt) {\normalsize $\theta_{1} $};
\node at (24pt,17pt) {\normalsize $\theta_{3} $};
\node at (-23pt,-17pt) {\normalsize $\theta_{2} $};
\node at (24pt,-16pt) {\normalsize $\theta_{4} $};
\node at (16pt,25pt) {\tiny $-\nicefrac{1}{2} $};
\node at (-17pt,25pt) {\tiny $-\nicefrac{1}{2} $};
\node at (0pt,12pt) {\tiny $\frac{1}{2}+\Delta-\eta$};
\node at (0pt,-22pt) {\tiny $\Delta$};
\node at (-29pt,0pt) {\tiny $1-\Delta$};
\node at (29pt,0pt) {\tiny $1-\Delta$};
\node at (0pt,-35pt) {\normalsize $(\textrm{B}) $};
\end{tikzpicture}-\begin{tikzpicture}[baseline={([yshift=-4pt]current bounding box.center)}]
\draw[thick,  mid arrow](-17pt,-17pt)-- (17pt,-17pt);
\draw[thick,  mid arrow](-17pt,17pt)-- (17pt,17pt);
\draw[thick, mid arrow](-17pt,17pt)--(-17pt,-17pt);
\draw[thick, mid arrow](17pt,17pt)--(17pt,-17pt);
\draw[thick](-17pt,17pt)--(0pt,30pt);
\draw[thick](17pt,17pt)--(0pt,30pt);
\draw[thick,dashed] (-17pt,17pt)..controls (-15pt,-10pt) and (15pt,-15pt)..(17pt,-17pt);
\node at (2pt,35pt) {\normalsize $\theta_{0} $};
\node at (-23pt,17pt) {\normalsize $\theta_{1} $};
\node at (24pt,17pt) {\normalsize $\theta_{3} $};
\node at (-23pt,-17pt) {\normalsize $\theta_{2} $};
\node at (24pt,-16pt) {\normalsize $\theta_{4} $};
\node at (16pt,25pt) {\tiny $-\nicefrac{1}{2} $};
\node at (-17pt,25pt) {\tiny $-\nicefrac{1}{2} $};
\node at (0pt,12pt) {\tiny $\frac{1}{2}+\Delta-\eta$};
\node at (0pt,-22pt) {\tiny $\Delta$};
\node at (-29pt,0pt) {\tiny $1-\Delta$};
\node at (29pt,0pt) {\tiny $1-\Delta$};
\node at (0pt,-35pt) {\normalsize $(\textrm{C}) $};
\end{tikzpicture}-\begin{tikzpicture}[baseline={([yshift=-4pt]current bounding box.center)}]
\draw[thick,  mid arrow](-17pt,-17pt)-- (17pt,-17pt);
\draw[thick,  mid arrow](-17pt,17pt)-- (17pt,17pt);
\draw[thick, mid arrow](-17pt,17pt)--(-17pt,-17pt);
\draw[thick, mid arrow](17pt,17pt)--(17pt,-17pt);
\draw[thick](-17pt,17pt)--(0pt,30pt);
\draw[thick](17pt,17pt)--(0pt,30pt);
\draw[thick,dashed] (17pt,17pt)..controls (15pt,-10pt) and (-15pt,-15pt)..(-17pt,-17pt);
\node at (2pt,35pt) {\normalsize $\theta_{0} $};
\node at (-23pt,17pt) {\normalsize $\theta_{1} $};
\node at (24pt,17pt) {\normalsize $\theta_{3} $};
\node at (-23pt,-17pt) {\normalsize $\theta_{2} $};
\node at (24pt,-16pt) {\normalsize $\theta_{4} $};
\node at (16pt,25pt) {\tiny $-\nicefrac{1}{2} $};
\node at (-17pt,25pt) {\tiny $-\nicefrac{1}{2} $};
\node at (0pt,12pt) {\tiny $\frac{1}{2}+\Delta-\eta$};
\node at (0pt,-22pt) {\tiny $\Delta$};
\node at (-29pt,0pt) {\tiny $1-\Delta$};
\node at (29pt,0pt) {\tiny $1-\Delta$};
\node at (0pt,-35pt) {\normalsize $(\textrm{D}) $};
\end{tikzpicture}
\right) \notag
\end{equation}
where the dashed lines represent $\sin^{2}(n\theta_{ij}/2)$. Applying the integral (\ref{delint}) it is easy to see that parts (C) and (D) vanish.
For the part (A) we also can use the integral (\ref{delint}) and  get 
 \begin{align}
\delta k_{\textrm{rail}, \textrm{A}}^{\rS} = \frac{1}{8\pi^{2} n}\int_{-\pi}^{+\pi} d\theta_{1}d\theta_{3}f_{0}^{\eta}(\theta_{13}) \frac{\sin^{2}\frac{n\theta_{13}}{2}}{\sin^{2}\frac{\theta_{13}}{2}}= \frac{1}{q-2}\delta k_{\textrm{rung}}^{\rS}\,,
\end{align}
thus we find 
 \begin{align}
\delta k_{\textrm{rail}, \textrm{A}}^{\rS}(k) = n\Big(\frac{1}{2\eta}-\log 2 -H_{n}+\frac{1}{2n}+\frac{3}{2}\Big)\,.
\end{align}
To compute the part (B) in $\delta k_{\textrm{rail}}^{\rS}$ we use the formula (\ref{multangl})
 \begin{align}
\delta k_{\textrm{rail},  \textrm{B}}^{\rS} = \sum_{k=1}^{n}c_{k} \delta k_{\textrm{rail}, \textrm{B}}^{\rS}(k)\,.
\end{align}
We notice that for $k> 1$ the integrals  $\delta k_{\textrm{rail}, \textrm{B}}^{\rS}(k)$ are convergent, so we can set the regulator to zero $\eta=0$ and take the integrals using the star-triangle identities and 
the integral (\ref{b3int})
\begin{equation}
\begin{tikzpicture}[baseline={([yshift=-4pt]current bounding box.center)}]
\draw[thick,  mid arrow](-17pt,-17pt)-- (17pt,-17pt);
\draw[thick,  mid arrow](-17pt,17pt)-- (17pt,17pt);
\draw[thick, mid arrow](-17pt,17pt)--(-17pt,-17pt);
\draw[thick, mid arrow](17pt,17pt)--(17pt,-17pt);
\draw[thick](-17pt,17pt)--(0pt,30pt);
\draw[thick](17pt,17pt)--(0pt,30pt);
\node at (2pt,35pt) {\normalsize $\theta_{0} $};
\node at (-23pt,17pt) {\normalsize $\theta_{1} $};
\node at (24pt,17pt) {\normalsize $\theta_{3} $};
\node at (-23pt,-17pt) {\normalsize $\theta_{2} $};
\node at (24pt,-16pt) {\normalsize $\theta_{4} $};
\node at (16pt,25pt) {\scriptsize $-\nicefrac{1}{2} $};
\node at (-18pt,25pt) {\scriptsize $-\nicefrac{1}{2} $};
\node at (0pt,10pt) {\scriptsize $\frac{1}{2}+\Delta$};
\node at (0pt,-22pt) {\scriptsize $\Delta-k$};
\node at (-29pt,0pt) {\scriptsize $1-\Delta$};
\node at (29pt,0pt) {\scriptsize $1-\Delta$};
\end{tikzpicture} = f_{\frac{1}{2}+\Delta,1-\Delta}
\begin{tikzpicture}[baseline={([yshift=-4pt]current bounding box.center)}]
\draw[thick,  mid arrow](-17pt,-17pt)-- (17pt,-17pt);
\draw[thick, mid arrow](-17pt,17pt)--(-17pt,-17pt);
\draw[thick, mid arrow](-17pt,17pt)--(0pt,30pt);
\draw[thick](-17pt,17pt)--(17pt,-17pt);
\draw[thick, mid arrow](0pt,30pt)--(17pt,-17pt);
\node at (2pt,35pt) {\normalsize $\theta_{0} $};
\node at (-23pt,17pt) {\normalsize $\theta_{1} $};
\node at (-23pt,-17pt) {\normalsize $\theta_{2} $};
\node at (24pt,-16pt) {\normalsize $\theta_{4} $};
\node at (-22pt,26pt) {\scriptsize $\Delta-1 $};
\node at (0pt,-22pt) {\scriptsize $\Delta-k$};
\node at (-29pt,0pt) {\scriptsize $1-\Delta$};
\node at (17pt,10pt) {\scriptsize $-\Delta$};
\node at (-3pt,-3pt) {\scriptsize $1$};
\end{tikzpicture}=  f_{\frac{1}{2}+\Delta,1-\Delta}  f_{1-\Delta,\Delta-1}
\begin{tikzpicture}[baseline={([yshift=-4pt]current bounding box.center)}]
\draw[thick](-17pt,-17pt)-- (17pt,-17pt);
\draw[thick](0pt,30pt)--(17pt,-17pt);
\draw[thick](0pt,30pt)--(-17pt,-17pt);
\node at (2pt,35pt) {\normalsize $\theta_{0} $};
\node at (-23pt,-17pt) {\normalsize $\theta_{2} $};
\node at (24pt,-16pt) {\normalsize $\theta_{4} $};
\node at (0pt,-22pt) {\scriptsize $\nicefrac{3}{2}-k$};
\node at (17pt,10pt) {\scriptsize $-\nicefrac{1}{2}$};
\node at (-18pt,10pt) {\scriptsize $-\nicefrac{1}{2}$};
\end{tikzpicture}
\end{equation}
We find 
 \begin{align}
\sum_{k=2}^{n}c_{k}  \delta k_{\textrm{rail}, \textrm{B}}^{\rS}(k) &= -\sum_{k=2}^{n}c_{k} \frac{f_{\frac{1}{2}+\Delta,1-\Delta}f_{1-\Delta,\Delta-1}b_{-\frac{1}{2},-\frac{1}{2},\frac{3}{2}-k}}{8\pi^{2}n f_{\Delta}f_{1-\Delta}} =\frac{(1-q)}{8\pi^{2}n}\sum_{k=2}^{n}c_{k}b_{-\frac{1}{2},-\frac{1}{2},\frac{3}{2}-k} \notag\\
&= n(1-q)\Big(-H_{n}+\frac{1}{2n}+\frac{1}{2}\Big)\,.
\end{align}
Now we compute the remaining integral $ \delta k_{\textrm{rail}, \textrm{B}}^{\rS}(k)$ for $k=1$. We have
\begin{equation}
\delta k_{\textrm{rail},\textrm{B}}^{\rS}(1)= \frac{-1}{8\pi^{2}n f_{1-\Delta}f_{\Delta}}\begin{tikzpicture}[baseline={([yshift=-4pt]current bounding box.center)}]
\draw[thick,  mid arrow](-17pt,-17pt)-- (17pt,-17pt);
\draw[thick,  mid arrow](-17pt,17pt)-- (17pt,17pt);
\draw[thick, mid arrow](-17pt,17pt)--(-17pt,-17pt);
\draw[thick, mid arrow](17pt,17pt)--(17pt,-17pt);
\draw[thick](-17pt,17pt)--(0pt,30pt);
\draw[thick](17pt,17pt)--(0pt,30pt);
\node at (2pt,35pt) {\normalsize $\theta_{0} $};
\node at (-23pt,17pt) {\normalsize $\theta_{1} $};
\node at (24pt,17pt) {\normalsize $\theta_{3} $};
\node at (-23pt,-17pt) {\normalsize $\theta_{2} $};
\node at (24pt,-16pt) {\normalsize $\theta_{4} $};
\node at (16pt,25pt) {\tiny $-\nicefrac{1}{2} $};
\node at (-17pt,25pt) {\tiny $-\nicefrac{1}{2} $};
\node at (0pt,12pt) {\tiny $\frac{1}{2}+\Delta-\eta$};
\node at (0pt,-22pt) {\tiny $\Delta-1$};
\node at (-29pt,0pt) {\tiny $1-\Delta$};
\node at (29pt,0pt) {\tiny $1-\Delta$};
\end{tikzpicture}
\end{equation}
Since this integral is symmetric under $\theta_{1},\theta_{2}\leftrightarrow \theta_{3},\theta_{4}$ we can make a trick  introducing additional regulator $\epsilon$,
by multiplying the integrand by $(\sin^{2}\frac{\theta_{01}}{2})^{-\epsilon}(\sin^{2}\frac{\theta_{03}}{2})^{\epsilon}$ \cite{Ciuchini:1999wy}. Such obtained integral is even under the change $\epsilon \to -\epsilon$,
therefore at the limit of $\epsilon\to0$ it only acquires  correction of order $\epsilon^{2}$. So now if we set $\epsilon=\eta$ this will not affect the  result for $1/\eta$ and the constant term.
This trick allows to apply the star-triangle identity, and we obtain
\begin{equation}
\delta k_{\textrm{rail},\textrm{B}}^{\rS}(1)= \frac{-1}{8\pi^{2}n f_{1-\Delta}f_{\Delta}}\begin{tikzpicture}[baseline={([yshift=-4pt]current bounding box.center)}]
\draw[thick,  mid arrow](-17pt,-17pt)-- (17pt,-17pt);
\draw[thick,  mid arrow](-17pt,17pt)-- (17pt,17pt);
\draw[thick, mid arrow](-17pt,17pt)--(-17pt,-17pt);
\draw[thick, mid arrow](17pt,17pt)--(17pt,-17pt);
\draw[thick](-17pt,17pt)--(0pt,30pt);
\draw[thick](17pt,17pt)--(0pt,30pt);
\node at (2pt,35pt) {\normalsize $\theta_{0} $};
\node at (-23pt,17pt) {\normalsize $\theta_{1} $};
\node at (24pt,17pt) {\normalsize $\theta_{3} $};
\node at (-23pt,-17pt) {\normalsize $\theta_{2} $};
\node at (24pt,-16pt) {\normalsize $\theta_{4} $};
\node at (22pt,25pt) {\tiny $-\nicefrac{1}{2}+\eta $};
\node at (-25pt,25pt) {\tiny $-\nicefrac{1}{2}-\eta $};
\node at (0pt,12pt) {\tiny $\frac{1}{2}+\Delta-\eta$};
\node at (0pt,-22pt) {\tiny $\Delta-1$};
\node at (-29pt,0pt) {\tiny $1-\Delta$};
\node at (29pt,0pt) {\tiny $1-\Delta$};
\end{tikzpicture} = -\frac{f_{\frac{1}{2}+\Delta-\eta,1-\Delta}}{8\pi^{2}n f_{1-\Delta}f_{\Delta}}\begin{tikzpicture}[baseline={([yshift=-4pt]current bounding box.center)}]
\draw[thick,  mid arrow](-17pt,-17pt)-- (17pt,-17pt);
\draw[thick, mid arrow](-17pt,17pt)--(-17pt,-17pt);
\draw[thick, mid arrow](-17pt,17pt)--(0pt,30pt);
\draw[thick](-17pt,17pt)--(17pt,-17pt);
\draw[thick, mid arrow](0pt,30pt)--(17pt,-17pt);
\node at (2pt,35pt) {\normalsize $\theta_{0} $};
\node at (-23pt,17pt) {\normalsize $\theta_{1} $};
\node at (-23pt,-17pt) {\normalsize $\theta_{2} $};
\node at (24pt,-16pt) {\normalsize $\theta_{4} $};
\node at (-27pt,26pt) {\tiny $\Delta-1-\eta $};
\node at (0pt,-22pt) {\tiny $\Delta-1$};
\node at (-29pt,0pt) {\tiny $1-\Delta$};
\node at (21pt,10pt) {\tiny $-\Delta+\eta$};
\node at (-5pt,-6pt) {\tiny $1-\eta$};
\end{tikzpicture} \label{krailBst2}
\end{equation}
Next we are going to use two  integrals assuming that $\theta_{1,2}\in [0,2\pi]$
\begin{equation}
 \begin{aligned}
 &\int_{0}^{2\pi}d\theta_{0}\frac{\textrm{sgn}(\theta_{10})\textrm{sgn}(\theta_{02})}{(\sin^{2}\frac{\theta_{10}}{2})^{\alpha-1}(\sin^{2}\frac{\theta_{02}}{2})^{-\alpha}} = -\pi \frac{(1-\alpha)\sin(\alpha(\pi -|\theta_{12}|))+\alpha\sin((1-\alpha)(\pi -|\theta_{12}|))}{\sin (\pi \alpha)}\,,  \\
&\int_{0}^{2\pi}d\theta_{0}\frac{\textrm{sgn}(\theta_{10})\textrm{sgn}(\theta_{02})}{(\sin^{2}\frac{\theta_{10}}{2})^{\alpha}(\sin^{2}\frac{\theta_{02}}{2})^{-\alpha}} = -2\pi \frac{\sin(\alpha(\pi -|\theta_{12}|))}{\sin (\pi \alpha)}\,, \label{usetwoint}
\end{aligned}
\end{equation}
which can be computed using the Fourier transform.  We notice that the integrand in (\ref{krailBst2}) is periodic under the shift of any variable $\theta_{j}\to \theta_{j}+2\pi$, therefore we can set one variable to zero and multiply the whole integral by $2\pi$. We set $\theta_{4}=0$, then integrate over $\theta_{0}$ and $\theta_{2}$ using (\ref{usetwoint}). Finally we integrate over $\theta_{1}$ and obtain
 \begin{equation}
 \begin{aligned}
\delta k_{\textrm{rail},\textrm{B}}^{\rS}(1)=  -\frac{f_{\frac{1}{2}+\Delta-\eta,1-\Delta}}{8\pi^{2}n f_{1-\Delta}f_{\Delta}} &\frac{ (2\pi)(2\pi^{2})}{\sin (\pi(1-\Delta))\sin (\pi(\Delta-\eta))} \bigg[\frac{1}{2}(1-\Delta+\eta)\big(I(1-2\Delta+\eta)\\
&-I(1-\eta)\big)+\frac{1}{2}(\Delta-\eta)\big(I(-\eta)-I(2-2\Delta+\eta)\big)\bigg]\,,
 \end{aligned}
 \end{equation}
where we denoted 
 \begin{equation}
I(\alpha) = \int_{-\pi}^{\pi} d\theta \frac{\cos (\alpha \theta)}{(\cos^{2}\frac{\theta}{2})^{1-\eta}} = \frac{2^{3-2\eta}\pi \Gamma(2\eta-1)}{\Gamma(\eta+\alpha)\Gamma(\eta-\alpha)}\,.
 \end{equation}
Taking the limit $\eta \to 0$ we find 
 \begin{equation}
\delta k_{\textrm{rail}, \textrm{B}}^{\rS}(1) = \frac{1}{n}(1-q)\Big(\frac{1}{2\eta}-\log 2+\frac{q^3-4 q^2+4}{2 (q-2) (q-1)}-\frac{\pi }{\sin (\frac{2 \pi }{q})}\Big)\,.
 \end{equation}
We notice that this can be written as 
\begin{equation}
\delta k_{\textrm{rail}, \textrm{B}}^{\rS}(k) = \frac{1}{n}(1-q)\Big(\frac{1}{2\eta}-\log 2+ {k^{\rA}_{c}}'(2)+1\Big)\,,
\end{equation}
where $k^{\rA}_{c}(h)$ is defined in \eqref{eqn: KSKA Q=0 v2}. 
So  finally we find for the rail integral  
 \begin{equation}
\delta k_{\textrm{rail}}^{\rS} =-\delta k_{\textrm{rung}}^{\rS}-n(q-1){k^{\rA}_{c}}'(2)\,.
 \end{equation}
Combining the rung and rail integrals and  restoring the factor $-\frac{2\alpha_{G}}{\beta \mathcal{J}}$ we  find for the corrected symmetric kernel eigenvalue 
 \begin{equation}
k^{\rS}(1,n)=1+\delta k_{\textrm{rail}}^{\rS} +\delta k_{\textrm{rung}}^{\rS}=1 +\frac{2 \alpha_{G} (q-1) {k^{\rA}_{c}}'(2)}{\beta \mathcal{J}}|n|\,. \label{lowKres}
 \end{equation}
Using equations (\ref{KrelalS}), (\ref{cortokA}) and (\ref{lowKres}) we find 
 \begin{equation}
 \label{Kovergamma}
\frac{K_{\textrm{linear}}}{\gamma}=\frac{1}{4\pi^{2}}\frac{3q^{2}}{q-1}\frac{\alpha_{K}^{\rS}}{\alpha_{K}^{\rA}} = \frac{3q}{2\pi^{2}}\,,
 \end{equation}
where $\gamma=4\pi^{2}\alpha_{\textrm{S}}$.

Unfortunately, the result (\ref{lowKres}) gives only a partial contribution for corrected symmetric kernel eigenvalue at order $1/\beta \mathcal{J}$, and (\ref{Kovergamma}) agrees with the $K_{\textrm{linear}}$ value in (\ref{Kgamma relation}). Presumably integrals obtained from higher corrections to the Green function have uncompensated power-law divergences which then contribute to the first order $1/\beta \mathcal{J}$ term, changing the final result. 

We also notice that the computation we did can be drastically simplified if we take large $n$ limit as in \cite{MS16-remarks}. Doing this we indeed find the result (\ref{lowKres}). The caveat of such approach in our case is that a priori we can not guarantee that the final result contains only linear in $n$ terms, because each integral is divergent, so in principle we could miss some other corrections in $n$.

\section{Large $q$ for symmetric kernel}
\label{largeqsymkern}
In the large $q$ limit, we take $q\to \infty $, keeping $\mathcal{J} = \sqrt{q}J/2^{\frac{q-1}{2}}$ fixed. Then in the first order 
the Green function reads
\begin{equation}
G(\theta) = \frac{\textrm{sgn}(\theta)}{2}\Big(1+\frac{g(\theta)}{q}+\dots\Big), \quad e^{\frac{1}{2}g(\theta)} = \frac{\cos \frac{\pi v}{2}}{\cos (\frac{\pi v}{2}(1-\frac{|\theta|}{\pi}))} \,, \label{lqgreenf}
\end{equation}
where  we set $\beta =2\pi$ and $v$ is a function of $\beta \mathcal{J}$, which is  found from the equation $\pi v/\cos \frac{\pi v}{2}=\beta \mathcal{J}$, so $v\to 1$, when $\beta \mathcal{J}\to \infty$. Now consider the symmetric kernel 
\begin{equation}
K^{\rS}(\theta_{1},\theta_{2};\theta_{3},\theta_{4}) = -J^{2}|G(\theta_{12})|^{\frac{q-2}{2}} G(\theta_{13})G(\theta_{24})|G(\theta_{34})|^{\frac{q-2}{2}}\,.
\end{equation}
Using the large $q$ Green function (\ref{lqgreenf}) we find in the leading $q$ order
\begin{equation}
K^{\rS}_{q=\infty}(\theta_{1},\theta_{2};\theta_{3},\theta_{4}) =  -\frac{1}{2q}\mathcal{J}^{2}e^{\frac{1}{2}g(\theta_{12})} \textrm{sgn}(\theta_{13})\textrm{sgn}(\theta_{24}) e^{\frac{1}{2}g(\theta_{34})}\,.
\end{equation}
In this expression  the dependence on $q$ remained only in $1/q$ prefactor, the rest depends only on $\beta \mathcal{J}$, which is our parameter. Therefore, evidently, the eigenvalues of the large $q$ kernel
\begin{equation}
\int_{0}^{2\pi} d\theta_{3}d\theta_{4} K^{\rS}_{q=\infty}(\theta_{1},\theta_{2};\theta_{3},\theta_{4}) \Psi^{\rS}_{h,n}(\theta_{3},\theta_{4})= k^{\rS}_{q=\infty} \Psi^{\rS}_{h,n}(\theta_{1},\theta_{2})
\end{equation}
will be proportional to $1/q$: $k^{\rS}_{q=\infty}\propto 1/q $. Parametrizing $k^{\rS}_{q=\infty}=\frac{2}{qh(h-1)}$, which is consistent with the large $q$ limit of the conformal eigenvalues $k^{\rS}_{c}(h)$
we find\footnote{We notice that the large $q$ limit does not commute with  setting $h=1$ first, because $k^{\rS}_{c}(1)=1$ for all $q$.} 
\begin{equation}
-\frac{\mathcal{J}^{2}}{4}h(h-1)\int_{0}^{2\pi} d\theta_{3}d\theta_{4}\textrm{sgn}(\theta_{13})\textrm{sgn}(\theta_{24}) e^{\frac{1}{2}g(\theta_{34})} \Psi^{\rS}_{h,n}(\theta_{3},\theta_{4})= e^{-\frac{1}{2}g(\theta_{12})}  \Psi^{\rS}_{h,n}(\theta_{1},\theta_{2})\,, \label{lqeigval}
\end{equation}
where dependence on $q$ is gone and the eigenvalues $h$ depend only on $\beta \mathcal{J}$. The advantage of the large $q$ limit is the possibility to reduce the integral equation (\ref{lqeigval})
to the second order differential equation. To do this, one uses  $\partial_{\theta}\textrm{sgn}(\theta)=2\delta(\theta)$. Differentiating the expression (\ref{lqeigval}) by $\theta_{1}$ and $\theta_{2}$ we find 
\begin{equation}
e^{-\frac{1}{2}g(\theta_{12})} \partial_{\theta_{1}}\partial_{\theta_{2}}\big(e^{-\frac{1}{2}g(\theta_{12})} \Psi^{\rS}_{h,n}(\theta_{1},\theta_{2})\big) = -\mathcal{J}^{2}h(h-1) \Psi^{\rS}_{h,n}(\theta_{1},\theta_{2})\,.\label{lqdifeq1}
\end{equation}
Now changing   variables to $x=\theta_{12}$ and $y= (\theta_{1}+\theta_{2})/2$ and using the anzats 
\begin{equation}
\Psi_{h,n}^{\rS}(x,y)= e^{-iny} \frac{\psi_{h,n}^{\rS}(x)}{\sin \frac{\tilde{x}}{2}}, \quad \tilde{x} = \pi(1-v)+v |x|\,,
\end{equation}
after some computations we reduce (\ref{lqdifeq1}) to a simple   ordinary differential equation for $\psi_{h,n}^{\rS}(x)$:
\begin{equation}
\left(4\partial_{x}^{2}+n^{2}-\frac{h(h-1) v^{2}}{\sin^{2}\frac{\tilde{x}}{2}}\right)\psi_{h,n}^{\rS}(x) = 0\,. \label{lqordeq}
\end{equation}
Since we are diagonalizing the symmetric kernel the wave-functions $\psi_{h,n}^{\rS}(x)$ must obey the symmetric boundary conditions 
\begin{equation}
\psi_{h,n}^{\rS}(-x) =\psi_{h,n}^{\rS}(x), \quad  \psi_{h,n}^{\rS}(2\pi - x) = (-1)^{n+1} \psi_{h,n}^{\rS}(x)\,, \label{lqbc}
\end{equation}
where in the second condition $x\in[0,2\pi]$. The first condition reduces $x$ domain to $x\in[0,2\pi]$ and also implies that $\partial_{x}\psi_{h,n}^{\rS}(0)=0$, 
which leads to quantization of $h$. A general solution of (\ref{lqordeq}) which obeys the second boundary condition (\ref{lqbc}) reads
\begin{equation}
 \psi_{h,n}^{\rS}(x)= 
\begin{cases}
(-1)^{\frac{n-1}{2}}\frac{1}{2\pi \sqrt{n}}(\sin\frac{\tilde{x}}{2})^{h} \,_{2}F_{1}\Big(\frac{h-n/v}{2}, \frac{h+n/v}{2}, \frac{1}{2}, \cos^{2}\frac{\tilde{x}}{2}\Big),\qquad\qquad\;\; n = \textrm{odd} \\
(-1)^{\frac{n}{2}+1}\frac{\sqrt{n}}{2\pi }\cos\frac{\tilde{x}}{2}(\sin\frac{\tilde{x}}{2})^{h} \,_{2}F_{1}\Big(\frac{1+h-n/v}{2}, \frac{1+h+n/v}{2}, \frac{3}{2}, \cos^{2}\frac{\tilde{x}}{2}\Big)\,, \qquad n = \textrm{even}
\end{cases} \label{psihnsans}
\end{equation}
The normalization is chosen in such a way that for $v\to 1$ in the IR region these functions coincide with the conformal ones.
Indeed, setting $v=1$ and $h=1$ in (\ref{psihnsans})  one  reproduces conformal eigenfunctions $ \psi_{h,n}^{\rS}(x) =  \frac{1}{2\pi n^{1/2}}\sin \frac{n x}{2}$. But this already contradicts the first boundary condition 
$\partial_{x}\psi_{h,n}^{\rS}(0)=0$, which is obtained assuming that the functions $\psi_{h,n}^{\rS}(x)$ are differentiable everywhere. This clash of limits is  a sign that 
UV domain is important for spectrum of the symmetric kernel. 

Using  properties of the hypergeometric functions one finds  $h$ from the first boundary condition  $\partial_{x}\psi_{h,n}^{\rS}(0)=0$ as  series in $1-v\to0$: 
\begin{equation}
\begin{aligned}
h&= 1+n^{\frac{1}{2}} (1-v)^{\frac{1}{2}}+n (1-v) \log (1-v)+ n \Big(H_{n-1}+\log \pi +\frac{1}{2n}\Big)(1-v)\\
&+\frac{3}{2} n^{\frac{3}{2}} (1-v)^{\frac{3}{2}} \log^2(1-v)+n^{\frac{3}{2}}\Big(3 \Big(H_{n-1}+\log \pi +\frac{1}{2n}\Big)-1\Big) (1-v)^{\frac{3}{2}} \log (1-v) +\dots\,.
\end{aligned}
\end{equation}
\begin{figure}[th]
\center
\includegraphics[width=0.55\textwidth]{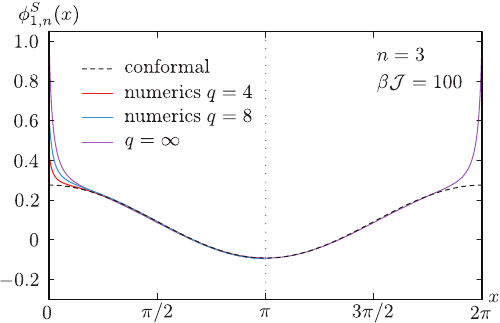}
\caption{\label{psi1nq} Plot of the wave functions $\phi^{\rS}_{1,n}(x)$ for $n=3$, $\beta\mathcal{J}=100$. The red and blue lines correspond to $q=4$ and $q=8$ wave functions, obtained by numerical diagonalization of the symmetric kernel (see section \ref{kerneldiag}). The dashed line is the conformal function (\ref{confeigfun}). The purple line is the large $q$ result (\ref{phihnsans}).  }
\end{figure}
In the leading order in $\beta \mathcal{J}$ this gives for the kernel eigenvalues $ k^{\rS}_{q=\infty}= \frac{\sqrt{2}}{q\sqrt{n}}\sqrt{\beta \mathcal{J}}+\dots$.
Finally the wave functions $\phi_{h,n}^{\rS}(x) = \psi_{h,n}^{\rS}(x)/\sin \frac{\tilde{x}}{2}$ are simply
\begin{equation}
 \phi_{h,n}^{\rS}(x)= 
\begin{cases}
(-1)^{\frac{n-1}{2}}\frac{1}{2\pi \sqrt{n}}(\sin\frac{\tilde{x}}{2})^{h-1} \,_{2}F_{1}\Big(\frac{h-n/v}{2}, \frac{h+n/v}{2}, \frac{1}{2}, \cos^{2}\frac{\tilde{x}}{2}\Big),\qquad\quad\;\;\;\;\; n = \textrm{odd} \\
(-1)^{\frac{n}{2}+1}\frac{\sqrt{n}}{2\pi }\cos\frac{\tilde{x}}{2}(\sin\frac{\tilde{x}}{2})^{h-1} \,_{2}F_{1}\Big(\frac{1+h-n/v}{2}, \frac{1+h+n/v}{2}, \frac{3}{2}, \cos^{2}\frac{\tilde{x}}{2}\Big)\,, \quad n = \textrm{even}
\end{cases} \label{phihnsans}
\end{equation}
and they are indeed diverging near the boundaries for large $\beta \mathcal{J}$:
\begin{equation}
\phi_{h,n}^{\rS}(x)\to \frac{1}{4} \frac{(1-v)^{1/2}}{\sin \Big(\frac{\pi(1-v)+\pi x}{2}\Big)}+\dots\,.
\end{equation}
We compare the large $q$ wave functions (\ref{phihnsans}) and numerical results for $h=1$ and $n=3$ in Fig.~\ref{psi1nq}.

\section{Zero temperature entropy and $\ln \det$ term}
\label{subsec: ln det}

For the Majorana model at large $N$, the free energy can be obtained by extremizing the $(G,\Sigma)$ action:
\begin{equation}
F  \approx \min_{\Sigma}\max_{G} \beta^{-1} I(G,\Sigma) \,.
\end{equation}
Inserting the solution of the Schwinger-Dyson equation $G$ and $\Sigma$, we have
\begin{equation}
    \frac{F}{N}= \underbrace{- \frac{1}{2\beta} \ln \det (-\partial_\tau -\Sigma)}_{F_1} + \underbrace{ \frac{1}{2} \int d\tau \left( \Sigma(\tau) G(\tau) - \frac{1}{q} |G(\tau)|^q \right)}_{F_2}\,.
\end{equation}
At low temperature, 
$
    \frac{F}{N} = \frac{F_0}{N} - \beta^{-1} \sz + O(\beta^{-2})\,
$. 
We would like to calculate $F$ to the first order in $\beta^{-1}$ and extract the linear coefficient $\sz$, namely the zero temperature entropy. Moreover, the linear term is only present in $F_1$, not in $F_2$. One way to see this is to calculate the $F_2$ integral in the conformal limit $\beta \rightarrow \infty$, and try to extract the finite piece in $\beta F_2$:
\begin{equation}
    \beta F_2 \sim \frac{q-1}{2q} 2\pi b \int_0^{2\pi}  \frac{d \theta}{\left(2\sin \frac{\theta}{2} \right)^2}   \qquad \text{at} \quad \beta \rightarrow \infty, \quad \theta = \frac{2\pi}{\beta} \tau\,.
\end{equation}
The integral has a UV divergence that contributes to the ground state energy $F_0$. We are interested in the finite piece, which can be obtained after a regularization. For instance, one can use a cut-off $\epsilon$ and evaluate the integral
\begin{equation}
    \int_{\epsilon}^{2\pi-\epsilon} \frac{d \theta}{(2\sin \frac{\theta}{2})^2} = \frac{2}{\epsilon} - \frac{\epsilon}{6} + O(\epsilon^3) \,,
\end{equation}
which has vanishing constant piece. The absence of even power in $\epsilon$ terms in the above integral is due to the $\theta \rightarrow - \theta$ symmetry of the integrand. Thus, we conclude that the zero temperature entropy is only contained in $F_1$, i.e. the $\ln \det$ term. 

To actually calculate $\sz= \frac{1}{2} \ln \det(-\partial_\tau -\Sigma) $, we substitute $-\partial_\tau-\Sigma = G^{-1}$ by the conformal solution and properly regularize the following sum
\begin{equation}
    \sz =- \sum_{n=0}^{\infty} \ln G(i\omega_n) , \quad G(i\omega_n) = \const \cdot \beta^{1-2\Delta} \cdot \frac{\Gamma (n+ \frac{1}{2}+\Delta)}{\Gamma(n+\frac{3}{2}-\Delta)} \,.
\end{equation}
More explicitly, we consider the partial sum with a cutoff $n_{\Lambda} \sim \beta \rightarrow \infty$ and single out the $n_\Lambda$ independent term as the zero temperature entropy
\begin{equation}
    - \sum_{n=0}^{n_{\Lambda}}  \ln G(i\omega_n) \approx \const \cdot n_{\Lambda}  + \sz\,.
\end{equation}
The analytic formula for the finite piece $\sz$ turns out to be easier to be found by evaluating its $\Delta$ derivative $\sz'(\Delta)$, which amounts to summing digamma functions $\psi(x)= \Gamma'(x)/\Gamma(x)$ by the following formula
\begin{equation}
    \sum_{n=0}^{n_{\Lambda}} \psi(n+x) = n_{\Lambda}(\ln n_{\Lambda} -1) + x \ln n_{\Lambda} + \left[ x- \frac{1}{2}+(1-x)\psi(x)  \right]+ O\left(n_{\Lambda}^{-1}\right) \,.
\end{equation}
Finally, integrating $\sz'(\Delta)$ to the desired position with boundary value, we have the entropy formula for the Majorana model
\begin{equation}
    \sz(\Delta) = \int_0^{\frac{1}{2}-\Delta} \frac{\pi x dx}{\tan (\pi x)}\,.
\end{equation}
This procedure has been described in \cite{Kit.KITP.1,Kit.KITP.2}. The emphasis here is to support the claim $\sz = \frac{1}{2} \ln \det (-\partial_\tau - \Sigma)$, which will be given a bulk interpretation.

\section{Dirac fermion on ${\rm H^2}$}
\label{sec:app_Dirac}

\subsection{Dirac operator and spinors in two dimensions}

In a Lorentzian space, the Dirac Lagrangian is $\calL = - i \bar{\psi} \left(  \gamma^c \nabla_c + M \right)\psi$, where 
\begin{equation}
\psi = \begin{pmatrix}
\psi_{\downarrow} \\
\psi_{\uparrow}
\end{pmatrix} \,, \quad \bar{\psi}= \psi^\dagger \gamma^0 \,, \quad \gamma^0 = \begin{pmatrix}
0 & 1 \\
-1 & 0
\end{pmatrix} \,, \quad \gamma^1 = \begin{pmatrix}
0 & 1 \\
1 & 0 
\end{pmatrix}\,.
\end{equation}
We will work in Euclidean signature. In the case of flat space, the Wick rotation takes $x^0$ to $x^2=ix^0$. The Dirac matrices $\gamma^c$ and spin matrices $\Sigma^{ab}= \frac{1}{4} \left[ \gamma^a, \gamma^b\right]$ are 
\begin{equation}
\gamma^1 = \begin{pmatrix}
0 & 1 \\
1 & 0
\end{pmatrix} \,, \quad \gamma^2 = i \gamma^0 = \begin{pmatrix}
0 & i \\
-i & 0
\end{pmatrix}\,, \quad 
\Sigma^{21}=-\Sigma^{12}= \frac{i}{2} \begin{pmatrix}
1 & 0 \\
0 & -1
\end{pmatrix} = \Lambda_0 
\end{equation}
($\Lambda_0$ represents an infinitesimal counterclockwise rotation.) The Euclidean action is
\begin{equation}
I_{\text{Dirac}}
=\int i \bar{\psi} \left(\gamma^c \nabla_c + M \right)\psi\,\sqrt{g}\,d^2x\,.
\end{equation}
The Majorana case differs in that $\psi_{\downarrow}$, $\psi_{\uparrow}$ are real and that the action has an overall factor $\frac{1}{2}$.

Note that the Dirac spinor $\psi$ splits into two irreducible representations of the universal cover of $\SO(2)$ (or the Lorentz group): $\psi_{\downarrow}$ has spin $-\frac{1}{2}$, and $\psi_{\uparrow}$ has spin $\frac{1}{2}$. A general $\nu$-spinor $\xi$ is one-dimensional and transforms as $\Lambda_0 \xi = -i \nu \xi$. In the absence of electromagnetic field, the covariant derivative may be written as 
\begin{equation}
\nabla_{\alpha} \psi = \left(\partial_{\alpha} + \frac{1}{2} \omega_{\alpha bc} \Sigma^{bc}  \right) \psi = \left( \partial_{\alpha} + \omega_{\alpha}\Lambda_0  \right) \psi \,,
\end{equation}
\begin{equation}
\begin{pmatrix}
\omega_{\alpha 11} & \omega_{\alpha 12} \\
\omega_{\alpha 21} & \omega_{\alpha 22} 
\end{pmatrix} = \omega_{\alpha} \begin{pmatrix}
0 & -1 \\
1 & 0
\end{pmatrix} \,.
\end{equation}
Here, $(\omega_{\alpha bc})$ is a spin connection defined relative to some local orthonormal frame (vielbein), whereas $(\omega_\alpha)$ may be regarded as a vector potential such that
\begin{equation}
\partial_{\alpha}\omega_\beta - \partial_{\beta} \omega_\alpha = - \frac{R}{2} \epsilon_{\alpha \beta}\,.
\end{equation}

To take advantage of the splitting of the tangent space under $\SO(2)$, let us replace the local orthonormal frame $(\vec{e}_1,\vec{e}_2)$ with 
\begin{equation}
\vec{e}_+ = \frac{1}{2} \left( \vec{e}_1 - i \vec{e}_2 \right) \,, \quad \vec{e}_- = \frac{1}{2} \left( \vec{e}_1 + i \vec{e}_2 \right)\,.
\end{equation}
This transformation is modeled on the transition from the Cartesian coordinates $(x^1,x^2)$ to $(x^+,x^-)=(z,\bar{z})=(x^1+ix^2,x^1-ix^2)$, in which case
\begin{equation}
\begin{pmatrix}
x^1 \\
x^2
\end{pmatrix} = \frac{1}{2} \begin{pmatrix}
1 & 1 \\
-i & i
\end{pmatrix} 
\begin{pmatrix}
x^+ \\
x^-
\end{pmatrix} = \begin{pmatrix}
e^1_+ & e_-^1 \\
e^2_+ & e_-^2 
\end{pmatrix}\begin{pmatrix}
x^+ \\
x^-
\end{pmatrix}
\end{equation}
In the new frame, the metric $\eta_{ab}=\left(\vec{e}_a,\vec{e}_b\right) $ and other relevant matrices are as follows:
\begin{equation}
\begin{pmatrix}
\eta_{++} & \eta_{+-} \\
\eta_{-+} & \eta_{--} 
\end{pmatrix} = \frac{1}{2} \begin{pmatrix}
0 & 1 \\
1 & 0
\end{pmatrix} \,, \quad 
\begin{pmatrix}
\omega_{\alpha ++} & \omega_{\alpha +-} \\
\omega_{\alpha -+}& \omega_{\alpha --}
\end{pmatrix} = \omega_{\alpha} \cdot \frac{i}{2} \begin{pmatrix}
0 & -1 \\
1 & 0
\end{pmatrix} \,,
\end{equation}
\begin{equation}
\gamma^+ = \begin{pmatrix}
0 & 0 \\
2 & 0
\end{pmatrix}\,, \quad 
\gamma^- = \begin{pmatrix}
0 & 2 \\
0 & 0
\end{pmatrix}\,, \quad \Sigma^{+-} = - \Sigma^{-+} = 2i \Lambda_0 \,.
\end{equation}

The individual components $\nabla_+$, $\nabla_-$ of the covariant derivative are themselves covariant, i.e. they commute with gauge transformations. More generally, let us define the operator 
\begin{equation}
\nabla_{\pm} = \partial_{\pm} - i \nu \omega_{\pm}
\label{eqn: Dirac no gauge}
\end{equation}
taking a $\nu$-spinor to a $(\nu\pm 1)$-spinor. The operators $\nabla_+$, $\nabla_-$, $\nu$ satisfy the following commutation relations:
\begin{equation}
\left[ \nu, \nabla_{\pm} \right] = \pm \nabla_{\pm} \,, \quad \left[ \nabla_+,\nabla_- \right] = - \frac{R}{4} \nu \,.
\label{eqn: Dirac commutation}
\end{equation}
In this notation, the Dirac operator becomes
\begin{equation}
D = \gamma^c \nabla_c +M = \begin{pmatrix}
M & 2 \nabla_- \\
2 \nabla_+ & M
\end{pmatrix}\,,
\label{eqn: Dirac operator}
\end{equation}
where $\nabla_+$ acts on the $\nu= - \frac{1}{2}$ spinor $\psi_\downarrow$, and $\nabla_-$ acts on the $\nu=\frac{1}{2}$ spinor $\psi_{\uparrow}$. 

If a $\UU(1)$ gauge field is present, \eqref{eqn: Dirac no gauge} should be replaced with 
\begin{equation}
\nabla_{\pm} = \partial_{\pm} - i\nu \omega_{\pm} - i A_{\pm} \,.
\end{equation}
Let us consider a special case where the field strength is imaginary and proportional to the curvature: 
\begin{equation}
A_{\alpha} = - i \calE \omega_{\alpha}\,,\quad \calE = \text{const} \,.
\end{equation}
This is equivalent to changing the spin by $-i\calE$. Thus, $\psi_{\downarrow}$ and $\psi_{\uparrow}$ have spins $-\frac{1}{2}-i\calE$ and $\frac{1}{2}-i\calE$, respectively.

\subsection{Spinors on ${\rm H^2}$ and $\tSL(2,\RR)$ symmetry}

 The hyperbolic plane ${\rm H^2}$ is described by the Poincare disk model with the metric
 \begin{equation}
 ds^2 = \frac{4\, dz\, d\bar{z}}{(1-z\bar{z})^2} \,, \quad z=x^1+ix^2 \,.
 \end{equation}
 The local frame is chosen to be proportional to the coordinate frame:
 \begin{equation}
 \begin{pmatrix}
 e^1_1 & e^1_2 \\
 e^2_1 & e^2_2 
 \end{pmatrix}=
 \begin{pmatrix}
 e^+_+ & e^+_- \\
 e^-_+ & e^-_- 
 \end{pmatrix} = f^{-\frac{1}{2}}
 \begin{pmatrix}
 1 & 0 \\
 0 & 1
 \end{pmatrix}\,, \quad f= \frac{4}{(1-z\bar{z})^2} \,.
 \end{equation}
This choice is called the ``disk gauge'' in Ref.~\cite{Kitaev:2017hnr}. It is a special case of conformal gauge, which is defined for any metric in the conformal form, $ds^2 = f(z,\bar{z})\, dz\, d\bar{z}$. In this gauge, 
\begin{equation}
\begin{aligned}
&\nabla_+ = f^{-\frac{1}{2}} \left( \partial_z - i \nu \omega_z  \right) \,, \quad \omega_z = - \frac{i}{2} \partial_z \ln f = \frac{-i \bar{z}}{1-z\bar{z}} \,, \\
&\nabla_- = f^{-\frac{1}{2}} \left( \partial_{\bar{z}} - i \nu \omega_{\bar{z}}  \right) \,, \quad \omega_{\bar{z}} =  \frac{i}{2} \partial_{\bar{z}} \ln f = \frac{i {z}}{1-z\bar{z}} \,.
\end{aligned}
\label{eqn: disk gauge}
\end{equation}
The operators $\nabla_+$, $\nabla_-$ commute with isometries of the underlying manifold.

Let us introduce modified polar coordinates $(u,\varphi)$ such that
\begin{equation}
z = \sqrt{u} e^{i\varphi} \,, \quad \bar{z} = \sqrt{u} e^{-i\varphi} \,.
\end{equation}
A $\nu$-spinor with angular momentum $m$ is proportional to $e^{i(m-\nu)\varphi}$ (in the disk gauge), hence,
\begin{equation}
m = - i \partial_{\varphi} + \nu \,.
\end{equation}

We now discuss $\tilde{\SL}(2,\RR)$ symmetry following Ref.~\cite{Kitaev:2017hnr}. The abstract symmetry generators $L_{-1}$, $L_{0}$, $L_{1}$, have two different realizations. The more natural one is by Killing vector fields, acting on spinors as follows:
\begin{equation}
L_{0}^L=-m \,, \quad L^L_{\pm 1} = e^{\pm i\varphi} \left(
\pm (1-u) u^{\frac{1}{2}} \partial_u + \frac{-m - \nu}{2} u^{\frac{1}{2}} + \frac{-m + \nu}{2} u^{-\frac{1}{2}}
 \right)\,.
\end{equation}
The second set of operators (commuting with the first) is denoted by $L_n^R$; they change the spin by $-n$: 
\begin{equation}
L_0^{R} = \nu\,, \quad   L_{\pm 1}^R =  e^{\pm i \varphi} \left(
\mp (1-u) u^{\frac{1}{2}} \partial_u + \frac{-m - \nu}{2} u^{\frac{1}{2}} + \frac{m - \nu}{2} u^{-\frac{1}{2}}
 \right) \,.
\end{equation}
Both sets have the same commutators and Casimir operator $Q$:
\begin{equation}
\left[ L_n , L_k \right] = (n-k) L_{n+k}\,, \qquad Q= - L_0^2 + \frac{1}{2} \left(  L_{-1} L_1 + L_1 L_{-1} \right) \,.
\end{equation}
Note that
\begin{equation}
L_1^R = - 2\nabla_- \,, \quad L_{-1}^R =  2\nabla_+ \,
\label{eqn: Dirac component}
\end{equation}
and that the commutation relations between these operators are just a special case of \eqref{eqn: Dirac commutation}. 
This is an explicit expression for the Casimir operator:
\begin{align}
Q &= -4 \nabla_- \nabla_+ - \nu -\nu^2 = -4 \nabla_+ \nabla_- + \nu -\nu^2
\label{eqn: Casimir}\\
&= - (1-u)^2 \left( u \partial_u^2 +\partial_u \right) + \frac{1-u}{4u} \left(
(m-\nu)^2 - (m+\nu)^2 u
 \right)\,.
\end{align}

The eigenvalues of $Q$ may be parametrized as $\Delta(1-\Delta)$. The joint eigenspace of this and the angular momentum operator, $m=-i\partial_\varphi+\nu$, is spanned by the $\nu$-spinors $\xi^{\nu}_{\Delta,m}$, $\xi^{\nu}_{1-\Delta,m}$ defined by the formula
\begin{equation}
\xi^{\nu}_{\Delta,m}(u,\varphi) = e^{i (m-\nu) \varphi} u^{\frac{m-\nu}{2}} (1-u)^{\Delta} {\rm \bf F}(\Delta+m,\Delta-\nu,2\Delta;1-u)\,,
\label{eqn: xi_nuDm}
\end{equation}
where ${\rm \bf F}(a,b,c;x)= \Gamma(c)^{-1} ~_2{\rm F}_1(a,b,c;x)$ is the scaled hypergeometric function. The asymptotic behavior near the boundary is as follows:
\begin{equation}
\xi^{\nu}_{\Delta,m} (u,\varphi) \approx \frac{1}{\Gamma(2\Delta)} e^{i (m-\nu) \varphi} (1-u)^\Delta \quad \text{for} \quad u\rightarrow 1 \,.
\end{equation}
We will also need the asymptotics at the origin, 
\begin{equation}
\xi^{\nu}_{\Delta,m} (Z) \approx 
\frac{\Gamma(\nu-m)}{\Gamma(\Delta+\nu)\,\Gamma(\Delta-m)} z^{m-\nu} +
\frac{\Gamma(m-\nu)}{\Gamma(\Delta-\nu)\,\Gamma(\Delta+m)} {\bar{z}}^{\nu-m}
\quad \text{for} \quad z \rightarrow 0 \,,
\label{eqn: asymptotic near 0}
\end{equation}
where one term usually dominates. In the special case $m=\nu$, we have
\begin{equation}
\xi^{\nu}_{\Delta,\nu} (Z) \approx 
\frac{-2}{\Gamma(\Delta+\nu)\,\Gamma(\Delta-\nu)} \ln |z|
\quad \text{for} \quad z \rightarrow 0 \,.
\end{equation}
Here and below, $Z$ stands for a point in ${\rm H^2}$ with coordinate $(z,\bar{z})$. This way, we distinguish general functions like $\xi^\nu_{\Delta,m}$ from analytic functions of $z$. 

The operators $L^R_{\pm 1}$ act on the basis spinors as follows:
\begin{equation}
L_{\pm 1}^R \xi_{\Delta,m}^\nu = (-\nu \pm \Delta) \xi^{\nu\mp 1}_{\Delta,m} \,.
\end{equation}
Using this fact, we can construct solutions of the Dirac equation $D \psi =0$ away from the origin. Indeed, it follows from \eqref{eqn: Dirac operator}, \eqref{eqn: Dirac component} that
\begin{equation}
D = \begin{pmatrix}
M & -L_1^R \\
L_{-1}^R & M
\end{pmatrix} \,.
\end{equation}
Recall that the Dirac operator $D$ acts on a vector whose components have spins $-\frac{1}{2}-i\calE$ and $\frac{1}{2}-i\calE$. A simple calculation yields a pair of fundamental solutions with angular momentum $m$:
\begin{equation}
\psi^M_{\pm, m} = \begin{pmatrix}
e^{\pm i \frac{\gamma}{2}} \xi^{-\frac{1}{2}-i\calE}_{\Delta_{\pm},m}
\vspace{2pt}\\
\pm e^{\mp i \frac{\gamma}{2}} \xi^{\frac{1}{2}-i\calE}_{\Delta_{\pm},m}
\end{pmatrix} \,, \quad \Delta_{\pm} = \frac{1}{2} \pm \sqrt{M^2-\calE^2}\,, \quad \gamma=\arcsin \frac{\calE}{M} \,.
\label{eqn: fundamental solutions}
\end{equation}
The  solutions $\psi^M_{+, m} $ and $\psi^M_{-, m} $ correspond to the ``Dirichlet'' and ``Neumann'' boundary conditions, respectively. More exactly, they have the following asymptotic form:
\begin{equation}
\psi(u,\varphi) \approx \psi(\varphi) \eta_{\pm} (u,\varphi) \quad \text{for} \quad u \rightarrow 1\,,  \qquad 
\begin{aligned}
&+: \quad \text{Dirichlet} \\
&-:  \quad \text{Neumann} 
\end{aligned}
\end{equation}
where
\begin{equation}
\eta_{\pm} (u,\varphi)= (1-u)^{\Delta_{\pm}}
\begin{pmatrix}
e^{i \frac{(\varphi\pm\gamma)}{2}} \\
\pm e^{-i \frac{(\varphi\pm\gamma)}{2}}
\end{pmatrix} \,.
\label{eqn: eta}
\end{equation}

\subsection{The propagator}

For a general Fermi system described by Grassmann variables $\psi_j$, $\psi_j^*$, the Euclidean propagator is simply the correlation function $C^{jk}= \langle \psi_j \psi_k^* \rangle$. If the action is quadratic, $I=-\sum_{j,k} B_{jk} \psi_j^* \psi_k$, the propagator is given by $C=-B^{-1}$. For the Dirac fermion, it is convenient to multiply $B$ by $(\gamma^0)^{-1}$ on the left and $C$ by $\gamma^0$ on the right so that
\begin{gather}
I = -\bar{\psi} B \psi \,,
\displaybreak[0]\\[3pt]
C=\langle \psi \bar{\psi} \rangle = \begin{pmatrix}
-C^{\downarrow \uparrow} &  C^{\downarrow \downarrow} \\
-C^{\uparrow\uparrow} & C^{\uparrow \downarrow} 
\end{pmatrix}\,, \quad C^{jk} = \langle \psi_j \psi_k^* \rangle \,,
\end{gather}
where the indices $j$, $k$ include both spin and spatial coordinates. (The relation $C=-B^{-1}$ still holds.) In this case, $B=-iD$, where $D$ is the Dirac operator. Thus,
\begin{equation}
i D^{(1)}C(Z_1,Z_0) = \delta (Z_1,Z_0) \begin{pmatrix}
1 & 0 \\
0 & 1
\end{pmatrix}\,.
\label{eqn: Dirac equation}
\end{equation}
Here, the superscript$~^{(1)}$ indicates that $D$ acts on $Z_1=(z_1,\bar{z}_1)$, whereas
\[
\delta(Z_1,Z_0)
= \frac{\delta(\Re (z_1-z_0))\,\delta(\Im (z_1-z_0))}{\sqrt{|g(Z_0)|}}\,.
\]

To solve \eqref{eqn: Dirac equation}, we first determine the asymptotic behavior of $C(Z_1,Z_0)$ for $Z_1-Z_0\rightarrow0$. In this limit, the mass term and spin connection in \eqref{eqn: Dirac operator}, \eqref{eqn: disk gauge} may be neglected so that
\begin{equation}
D \approx 2 f^{-\frac{1}{2}} \begin{pmatrix}
0 & \partial_{\bar{z}} \\
\partial_z & 0
\end{pmatrix}\,, \quad f= \sqrt{|g|} = \frac{4}{(1-z\bar{z})^2}\,.
\end{equation}
Hence, 
\begin{equation}
C(Z_1,Z_0) \approx \frac{1-z_0 \bar{z}_0}{4\pi i}
\begin{pmatrix}
0 & (\bar{z}_1-\bar{z}_0)^{-1} \\
(z_1-z_0)^{-1}& 0
\end{pmatrix} \quad \text{for} \quad z_1-z_0 \rightarrow 0 \,.
\label{eqn: propagator near 0}
\end{equation}
 Next, we consider the case $Z_0=0$. The columns of the $C$ matrix must be proportional to the fundamental solutions  \eqref{eqn: fundamental solutions} with suitable values of $m$. Matching the $z\rightarrow 0$ asymptotics \eqref{eqn: asymptotic near 0} with \eqref{eqn: propagator near 0}, we get this result:
 \begin{equation}
 C_{\pm}(Z;0) = \frac{\Gamma(\Delta_{\pm}-\downarrow)\,
\Gamma(\Delta_{\pm}+\uparrow)}{4\pi i} 
 \begin{pmatrix}
\pm e^{\pm i \gamma} \xi^{\downarrow}_{\Delta_\pm, \downarrow} (Z) &  \xi^{\downarrow}_{\Delta_\pm, \uparrow} (Z)
\vspace{2pt}\\
\xi^{\uparrow}_{\Delta_\pm, \downarrow} (Z)  & \pm e^{\mp i \gamma} \xi^{\uparrow}_{\Delta_\pm, \uparrow} (Z)
 \end{pmatrix}
 \label{eqn: propagator Z0}
 \end{equation}
 where ${\downarrow} = -\frac{1}{2}-i\calE$ and ${\uparrow} = \frac{1}{2}-i\calE$.

 Now, let us compute the propagator $C_{\pm}(Z_1,Z_0)$ for arbitrary  $Z_1,Z_0$. To this end, let $V$ be the conformal map of the unit disk such that $V(z_0)$ is $0$ and $V(z_1)$ is some positive real number, denoted as $z_{10}$:
 \begin{gather}
 V(z) = \frac{w}{v} \cdot \frac{z-z_0}{1-z\bar{z}_0}\,,
 \displaybreak[0]\\[2pt]
 v= \sqrt{\frac{z_1-z_0}{\bar{z}_1-\bar{z}_0}} \,, \quad w= \sqrt{\frac{1-z_1\bar{z}_0}{1-\bar{z}_1z_0}}\,, \quad z_{10} = \left| 
 \frac{z_1-z_0}{1-z_1 \bar{z}_0}
 \right|\,.
\label{eqn: bcz10}
 \end{gather}
 Conformal maps of the Poincare disk transform a $\nu$-spinor $\xi$ to
 \begin{equation}
 (V^{-1} \xi)(z,\bar{z})
= a(z,\bar{z})^{\nu}\, \xi\bigl(V(z),\bar{V(z)}\bigr)\,, \quad a(z,\bar{z})= \sqrt{\frac{d V(z)/dz}{d\bar{V(z)}/d\bar{z}}}\,.
 \end{equation}
 Therefore, 
 \begin{equation}
 C_{\pm}(Z_1,Z_0) = \begin{pmatrix}
 a(Z_1)^{-\frac{1}{2}-i\calE} & 0 \\
 0 & a(Z_1)^{\frac{1}{2}-i\calE}
 \end{pmatrix}
 C_{\pm}(Z_{10},0)
 \begin{pmatrix}
 a(Z_0)^{\frac{1}{2}+i\calE} & 0 \\
 0 & a(Z_0)^{-\frac{1}{2}+i\calE}
 \end{pmatrix}\,.
 \end{equation}
 Plugging the concrete expression \eqref{eqn: propagator Z0} for $C(Z,0)$ together with 
 \begin{equation}
 a(Z_1) = \frac{1}{vw}\,, \quad a(Z_0) = \frac{w}{v}\,,
 \end{equation}
 we obtain the final answer:
\begin{equation}
C_{\pm}(Z_1;Z_0) = \frac{\Gamma(\Delta_{\pm}-\downarrow)\, \Gamma(\Delta_{\pm}+\uparrow)}{4\pi i}\,
w^{2i\calE}
\begin{pmatrix}
\pm w e^{\pm i \gamma} \xi^{\downarrow}_{\Delta_\pm, \downarrow} (Z_{10}) &
v \xi^{\downarrow}_{\Delta_\pm, \uparrow} (Z_{10})
\vspace{2pt}\\
v^{-1} \xi^{\uparrow}_{\Delta_\pm, \downarrow} (Z_{10})  &
\pm w^{-1} e^{\mp i \gamma} \xi^{\uparrow}_{\Delta_\pm, \uparrow} (Z_{10})
\end{pmatrix}\,,
\label{eqn: propagator Z1Z0}
\end{equation}
where ${\downarrow} = -\frac{1}{2}-i\calE$,\, ${\uparrow} = \frac{1}{2}-i\calE$,\, $\gamma=\arcsin(\calE/M)$;  the numbers $v$, $w$, $z_{10}$ are defined in \eqref{eqn: bcz10}, and the functions $\xi^{\nu}_{\Delta,m}$ in \eqref{eqn: xi_nuDm}.
 
 Finally, we examine the near-boundary asymptotics of the propagator. Let 
 \begin{equation}
 z_0=\sqrt{u_0} e^{i\varphi_0}\,, \quad z_1 =\sqrt{u_1} e^{i\varphi_1}\,, \quad 0 < \varphi_1-\varphi_0 < 2\pi\,, \quad u_0\,,  u_1 \rightarrow 1\,.
 \end{equation}
 Then
 \begin{equation}
 v \approx e^{i(\varphi_1+\varphi_0+\pi)/2} \,, \quad 
  w \approx e^{i(\varphi_1-\varphi_0-\pi)/2} \,, \quad 1-z_0^2 \approx \frac{(1-u_1)(1-u_0)}{4 \sin^2 \frac{\varphi_1-\varphi_0}{2}} \,.
 \end{equation}
 Hence, 
 \begin{equation}
 C_{\pm}(Z_1;Z_0) \approx \frac{\Gamma(\Delta_{\pm}-\downarrow)\,
\Gamma(\Delta_{\pm} +\uparrow)}{4\pi\,\Gamma(2\Delta_\pm)}\,
e^{\calE (\pi-\varphi_1+\varphi_0) } \biggl( 2\sin \frac{\varphi_1-\varphi_0}{2} \biggr)^{-2\Delta_{\pm}} \eta_{\pm}(Z_1) \bar{\eta}_{\pm}(Z_0)
 \end{equation}
 where $\eta_{\pm}(Z)$ is defined in \eqref{eqn: eta}. Importantly, the scalar factor 
$e^{\calE (\pi-\varphi_1+\varphi_0) } \left( 2\sin \frac{\varphi_1-\varphi_0}{2} \right)^{-2\Delta_{\pm}}$ coincides with 
$-\tilde{\Sigma}(\varphi_1,\varphi_0)$ (for Dirichlet b.c.) or $-G(\varphi_1,\varphi_0)$ (for Neumann b.c.) up to an overall constant.

\section{Higher dimensional black holes in asymptotically AdS space}
\label{app:em}

This appendix will begin by recalling the basic thermodynamic properties of charged spherical black holes in global AdS$_{d+2}$ \cite{Myers99,Faulkner09} in $d+2$ spacetime dimensions ($d \geq 2$). We denote the $T=0$ radius of the black hole by $R_h$. Then we will discuss the universal structure of the theory of such black holes at temperature $T \ll 1/R_h$ \cite{SS15,Gaikwad:2018dfc,Nayak:2018qej,Moitra:2018jqs,Chaturvedi:2018uov,Sachdev19,Moitra:2019bub}, where the effective action in Eq.~(\ref{Seff}) applies. See Ref.~\cite{Sachdev19} for more details.

In the AdS/CFT correspondence, AdS$_{d+2}$ spacetimes are dual to conformal field theories (CFTs) in $d+1$ spacetime dimensions. With an Einstein-Hilbert gravitational action, the CFT is the large $N$ maximally supersymmetric SU($N$) Yang-Mills theory for $d=3$, and a suitable large $N$ limit of the ABJM theory for $d=2$. We place the CFT on a sphere, $S_d$, and add a chemical potential conjugate to a global U(1) symmetry. This induces a total charge $Q$ on $S_d$. In the holographic description, the asymptotically AdS$_{d+2}$ spacetime crosses over to a charged black hole with a with a near-horizon AdS$_2 \times S_d$ geometry \cite{Myers99,Faulkner09,Hartnoll:2016apf}, which was identified with the physics of SYK models \cite{SS10}.

The Einstein-Maxwell theory of a metric $g$ and a U(1) gauge flux $F = dA$ has Euclidean action
\begin{equation}
I_{\text{EM}} =  \int \left[ -\frac{1}{2 \kappa^2} \left(\mathcal{R}_{d+2} + \frac{d(d+1)}{L^2} \right) +  \frac{1}{4g_F^2} F^2 \right] \sqrt{g} \ d^{d+2} x  \,, \label{EM}
\end{equation}
where $\kappa^2 = 8 \pi G_N$, $\mathcal{R}_{d+2}$ is the Ricci scalar, 
$L$ is the radius of AdS$_{d+2}$, and $g_F$ is a gauge coupling constant.
The properties of the black hole are fully specified by the temperature $T$ and the chemical potential $\mu$. The later is specified by a boundary condition on the time component of the $\UU(1)$ gauge potential $A$
\begin{equation}
\lim_{r \rightarrow \infty} A_\tau (r, \tau) = i \mu \,.
\end{equation}

At $T=0$, let the radius of the black hole horizon equal $R_h$, the total charge equal $Q$, and the chemical potential equal $\mu_0$. These quantities are related to each other by 
\begin{equation}
\begin{aligned}
Q &= \frac{s_d R_h^{d-1} \sqrt{d \left[ (d+1)R_h^2  + (d-1)L^2  \right]}}{L \kappa g_F} \,, \\
\mu_0 &=  \frac{g_F}{L \kappa (d-1)} \left[ d \left( (d+1) R_h^2 + (d-1) L^2 \right) \right]^{1/2}\,,
\end{aligned}
\end{equation}
where $s_d$ is the area of the $d$-dimensional surface of a unit sphere. We will treat $R_h$ as the independent variable below, the dependence on $Q$ and $\mu_0 = \mu(T=0)$ follows from the above.

Moving to non-zero $T \ll 1/R_h$, we find the entropy $S( Q, T)$ has the form in Eq.~(\ref{defgamma}) (we do not track factors of $N$ in this Appendix) with
\begin{equation}
\sz (Q) = \frac{2 \pi s_d}{\kappa^2} \, R_h^d  \,, \quad
\gamma = \left( \frac{\partial S}{\partial T} \right)_Q = 
\frac{4 \pi^2 d s_d L^2 R_h^{d+1}}{\kappa^2 (d(d+1) R_h^2 + (d-1)^2 L^2)}\,.
\label{S0bh}
\end{equation}
Note that the entropy is simply given by the area of the horizon in the higher-dimensional geometry. The contribution of fermion determinants to the action is subdominant in the large $N$ limit of the AdS/CFT correspondence, unlike the computation in Section~\ref{sec:bulk}.

The low $T$ behavior of the chemical potential is given as follows,
\begin{equation}
\mu = \mu_0 - 2 \pi \calE T \,, \qquad  T \rightarrow 0 ~\text{and}~~Q~~\text{fixed}  
\end{equation}
where
\begin{equation}
2 \pi \calE = - \left( \frac{\partial \mu}{\partial T} \right)_{Q} = 
\frac{2 \pi g_F R_h L \sqrt{d \left[ (d+1) R_h^2 + (d-1)L^2 \right]}}{\kappa \left[ d(d+1)R_h^2+  (d-1)^2 L^2 \right]}\,. \label{dmudt}
\end{equation}
We can now verify that the Maxwell relation
\begin{equation}
 - \left( \frac{\partial \mu}{\partial T} \right)_{Q} = \left(\frac{\partial S}{\partial Q} \right)_T = \frac{\partial \sz/\partial R_h}{\partial Q/\partial R_h}
\end{equation}
is obeyed as $T \rightarrow 0$, which then implies the fundamental identity in Eq.~(\ref{dSdQ}). Finally, we also note the value of the compressibility
\beq
K = \left. \frac{d Q}{d \mu} \right|_{T=0} =  \frac{d Q/dR_h}{d \mu_0/dR_h} = \frac{(d-1) s_d R_h^{d-3} \left[ d(d+1) R_h^2
+ (d-1)^2 L^2 \right]}{(d+1) g_F^2}\,.
\eeq

Now we turn to the universal structure for $T \ll 1/R_h$. The near-horizon metric takes the 
AdS$_2 \times S_d$ form with metric  and gauge field (at $T=0$, see Ref.~\cite{Sachdev19} for $T>0$)
\begin{equation}
ds^2 = R_2^2 \left[ \frac{-dt^2 + d \zeta^2}{\zeta^2} \right] + R_h^2 d \Omega_d^2 \,, \quad
A = \frac{\calE}{\zeta} dt \,, \label{defE}
\end{equation}
where
\begin{equation}
R_2  = \frac{L R_h}{\sqrt{d(d+1) R_h^2 + (d-1)^2 L^2}} \,, \label{valR2}
\end{equation}
and the dimensionless parameter $\calE$ determining the strength of the near-horizon electric field is the same as that in Eq.~(\ref{dmudt}). The Green function of a massive Dirac fermion at the AdS$_2$ boundary was computed in Ref.~\cite{Faulkner09,Iqbal:2011ae}, and was found to have the same spectral asymmetry as that in Eq.~(\ref{GE}), also determined by $\calE$.

Several works \cite{SS10,SS15,JMDS16b,KJ16,HV16,Davison17,Gaikwad:2018dfc,Nayak:2018qej,Moitra:2018jqs,Chaturvedi:2018uov,Sachdev19,Moitra:2019bub} have discussed the nature of the theory of AdS$_2$ horizons, which is applicable to the higher dimensional black holes at $T \ll 1/R_h$. Under these conditions, all modes which are non-constant on $S_d$ can be ignored, and we can focus on the fluctuations of the AdS$_2$ sector. 
These fluctuations are also described by the $0+1$ dimensional Schwarzian theory \cite{JMDS16b,KJ16,HV16,Davison17} in Eq.~(\ref{Seff}). A recent analysis \cite{Sachdev19} has shown that the parameters $K$, $\gamma$, and $\calE$ appearing in this Schwarzian theory are precisely those specified by the thermodynamic results recalled in this appendix.

\bibliography{ref.bib}

\end{document}